\documentclass[twoside, a4paper, notoc]{JHEP3}
\usepackage{amsfonts}
\usepackage{amssymb}
\usepackage{comment}
\usepackage{epsfig}
\usepackage{graphicx}
\usepackage{amsmath}

\usepackage{axodraw}

\newcommand{\be}{\begin{equation}}
\newcommand{\ee}{\end{equation}}
\newcommand{\bea}{\begin{eqnarray}}
\newcommand{\eea}{\end{eqnarray}}
\newcommand{\nn}{\nonumber}
\newcommand{\pd}{\partial}
\newcommand{\mbb}{\mathbb}

\newcommand{\half}{\frac{1}{2}}
\newcommand{\mc}{\mathcal}

\newcommand{\beqa}{\begin{eqnarray}}
\newcommand{\eeqa}{\end{eqnarray}}
\newcommand{\V}{{\cal{V}}}
\newcommand{\la}{\langle}
\newcommand{\ra}{\rangle}

\title{LARGE Volume String Compactifications at Finite Temperature}

\author{Lilia Anguelova\\ Dept. of Physics, University of Cincinnati\\
Cincinnati, OH 45221, USA\\
E-mail: \email{anguella@ucmail.uc.edu}}
\author{Vincenzo Cal\`o \\ Center for Research in String Theory, Queen Mary, \\ University of London, Mile end Road, London, E1 4NS. \\
E-mail: \email{v.calo@qmul.ac.uk}}
\author{Michele Cicoli \\ DAMTP, Centre for Mathematical Sciences\\
Wilberforce Road, Cambridge, CB3 0WA, UK. \\
E-mail: \email{M.Cicoli@damtp.cam.ac.uk}}

\abstract{We present a detailed study of the finite-temperature
behaviour of the LARGE Volume type IIB flux compactifications. We
show that certain moduli can thermalise at high temperatures.
Despite that, their contribution to the finite-temperature
effective potential is always negligible and the latter has a
runaway behaviour. We compute the maximal temperature $T_{max}$,
above which the internal space decompactifies, as well as the
temperature $T_*$, that is reached after the decay of the heaviest
moduli. The natural constraint $T_*<T_{max}$ implies a lower bound
on the allowed values of the internal volume ${\cal V}$. We find
that this restriction rules out a significant range of values
corresponding to smaller volumes of the order $\mathcal{V}\sim
10^{4}l_s^6$, which lead to standard GUT theories. Instead, the
bound favours values of the order $\mathcal{V}\sim 10^{15}l_s^6$,
which lead to TeV scale SUSY desirable for solving the hierarchy
problem. Moreover, our result favours low-energy inflationary
scenarios with density perturbations generated by a field, which
is not the inflaton. In such a scenario, one could achieve both
inflation and TeV-scale SUSY, although gravity waves would not be
observable. Finally, we pose a two-fold challenge for the solution
of the cosmological moduli problem. First, we show that the heavy
moduli decay before they can begin to dominate the energy density
of the Universe. Hence they are not able to dilute any unwanted
relics. And second, we argue that, in order to obtain thermal
inflation in the closed string moduli sector, one needs to go
beyond the present EFT description.}
\preprint{QMUL-PH-09-09 \\ DAMTP-2008-29}

\begin{document}

\tableofcontents

\bigskip

%\newpage

\section{Introduction}

The low energy effective action of string compactifications on
Calabi-Yau (CY) 3-folds typically has a large number of uncharged
massless scalar fields with a flat potential, called moduli. This
has been a long-standing problem for string phenomenology, as the
values of those moduli determine the various parameters (like
masses and coupling constants) of the four-dimensional effective
description. Hence, the presence of these massless scalars with
effective gravitational coupling would lead to unobserved long
range fifth forces, as well as a lack of predictability of the
theory.

The last decade has seen a lot of progress towards the resolution
of this problem. A major ingredient in these developments was the
realization of \cite{gvw,drs,gkp} that nonzero background fluxes
induce potentials for some of the moduli. In fact, in type IIA all
geometric moduli can be stabilized in this way \cite{IIAmodst}. In
type IIB, on the other hand, one also needs to take into account
various perturbative and non-perturbative effects \cite{kklt,LVS}.
Hence it may seem that IIA compactifications are under better
control. The reason this is not so is that the backreaction of the
fluxes on the geometry is more severe in type IIA than in type
IIB. As a result, generically in type IIA one has to consider as
internal spaces manifolds with $SU(3)\times SU(3)$ structure (see
\cite{Grana} for a comprehensive review). The latter are
mathematically much more involved than a Calabi-Yau and so are, in
principle, much harder to study. In contrast, in type IIB there is
a huge class of solutions, such that the backreaction of the
fluxes is entirely encoded by a warp factor. Naturally then, it is
within this class of IIB CY flux compactifications that moduli
stabilization is best understood at present.

An excellent example of IIB compactifications with stabilized
moduli, which is very appealing both for particle physics
phenomenology and for cosmology, is given by the LARGE Volume
Scenarios (LVS) originally proposed in \cite{LVS}. According to
the general analysis of \cite{ccq2}, in these compactifications,
$\alpha'$ and $g_s$ corrections are combined with non-perturbative
effects to generate a potential for the K\"{a}hler moduli, whereas
the background fluxes induce a potential for the dilaton and the
complex structure moduli. Unlike the KKLT set-up \cite{kklt}, the
moduli stabilisation is performed without fine tuning of the
values of the internal fluxes and the CY volume is fixed at an
exponentially large value (in string units). As a consequence, one
has a very reliable four-dimensional effective description, as
well as a tool for the generation of phenomenologically desirable
hierarchies.

\begin{figure}[t]
\begin{center}
\scalebox{0.9}{\includegraphics{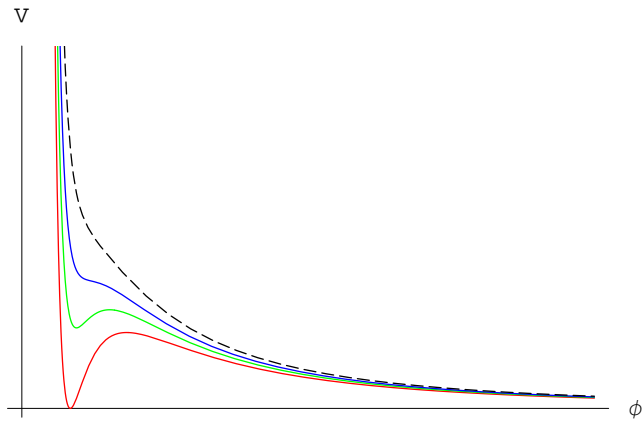}}
\end{center}
\vspace{-0.5cm} \caption{The effective potential $V$ versus the
volume modulus $\phi$ for a typical potential of KKLT or LARGE
Volume compactifications. The different curves show the effect of
various sources of energy that, if higher than the barrier of the
potential, can lead to a decompactification of the internal
space.} \label{Fig:decomp}
\end{figure}

The exponentially large volume minimum of LVS is AdS with broken
SUSY, even before any uplifting. In contrast, in KKLT
constructions the AdS minimum is supersymmetric and the uplifting
term is the source of SUSY breaking. In both cases, however, in
addition to this minimum there is always the supersymmetric one at
infinity. The two minima are separated by a potential barrier
$V_b$, whose order of magnitude is very well approximated by the
value of the potential at the AdS vacuum before uplifting. As is
well-known, the modulus, related to the overall volume of the
Calabi-Yau, couples to any possible source of energy, due to the
Weyl rescaling of the metric needed to obtain a 4D supergravity
effective action in the Einstein frame. Thus, in the presence of
any source of energy, greater then the height of the potential
barrier, the system will be driven to a dangerous
decompactification limit. For example, during inflation the energy
of the inflaton $\varphi$ could give an additional uplifting term
of the form $\Delta
V(\varphi,\mathcal{V})=V(\varphi)/\mathcal{V}^n$ for $n>0$, that
could cause a run-away to infinity \cite{linde}. Another source of
danger of decompactification is the following. After inflation,
the inflaton decays to radiation and, as a result, a
high-temperature thermal plasma is formed. This gives rise to
temperature-dependent corrections to the moduli potential, which
could again destabilise the moduli and drive them to infinity, if
the finite-temperature potential has a run-away behaviour. The
decompactification temperature, at which the finite-temperature
contribution starts dominating over the $T=0$ potential, is very
well approximated by $T_{max}\sim V_b^{1/4}$ since $V_T\sim T^4$.
Clearly, $T_{max}$ sets also an upper bound on the reheating
temperature after inflation. The discussion of this paragraph is
schematically illustrated on Figure \ref{Fig:decomp}.

On the other hand, if, instead of having a run-away behaviour, the
finite-temperature potential develops new minima, then there could
be various phase transitions, which might have played an important
role in the early Universe and could have observable signatures
today. The presence of minima at high $T$ could also have
implications regarding the question how natural it is for the
Universe to be in a metastable state at $T=0$. More precisely,
recent studies of various toy models
\cite{craig,abel,fischler,ART,papineau} have shown that, despite
the presence of a supersymmetric global minimum, it is
thermodynamically preferable for a system starting in a high $T$
minimum to end up at low temperatures in a (long-lived) local
metastable minimum with broken supersymmetry. Similar arguments,
if applicable for more realistic systems, could be of great
conceptual value given the present accelerated expansion of the
Universe.

For cosmological reasons then, it is of great importance to
understand the full structure of the finite temperature effective
potential. We investigate this problem in great detail for the
type IIB LVS of \cite{LVS}. Contrary to the traditional thought
that moduli cannot thermalize due to their Planck-suppressed
couplings to ordinary matter and radiation, we show that in LVS
some of the moduli {\it can} be in thermal equilibrium with MSSM
particles for temperatures well below the Planck scale. The main
reason is the presence of an additional large scale in this
context, namely the exponentially large CY volume, which enters
the various couplings and thus affects the relevant interaction
rates. The unexpected result, that some moduli can thermalize, in
principle opens up the possibility that the finite temperature
potential could develop new minima instead of just having a
run-away behaviour as, for example, in \cite{BHLR}. However, we
show that this is not the case since, for temperatures below the
Kaluza-Klein scale, the $T$-dependent potential still has a
run-away behaviour. Although it is impossible to find exactly the
decompactification temperature $T_{max}$, as it is determined by a
transcendental equation, we are able to extract a rather precise
analytic estimate for it. As expected, we find that $T_{max}$ is
controlled by the SUSY breaking scale: $T_{max}^4\sim m_{3/2}^3
M_P$. This expression gives also an upper bound on the temperature
in the early Universe. We show that this constraint can be
translated into a lower bound on the value of the CY volume, by
computing the temperature of the Universe $T_*$, just after the
heaviest moduli of LVS decay, and then imposing $T_*<T_{max}$.

Our lower bound implies that, for cosmological reasons, larger
values of the volume of the order $\mathcal{V}\sim 10^{15}l_s^6$,
which naturally lead to TeV scale supersymmetry, are favoured over
smaller values of the order $\mathcal{V}\sim 10^{4}l_s^6$, which
lead to standard GUT theories. More precisely, what we mean by
this is the following. Upon writing the volume as $\mathcal{V}\sim
10^{x}$ and encoding the fluxes and the Calabi-Yau topology in the
definition of a parameter $c$, we are able to rule out a
significant portion of the $(x,c)$-parameter space that
corresponds to small $x$ (for example, for $c=1$ we obtain $x>6$).
This is rather intriguing, given that other cosmological
considerations seem to favour smaller values of the volume.
Indeed, the recent inflationary model of \cite{fibInfl} requires
$\mathcal{V}\sim 10^{4}l_s^6$, in order to generate the right
amount of density perturbations. Despite that, our lower bound on
${\cal V}$ does not represent an unsurmountable obstacle for the
realization of inflation. The reason is that the Fibre Inflation
model can give rise to inflation even for large values of the
volume. Hence, a modification of it, such that the density
fluctuations are generated by a curvaton-like field different from
the inflaton, would be a perfectly viable model with large ${\cal
V}$. The large value of $\mathcal{V}$ would imply a low-energy
inflationary scale, and so, in turn, gravity waves would not be
observable. However, it is likely that both inflation and TeV
scale SUSY could be achieved at the same time, with also the
generation of a relevant amount of non-gaussianities in the CMB,
which is a typical feature of curvaton models.

On the other hand, we pose a challenge for the solution of the
cosmological moduli problem, that the overall breathing mode of
LVS with $\mathcal{V}> 10^{10}l_s^6$ is afflicted by \cite{CQ}.
This is so, because we show that unwanted relics cannot be diluted
by the entropy released by the decay of the heaviest moduli of
LVS, nor by a low-energy period of thermal inflation. More
precisely, we show that the heaviest moduli of LVS decay before
they can begin to dominate the energy density of the Universe and,
also, that in order to study thermal inflation in the closed
string moduli sector, it is necessary to go beyond our low energy
EFT description.

The present paper is organized as follows. In Section
\ref{Sec:Review}, we review the basic features of the type IIB
LVS. In Section \ref{EffPotFT}, we recall the general form of the
effective potential at finite temperature and discuss in detail
the issue of thermal equilibrium in an expanding Universe. In
Section \ref{Mmc}, we derive the masses and the couplings to
visible sector particles of the moduli and modulini in LVS. Using
these results, in Section \ref{Sec:ModuliTherm} we investigate
moduli thermalization and show that, generically, the moduli
corresponding to the small cycles can be in thermal equilibrium
with MSSM particles, due to their interaction with the gauge
bosons. In Section \ref{FTCLVS}, we study the finite temperature
effective potential in LVS. We show that it has a runaway
behaviour and find the decompactification temperature $T_{max}$.
Furthermore, we establish a lower bound on the CY volume, which
follows from the constraint that the temperature of the Universe
just after the small moduli decay should not exceed $T_{max}$. In
Section \ref{Dis} we discuss some open issues, among which the
question why thermal inflation does not occur within our
approximations. Finally, after a summary of our results in Section
\ref{Con}, Appendix \ref{App:ModuliCouplings} contains technical
details on the computation of the moduli couplings.

\section{Large Volume Scenarios}
\label{Sec:Review}

The distinguishing feature of the type IIB Large Volume Scenarios
(LVS) of \cite{LVS} is that, in addition to the non-perturbative
effects considered in \cite{kklt}, they also take into account
$\alpha'$ corrections, which lead to moduli stabilisation at an
exponentially large volume of the internal manifold. In
\cite{LVS}, the Calabi-Yau 3-fold was assumed to have a
characteristic topology with one exponentially large cycle and
several small del Pezzo 4-cycles. Including string loop
corrections, as in \cite{ccq2}, extends these scenarios to a
larger class of Calabi-Yau manifolds which can also have fibration
structure. The exponentially large volume allows to explain many
hierarchies observed in nature and guarantees that the low-energy
effective field theory is under good control.

In this Section we summarise the basic ingredients of LVS. We
begin by recalling necessary material about type IIB flux
compactifications. After that we turn to the relevant perturbative
and non-perturbative effects and the resulting scalar potential
with a LARGE volume minimum.

\subsection{Type IIB with fluxes}
\label{2}

We will be interested in type IIB CY orientifold compactifications
with background fluxes, which preserve $\mathcal{N} = 1$
supersymmetry in 4D \cite{gkp}. The effective low-energy
description is then given by a $d=4$, $\mathcal{N} = 1$
supergravity characterised by a K\"{a}hler potential $K$, a
superpotential $W$ and a gauge kinetic function $f_{ab}$, where
the indices $a,b$ run over the various vector multiplets. In
particular, the scalar potential of this theory has the standard
form:
\begin{equation}
V = e^{K/M_P^2} \left( K^{i\bar{j}} D_i W D_{\bar{j}}\bar{W} - 3
\frac{|W|^2}{M_P^2} \right), \label{pot}
\end{equation}
where $i,j$ run over all moduli of the compactification.
Generically, the latter consist of the axio-dilaton $S=e^{-\phi
}+i C_{0}$, $h_{1,1}$ K\"{a}hler ($T$-moduli) and $h_{2,1}$
complex structure moduli ($U$-moduli). The tree level K\"{a}hler
potential has the following form:
\begin{equation}
  \frac{K_{tree}}{M_P^2}=-\ln \left(S+\bar{S}\right) -2\ln \mathcal{V}
  -\ln \left( -i\int_{CY}\limits\Omega \wedge
  \bar{\Omega}\right), \label{eqtree}
\end{equation}
where $\mathcal{V}$ is the Einstein frame CY volume, in units of
the string length $l_{s} = 2\pi \sqrt{\alpha'}$, and $\Omega $ is
the CY holomorphic (3,0)-form.
%Clearly, the integral
%in (\ref{eqtree}) is over the internal six dimensions.
Note that the the $T$-moduli enter $K_{tree}$ only through ${\cal
V}$ and the $U$-moduli only through $\Omega$. For later purposes,
it will be useful to recall a couple of relations regarding the
K\"{a}hler moduli. Expanding the K\"{a}hler form $J =
\sum_{i=1}^{h_{1,1}} t^{i} D_{i}$ in a basis $\{ D_{i} \}$ of
$H^{1,1}(CY,\mathbb{Z})$ and considering orientifold projections
such that $h_{1,1}^{-} = 0 \, \Rightarrow \, h_{1,1}^{+} =
h_{1,1}$, we obtain:
\begin{equation}
\mathcal{V} = \frac{1}{6}\int_{CY} J\wedge J\wedge J = \frac{1}{6}
k_{ijk}t^{i}t^{j}t^{k} \, , \label{IlVolume}
\end{equation}
where $k_{ijk}$ are related to the triple intersection numbers of
the CY and the $t^i$ are 2-cycle volumes. The volumes of the
Poincar\'{e} dual 4-cycles are given by:
\begin{equation}
\tau _{i}=\frac{\partial \mathcal{V}}{\partial t^{i}}=
\frac{1}{2}\int\limits_{CY} D_{i}\wedge J\wedge J= \frac{1}{2}
k_{ijk}t^{j}t^{k} \, . \label{defOfTau}
\end{equation}
Finally, the scalar components of the chiral superfields,
corresponding to the K\"{a}hler moduli, that enter the 4D
effective action are $T_i = \tau_i + i b_i$, where the axions
$b_i$ are the components of the RR 4-form $C_4$ along the 4-cycle
Poincar\'{e} dual to $D_i$. Obviously, from (\ref{IlVolume}) and
(\ref{defOfTau}) one can express ${\cal V}$ in (\ref{eqtree}) as a
function of $\tau_i = \frac{1}{2} (T_i + \overline{T}_i)$.

Now, turning on background fluxes $G_3 = F_3 + i S H_3$, where
$F_3$ and $H_3$ are respectively the RR and NSNS 3-form fluxes of
type IIB supergravity (for recent reviews on flux
compactifications, see \cite{fluxes}), generates an effective
superpotential of the form \cite{gvw}\footnote{We neglect the
effect of warping, generated by non-zero background fluxes, since
we will be considering CY compactifications with large
internal volume.}:
\be W_{tree}\sim\int\limits_{CY}G_{3}\wedge \Omega.
\label{yujtree} \ee
As a result, one can stabilise the axio-dilaton $S$ and the
$U$-moduli. However, the K\"{a}hler moduli $T_i$ do not enter
$W_{tree}$ and therefore remain massless at leading semiclassical
level. One can obtain an effective description for these fields by
integrating out $S$ and $U$.\footnote{For more details on the
consistent supersymmetric implementation of this procedure, see
\cite{IntOut}.} Then the superpotential is constant, $W = \langle
W_{tree} \rangle\equiv W_0$, and the K\"{a}hler potential reads
$K=K_{cs} - \ln\left({2}/{g_{s}}\right) + K_0$ with
\be
K_0 = -2\ln \mathcal{V} \qquad \hbox{and} \qquad e^{-K_{cs}} =
\left\langle -i\int_{CY} \Omega \wedge \bar{\Omega}\right\rangle
\, . \label{K0expr}
\ee
Before concluding this Subsection, let us make a couple of
remarks. First, note that the background flux $G_3$ may or may not
break the remaining 4D $\mathcal{N}=1$ supersymmetry, depending on
whether or not $D_\alpha W = \partial_\alpha W + W \partial_\alpha
K$ vanishes at the minimum of the resulting scalar potential.
Second, the K\"{a}hler potential $K_0$ satisfies the no-scale
property $K_0^{i\bar\jmath} \partial_i K_0
\partial_{\bar\jmath} K_0 = 3$, which implies that the scalar
potential for the K\"{a}hler moduli vanishes in accord with the
statement above that the fluxes do not stabilise those moduli.

\subsection{Leading order corrections} \label{22}

As we recalled in the previous Subsection, at leading
semiclassical order the scalar potential for the $T$-moduli
vanishes. So, unlike the situation for the $S$ and $U$-moduli, in
order to stabilise $T_i$ one has to consider the leading order
corrections to the tree level action. The first such corrections
to be studied were non-perturbative contributions to $W$. Recall
that there is a non-renormalisation theorem forbidding $W$ to be
corrected perturbatively. On the other hand, the K\"{a}hler
potential $K$ does receive perturbative corrections both in
$\alpha'$ and in $g_s$. Therefore, non-perturbative effects are
subleading in $K$ and we shall neglect them in the following. Let
us now review briefly all these kinds of corrections.

\subsubsection{Non-perturbative corrections}

Non-perturbative corrections to the superpotential can be due to
Euclidean D3 brane (ED3) instantons wrapping 4-cycles in the extra
dimensions, or to gaugino condensation in the supersymmetric gauge
theories located on D7 branes that also wrap internal 4-cycles.
The superpotential that both kinds of effects generate is of the
form:\footnote{The prefactor in (\ref{yuj}) is due to careful
dimensional reduction, as can be seen in Appendix A of
\cite{LVS2}. However, the authors of \cite{LVS2} define the
Einstein metric via
$g_{\mu\nu,s}=e^{(\phi-\langle\phi\rangle)/2}g_{\mu\nu,E}$\,, so that
it coincides with the string frame metric in the physical vacuum.
On the contrary, we opt for the more traditional definition
$g_{\mu\nu,s}=e^{\phi/2}g_{\mu\nu,E}$, which implies no factor of
$g_s$ in the prefactor of $W$.}
\begin{equation}
  W=\frac{M_P^3}{\sqrt{4
\pi}}\left(W_{0}+\sum\limits_{i}A_{i}e^{-a_{i}T_{i}}\right),
 \label{yuj}
\end{equation}
where the coefficients $A_{i}$ correspond to threshold effects
and, in principle, can depend on $U$ and D3 position moduli, but
not on $T_i$. The constants $a_i$ are given by $a_{i}=2\pi $ for
ED3 branes, and by $a_{i}=2\pi /N$ for gaugino condensation in an
$SU(N)$ gauge theory. Note that one can neglect multi-instanton
effects in (\ref{yuj}), as long as $\tau_i$ are stabilised such
that $a_{i}\tau_{i}\gg 1$. From (\ref{pot}), the above
superpotential leads to the following $T_i$-dependent contribution
to the scalar potential (up to a numerical prefactor and powers of
$g_s$ and $M_P$, that we will be more precise about below):
\begin{eqnarray}
 \delta V_{(np)} &=& e^{K_0}  K_{0}^{j\bar\imath} \Bigl[
 a_{j}A_{j} \, a_{i} \bar{A}_{i} e^{-\left(
 a_{j}T_{j} + a_{i}\overline{T}_{i}\right) } \nonumber\\
 && \qquad\qquad  -\left(
 a_{j}A_{j}e^{-a_{j}T_{j}}\overline{W}
 \partial_{\bar\imath} K_{0}+a_{i}\bar{A}_{i} e^{-a_{i}\overline{T}_{i}} W
 \partial_j K_{0}\right) \Bigr].  \label{scalarWnp}
\end{eqnarray}

\subsubsection{$\alpha'$ corrections}

The K\"{a}hler potential receives corrections at each order in the
$\alpha'$ expansion. In the effective supergravity description
these correspond to higher derivative terms. The leading $\alpha'$
contribution comes from the $R^4$ term and it leads to the
following K\"{a}hler potential \cite{BBHL}:
\begin{equation}
  \frac{K}{M_P^2}= -2\ln \left( \mathcal{V}+\frac{\xi }{2g_{s}^{3/2}}\right)
  \simeq -2\ln\mathcal{V}-\frac{\xi}{g_s^{3/2}\mathcal{V}}.
 \label{eq}
\end{equation}
Here $\xi$ is given by $ \xi =-\frac{\chi \zeta (3)}{2(2\pi
)^{3}}$, where $\chi=2\left(h_{1,1}-h_{2,1}\right)$ is the CY
Euler number, and the Riemann zeta function is $\zeta(3) \simeq
1.2$. Denoting for convenience $\hat{\xi}\equiv \xi/g_s^{3/2}$,
eq.~(\ref{eq}) implies to leading order the following contribution
to $V$ (again, up to a prefactor containing powers of $g_s$ and $M_P$):
\begin{equation}
 \delta V_{(\alpha')} = 3e^{K_0}\hat{\xi}\frac{\left( \hat{\xi}^{2}+7\hat{\xi}
 \mathcal{V}+\mathcal{V}^{2}\right) }{\left( \mathcal{V}-\hat{\xi}
 \right) \left( 2\mathcal{V}+\hat{\xi} \right)^{2}} W_0^2
 \simeq \frac{3\xi W_0^2}{4 g_s^{3/2}\mathcal{V}^3},
 \label{scalaralpha'}
\end{equation}
where $\mathcal{V} \gg \hat\xi \gg 1$ in order for perturbation
theory to be valid.

\subsubsection{$g_s$ corrections} \label{5}

The K\"{a}hler potential receives also string loop corrections. At
present, there is no explicit derivation of these corrections from
string scattering amplitudes for a generic CY compactification.
Nevertheless, it has been possible to conjecture their form
indirectly \cite{07040737}:
\begin{equation} \label{Kgs}
 \delta K_{(g_{s})}=\delta K_{(g_{s})}^{KK}+\delta K_{(g_{s})}^{W} \,,
\end{equation}
where
\begin{equation}
 \delta K_{(g_{s})}^{KK}\sim \sum\limits_{i=1}^{h_{1,1}}
 \frac{g_s \mathcal{C}_{i}^{KK}(U,\bar{U})\left( a_{il}t^{l}\right)
 }{\mathcal{V}},  \label{UUU}
\end{equation}
and
\begin{equation}
 \delta K_{(g_{s})}^{W}\sim
 \sum\limits_{i}\frac{\mathcal{C} _{i}^{W}(U,\bar{U})}{\left(
 a_{il}t^{l}\right) \mathcal{V}}. \label{UUUU}
\end{equation}
In (\ref{Kgs}), $\delta K_{(g_{s})}^{KK}$ comes from the exchange
of closed strings, carrying  Kaluza-Klein momentum, between D7-
and D3-branes. The expression (\ref{UUU}) is valid for vanishing
open string scalars and is based on the assumption that all the
$h_{1,1}$ 4-cycles of the CY are wrapped by D7-branes. The other
term $\delta K_{(g_{s})}^{W}$ in (\ref{UUU}) is due, from the
closed string perspective, to the exchange of winding strings
between intersecting stacks of D7-branes.

In addition, in (\ref{UUU}) the linear combination
$\left(a_{il}t^l\right)$ of the volumes of the basis 2-cycles is
transverse to the 4-cycle wrapped by the $i$-th D7-brane, whereas
in (\ref{UUUU}) it gives the 2-cycle where two D7-branes
intersect. The functions $\mathcal{C}^{KK}_i (U, \bar{U})$ and
$\mathcal{C}^W_i (U, \bar{U})$ are, in principle, unknown.
However, for our purposes they can be viewed as $\mathcal{O}(1)$
constants\footnote{In the
$T^6/\left(\mbb{Z}_2\times\mbb{Z}_2\right)$ orientifold case,
where these constants can be computed explicitly \cite{Berg}, they
turn out to be, in our conventions, of $\mathcal{O}(1)$ for natural
values of the complex structure moduli:
$\textrm{Re}(U)\sim\textrm{Im}(U)\sim\mathcal{O}(1)$. Note that
\cite{Berg} uses conventions different from ours.} since the
complex structure moduli are already stabilised at the classical
level by background fluxes.

Comparing (\ref{UUU}) with (\ref{eq}), we notice that $\delta
K^{KK}_{(g_s)}$ is generically leading with respect to $\delta
K_{(\alpha')}$ due to the presence of the linear combination
$\left(a_{il}t^l\right)$ in the numerator of (\ref{UUU}) and the
fact that each 4-cycle volume $\tau_i=\frac{1}{2}k_{ijk}t^j t^k$
has to be fixed larger than the string scale in order to trust the
effective field theory. However, in ref. \cite{cicq} it was discovered that
for an arbitrary CY background, the leading contribution of
(\ref{UUU}) to the scalar potential is vanishing, so leading to an
\textit{extended no-scale structure}. This result renders $\delta
V_{(g_s)}$ generically subleading with respect to $\delta
V_{(\alpha')}$ and the final formula can be expressed as
(neglecting a prefactors with powers of $g_s$ and $M_P$):
\begin{equation}
 \delta V_{\left( g_{s}\right) }^{1-loop}=\left[
 \left( g_s \mathcal{C}_{i}^{KK} \right)^2 a_{ik} a_{ij}
 K_{k\bar\jmath}^{0} - 2 \delta
 K_{(g_{s})}^{W}\right] \frac{W_{0}^{2}}{\mathcal{V}^{2}}.
 \label{V at 1-loop}
\end{equation}
Notice that for branes wrapped only around the basis 4-cycles the
combination appearing in the first term degenerates to $a_{ik}
a_{ij} K^0_{k\bar\jmath} = K^0_{i\bar\imath}$. The fact that
$\delta V_{(g_s)}$ is generically subleading with respect to
$\delta V_{(\alpha')}$ can be easily seen by recalling the generic
expression of the tree-level K\"{a}hler metric for an arbitrary
CY:
$K^0_{i\bar{j}}=t_it_{\bar{j}}/(2\mc{V}^2)-(k_{i\bar{j}k}t^k)^{-1}/\mc{V}\sim
1/(\mc{V}t)$. Therefore the ratio between $\delta V_{(\alpha')}$
given by (\ref{scalaralpha'}) (with $\xi\sim\mc{O}(1)$ for known
CY three-folds) and the expression (\ref{V at 1-loop}) for $\delta
V_{(g_s)}$, scales as $\delta V_{(\alpha')}/\delta V_{(g_s)}\sim
g_s^{-3/2}t\gg 1$ due to the fact that the size of each two-cycle has to be
fixed at a value larger than $1$ (in string units) in order to trust the effective
field theory and, in addition, the string coupling has to be smaller than unity
in order to be in the perturbative regime.

\subsection{Moduli stabilisation in LVS} \label{ELV}

Combining (\ref{scalarWnp}), (\ref{scalaralpha'}) and (\ref{V at
1-loop}), we obtain the final form of the K\"{a}hler moduli
effective scalar potential, that contains all the leading order
corrections to the vanishing tree-level part (with all prefactors included):
\begin{eqnarray}
V &=&V_{tree}+\delta V_{(np)}+\delta V_{(\alpha ^{\prime
})}+\delta
V_{(g_{s})}^{1-loop}=  \notag \\
&=&\frac{g_s e^{K_{cs}}M_P^4}{8\pi\mathcal{V}^{2}}\left\{ K_{0}^{j\bar{\imath}%
}a_{j}A_{j}\,a_{i}\bar{A}_{i}e^{-\left( a_{j}T_{j}+a_{i}\overline{T}%
_{i}\right) }+4W_{0}\sum\limits_{i}a_{i}A_{i}\tau _{i}\cos
(a_{i}b_{i})e^{-a_{i}\tau _{i}}\right.   \notag \\
&&+\left. \left[ \frac{3\xi}{g_s^{3/2}}+\sum\limits_{i}\left( g_{s}^{2}\left( \mathcal{%
C}_{i}^{KK}\right) ^{2}\left( \frac{1}{2}\frac{t_{i}^{2}}{\mathcal{V}}%
-A^{ii}\right) -8\frac{\mathcal{C}_{i}^{W}}{\left( a_{il}t^{l}\right) }%
\right) \right] \frac{W_{0}{}^{2}}{4\mathcal{V}}\right\},
\label{totale}
\end{eqnarray}
where $A_{ij}\equiv \frac{\partial\tau_i}{\partial
t_j}=k_{ijk}t^k$. Ref.~\cite{ccq2} derived the topological
conditions that an arbitrary CY has to satisfy in order for the
general potential (\ref{totale}) to have a non-supersymmetric AdS
minimum at {\it exponentially large} volume. These conditions can
be summarised as follows:
\begin{enumerate}
\item{} $h_{2,1} > h_{1,1} > 1$ $\Leftrightarrow$
$\xi\sim(h_{2,1}-h_{1,1})>0$,
\item{} The CY 3-fold has to have at least one K\"{a}hler modulus
corresponding to a blow-up mode resolving a point-like singularity
(the volume of a del Pezzo 4-cycle).
\end{enumerate}
These conditions give rise to two different LVS. In the first
case, the CY has a typical Swiss-cheese topological structure
where there is just one LARGE cycle controlling the size of the
overall volume and all the other 4-cycles are blow-up modes. The
string loop corrections turn out to be negligible. On the other
hand, in the case of fibred CY manifolds, with the presence of
modes which do not resolve point-like singularities or correspond
to the overall volume modulus, $g_s$ corrections play a crucial
role to lift the fibration moduli that are left unfixed by
$\delta V_{(np)}+\delta V_{(\alpha')}$ (more precisely, the string
loop corrections for these modes are always dominant compared to
non-perturbative effects). Let us now describe these two cases in
more detail.

\subsubsection{Swiss-cheese Calabi-Yaus}
\label{Sec:Swiss-CheeseCY}

The original example of LVS of \cite{LVS}, and the one we will
%be most concerned with here,
study in most detail, is the degree 18 hypersurface in
$\mathbb{C}P^{4}_{[1,1,1,6,9]}$ whose volume is given by
\begin{equation} \label{Vbs}
 \mathcal{V} = \frac{1}{9\sqrt{2}}\left(\tau_{b}^{3/2}
 -\tau_{s}^{3/2}\right),
\end{equation}
where $\tau_b$ and $\tau_s$ are the two K\"{a}hler moduli and the
subscripts $b$ and $s$ stand for \textit{big} and \textit{small}
respectively. The general expression (\ref{totale}) for the scalar
potential, in this case takes the form
\begin{equation}
V =\frac{g_s e^{K_{cs}}M_P^4}{8\pi}\left( \lambda \sqrt{\tau_s}
\frac{e^{-2a_s \tau_s}}{\cal V} - \mu \frac{\tau_s e^{-a_s
\tau_s}}{{\cal V}^2} + \frac{\nu}{{\cal
V}^3}+\frac{\sigma}{\mathcal{V}^3\sqrt{\tau_s}}+\frac{\rho}{\mathcal{V}^{10/3}}\right),
\label{Vsimple}
\end{equation}
with
\begin{equation}
\lambda = 24 \sqrt{2} a_s^2 A_s^2,\text{ \ }\mu = 4 a_s A_s W_0,
\text{ \ }\nu = \frac{3\xi W_0^2}{4 g_s^{3/2}},\text{ \ }
\sigma=\left(g_s\mathcal{C}^{KK}_s W_0\right)^2,\text{ \ }
\rho=\frac{\left(g_s\mathcal{C}^{KK}_b
W_0\right)^2}{\left(9\sqrt{2}\right)^{1/3}}. \label{numbers}
\end{equation}
Also, in (\ref{Vsimple}) the minimisation with respect to the
axion $b_s$ has already been performed. For natural values of the
tree-level superpotential $W_{0}\sim\mathcal{O}(1)$, the scalar
potential (\ref{Vsimple}) admits a non-supersymmetric AdS minimum
at exponentially large volume due to the interplay of $\alpha'$
and non-perturbative effects. This minimum is located at
\begin{equation}
 \mathcal{V} \sim W_{0} \, e^{a_{s}\tau_{s}} \gg
 \tau_{s} \sim \hat{\xi}^{2/3} \gg 1 \,.
\end{equation}
The string loop corrections can be safely neglected since they are
subdominant relative to the other corrections due to inverse
powers of $\mathcal{V}$ and factors of $g_s$. There are several
ways to up-lift this minimum to Minkowski or dS: adding
$\overline{D3}$ branes \cite{kklt}, considering D-terms from
magnetised D7 branes \cite{bkq} or F-terms from a hidden sector
\cite{ss} etc.

An immediate generalisation of the
$\mathbb{C}P^{4}_{[1,1,1,6,9]}$ model is given by the so
called `Swiss-cheese' Calabi-Yaus, whose volume looks like
\begin{equation} \label{SCmsm}
 \mathcal{V}=\alpha\left(\tau_{b}^{3/2}
 -\sum_{i=1}^{N_{small}}\lambda_{i}\tau_{i}^{3/2}\right),\text{ \ }
 \alpha > 0,\text{ \ }\lambda_{i} > 0\text{ \ }\forall
 i=1,...,N_{small}.
\end{equation}
Examples having this form with $h_{1,1}=3$ are the the degree 15
hypersurface embedded in $\mathbb{C}P^{4}_{[1,3,3,3,5]}$ and the
degree 30 hypersurface in $\mathbb{C}P^{4}_{[1,1,3,10,15]}$
\cite{blumenhagen}. More generally, in ref. \cite{blum} it was proved that
examples of Swiss-cheese CY 3-folds with $h_{1,1}=n+2$, $0\leq
n\leq 8$, can be obtained by starting from elliptically fibred CY
manifolds over a del Pezzo $dP_n$ base\footnote{A del Pezzo $dP_n$
surface is obtained by blowing-up $\mathbb{C}P^2$ (or
$\mathbb{C}P^1\times\mathbb{C}P^1$ ) on $0\leq n\leq 8$ points.},
and then performing particular flop transitions that flop away all
$n$ $\mathbb{C}P^1$-cycles in the base.
In this case, assuming that all the small cycles get
non-perturbative effects, the 4-cycle $\tau_{b}$, controlling the
overall size of the CY, is stabilised exponentially large,
$\mathcal{V} \simeq \alpha \tau_b^{3/2} \sim W_0 e^{a_{i}
\tau_{i}}$, while the various 4-cycles, $\tau_{i}$, controlling
the size of the `holes' of the Swiss-cheese, get fixed at small
values $\tau_{i} \sim \mathcal{O}(10)$, $\forall
i=1,...,N_{small}$.
However, in  ref. \cite{collinucci} it was discovered that the Swiss-cheese
structure of the volume is not enough to guarantee that all the
rigid `small' cycles $\tau_i$ can indeed be stabilised small. In
fact, a further condition is that each rigid `small' cycle
$\tau_i$ must be del Pezzo. In \cite{collinucci}, there are 3
examples of Swiss-cheese CY 3-folds with $h_{1,1}=4$ where just
one 4-cycle has the topology $\mathbb{C}P^2$ (and so it is
$dP_0$).

\subsubsection{Fibred Calabi-Yaus}
\label{Sec:K3FibrCY}

The first examples of LVS with a topological structure more
complicated than the Swiss-cheese one, were discovered in
\cite{ccq2}. The authors focused on a K3 fibred CY with $h_{2,1}>
h_{1,1} = 3$, obtained by adding a blow-up mode to the geometry
$\mathbb{C}P^{4}_{[1,1,2,2,6]}$. The volume reads:
\begin{equation}
 \mathcal{V} = \alpha \left( \sqrt{\tau _{1}}\tau _{2} - \gamma \tau _{s}^{3/2}\right)
 = t_1\tau_1-\alpha\gamma\tau_s^{3/2},  \label{hhh}
\end{equation}
where the constants $\alpha $ and $\gamma $ are positive, and
$t_1$ is the volume of the $\mathbb{C}P^1$ base of the K3
fibration. Working in the parameter regime $\tau_2 > \tau_1 \gg
\tau_s$, where the volume of the CY is large, while the blow-up
cycle $\tau_s$ remains comparatively small\footnote{in this limit
$t_1 \sim \tau_2/\sqrt{\tau_1} > \sqrt{\tau_1}$, corresponding to
interesting geometries having the two dimensions of the base,
spanned by the cycle $t_1$, larger than the other four of the K3
fibre, spanned by $\tau_1$.}, the general expression
(\ref{totale}) for $V$ becomes (having already minimised $V$ with
respect to the axion $b_s = \hbox{Im} \, T_s$):
\begin{equation}
 V = \frac{g_s e^{K_{cs}}M_P^4}{8\pi}\left[\beta
 \sqrt{\tau _{s}}\frac{e^{-2a_{s}\tau_s}}{\mathcal{V}}
 -\mu \frac{\tau _{s}e^{-a_{s}\tau_s}}{\mathcal{V}^{2}}
  +\frac{\nu}{\mathcal{V}^{3}}+\left( \frac{A}{\tau_1^2} -
\frac{B}{{\cal V} \sqrt{\tau_1}} + \frac{C \tau_1}{{\cal V}^2}
\right) \frac{W_0^2}{{\cal V}^2}\right],
 \label{ygfdo}
\end{equation}
with $\lambda$, $\mu$ and $\nu$ as given in (\ref{numbers}) and
\begin{equation}
\beta=\frac{\lambda}{9\sqrt{2}\alpha\gamma} \, , \qquad A = (g_s
C_1^{KK})^2 \, , \qquad B = 4 \alpha C_{12}^W \, , \qquad C = 2
(\alpha g_s C_2^{KK})^2.
\end{equation}
It is evident that the leading $\delta V_{(\alpha')} + \delta
V_{(np)}$ part of the potential depends only on two K\"{a}hler
moduli, ${\cal V}$ and $\tau_s$, instead of all three. And, in
fact, it turns out to be of exactly the same form as
(\ref{Vsimple}) above. Hence, viewing ${\cal V}$, $\tau_s$ and
$\tau_1$ as the three independent moduli (instead of $\tau_1$,
$\tau_2$ and $\tau_s$), it is clear that, without
taking into account the subleading $g_s$ corrections, $\tau_1$ is
a flat direction of the scalar potential. Also, it is evident that
at this order, ${\cal V}$ and $\tau_s$ are stabilised as in the
$\mathbb{C}P^{4}_{[1,1,1,6,9]}$ model of Subsection
\ref{Sec:Swiss-CheeseCY}:
\begin{equation}
 \langle \tau _{s}\rangle = \left(
 \frac{\hat\xi}{2\, \alpha \gamma } \right) ^{2/3}
 \qquad \hbox{and} \qquad
 \langle
 \mathcal{V}\rangle = \left( \frac{ 3 \,\alpha \gamma }{4a_{s}A_s}
 \right) W_0 \, \sqrt{\langle \tau _{s}\rangle }
 \; e^{a_{s} \langle \tau_s
 \rangle }\text{\ .}  \label{x}
\end{equation}
Obviously, loop corrections shift insignificantly the VEVs of
these two moduli. However, $g_s$ corrections are crucial to
generate a potential for $\tau_1$ that admits a minimum at
\be
  \frac{1}{\tau_1^{3/2}} = \left( \frac{B}{8 A \V} \right)
  \left[ 1 + (\hbox{sign} \, B) \sqrt{1 + \frac{32 AC}{B^2}}
  \right] \label{tau1soln1}.
\ee

Some concrete numerical choices for the various underlying
parameters, without any fine-tuning, are listed in Table 1. The
`LV' case gives very large volumes, $\V \simeq 10^{13}$  and the
modulus $\tau_1$ is stabilised at hierarchically large values,
$\tau_2 > \tau_1 \gg \tau_s$. The string scale and the gravitino
mass turn out to be
\begin{equation}
M_s=\frac{M_P}{\sqrt{4\pi\mathcal{V}}}\sim 10^{11}\text{GeV, \ \ \
\ }m_{3/2}=e^{K/2M_P^2}\frac{W}{M_P^2}=\frac{g_s^{1/2}
e^{K_{cs}/2}W_0 M_P}{\sqrt{8\pi}\mathcal{V}}\sim 10\text{TeV}.
\end{equation}
This gives a solution of the hierarchy problem but the huge value
of the volume destroys the standard picture of gauge coupling
unification. The `SV' case instead has $\V \sim 10^3$ much smaller
(and so with $M_s \sim M_{GUT}$ and a very high gravitino mass).
This set of parameter values is not chosen just in relation to GUT
theories but also in order to provide observable density
fluctuations for the inflationary model of \cite{fibInfl}. In that
model, the inflaton is the modulus $\tau_1$, whose potential is
loop-generated, and the main feature of the model is that it
produces detectable gravity waves.
\TABULAR[ht]{c||c|c|c|c|c|c|c|c|c|c|c|c|c}
 { & $g_s$ & $\xi$ & $W_0$ & $a_3$ & $A_3$ & $\alpha$ & $\gamma$ & $C^{KK}_1$ & $C^{KK}_2$ & $C^W_{12}$ & $\langle\tau_s\rangle$ & $\langle\tau_1\rangle$ & $\langle\V\rangle$ \\
  \hline\hline
  LV & 0.1 & 0.4 & 1 & $\pi$ & 1 & 0.5 & 0.39 & 0.1 & 0.1 & 5 & 10.5 & $10^6$ & $3 \cdot 10^{13}$ \\
  SV & 0.3 & 0.9 & 100 & $\pi/5$ & 1 & 0.13 & 3.65 & 0.15 & 0.08 & 1 & 4.3 & 9  &
  1710} { Some model parameters.}

More general examples of this kind of LVS have been discovered in
\cite{blum}. These authors noticed that starting from an
elliptically fibred CY over a $dP_n$ base, and then flopping away
only $r<n$ (instead of all $n$) of the $\mathbb{C}P^1$-cycles in
the base, one obtains another elliptically fibred CY (instead of a
Swiss-cheese one), whose volume looks like:
\begin{equation}
 \mathcal{V}=Vol\left(X_{n-r}\right)-\sum_{i=1}^{r}\lambda_{i}\tau_{i}^{3/2},
 \text{ \ }\lambda_{i} > 0\text{ \ }\forall
 i=1,...,r,
\end{equation}
where $X_{n-r}$ is the resulting elliptical fibration over a
$dP_{n-r}$ base. It is natural to expect that the scalar potential
for these examples has an AdS minimum at exponentially large
volume, together with $( h_{1,1} - N_{small} - 1)=n-r$ flat
directions that will be lifted by $g_s$ corrections.

We should note that string loop corrections can play an important
role for compactifications on Swiss-cheese CY manifolds as
well. Namely, they can be crucial, even in this case, to achieve
full moduli stabilisation when the topological condition, that all
rigid 4-cycles be del Pezzo, is not satisfied or when one imposes
the phenomenological condition that the 4-cycles supporting chiral
matter do not get non-perturbative effects
\cite{blumenhagen}.\footnote{Also D-terms could play a significant
role as pointed out still in \cite{blumenhagen}.}

%\section{Decompactification problems for string vacua}
%\label{Sec:DecProbl}
\section{Effective potential at finite temperature}
\label{EffPotFT}

At nonzero temperature, the effective potential receives a
temperature-dependent contribution. The latter is determined by
the particle species that are in thermal equilibrium and, more
precisely, by their masses and couplings. In this Section, we
review the general form of the finite temperature effective
potential and discuss in detail the establishment of thermal
equilibrium in an expanding Universe. In particular, we elaborate
on the relevant interactions at the microscopic level. This lays
the foundation for the explicit computation, in Section
\ref{Sec:ExplicitTPot}, of the finite temperature effective
potential in LVS.

\subsection{General form of temperature corrections}
\label{Sec:GenTemp}

The general structure of the effective scalar potential is the
following one:
\begin{equation}
V_{TOT}=V_0+V_T,
\end{equation}
where $V_0$ is the $T=0$ potential and $V_T$ the thermal
correction. As discussed in Section \ref{Sec:Review}, $V_0$ has
the general form:
\begin{equation}
V_0=\delta V_{(np)}+\delta V_{(\alpha')}+\delta V_{(g_s)},
\end{equation}
where the tree level part is null due to the no-scale structure
(recall that we are studying the scalar potential for the K\"{a}hler
moduli), $\delta V_{(np)}$ arises due to non-perturbative effects,
$\delta V_{(\alpha')}$ are $\alpha'$ corrections and the
contribution $\delta V_{(g_s)}$ comes from string loops and, as
noticed in \cite{cicq}, matches the Coleman-Weinberg potential of
the effective field theory. In addition, $\delta V_{(g_s)}$ has an
extended no-scale structure, which is crucial for the
robustness of LVS since it renders $\delta V_{(g_s)}$ subleading with
respect to $\delta V_{(np)}$ and $\delta V_{(\alpha')}$.

On the other hand, the finite temperature corrections $V_T$ have
the generic loop expansion:
\begin{equation}
V_{T}=V_T^{1-loop}+V_T^{2-loops}+... \,\, .
\end{equation}
The first term $V_T^{1-loop}$ is a 1-loop thermal correction
describing an ideal gas of non-interacting particles. It has been
derived for a renormalisable field theory in flat space in
\cite{DJ}, using the zero-temperature functional integral method
of \cite{Jackiw}, and reads:
\begin{equation}
V_T^{1-loop}=\pm \frac{T^4}{2\pi^2}\int_0^{\infty}dx \,x^2
\ln\left(1\mp e^{-\sqrt{x^2+m(\varphi)^2/T^2}}\right), \label{klo}
\end{equation}
where the upper (lower) signs are for bosons (fermions)
and $m$ is the background field dependent mass
parameter. At temperatures much higher than the mass of the particles
in the thermal bath, $T\gg m(\varphi)$, the 1-loop finite temperature
correction (\ref{klo}) has the following expansion:
\begin{equation}
V_T^{1-loop}=-\frac{\pi^2 T^4}{90}\alpha +\frac{T^2
m(\varphi)^2}{24}+\mathcal{O}\left(T m(\varphi)^3\right),
\label{1-LOOP}
\end{equation}
where for bosons $\alpha=1$ and for fermions $\alpha=7/8$.
The generalisation of (\ref{1-LOOP}) to supergravity,
coupled to an arbitrary number of chiral superfields,
takes the form \cite{BG1}:
\begin{equation}
V_T^{1-loop}=-\frac{\pi^2 T^4}{90} \left( g_B+ \frac{7}{8} g_F
\right)+\frac{T^2}{24}\left(Tr M_b^2+Tr M_f^2
\right)+\mathcal{O}\left(T M_b^3\right), \label{1-loop}
\end{equation}
where $g_B$ and $g_F$ are, respectively, the numbers of bosonic and
fermionic degrees of freedom and $M_b$ and $M_f$ are the
moduli-dependent bosonic and fermionic
mass matrices of all the particles forming the thermal plasma.

If the particles in the thermal bath interact among themselves, we
need to go beyond the ideal gas approximation. The effect of the
interactions is taken into account by evaluating higher thermal
loops. The high temperature expansion of the 2-loop contribution
looks like:
\begin{equation}
V_T^{2-loops}=\alpha_2 T^4\left(\sum_i f_i(g_i) \right)+\beta_2
T^2\left(Tr M_b^2+Tr M_f^2 \right)\left(\sum_i f_i(g_i)
\right)+... \,\, , \label{2-loop}
\end{equation}
where $\alpha_2$ and $\beta_2$ are known constants, $i$ runs over
all the interactions through which different species reach thermal
equilibrium, and the functions $f_i$ are determined by the
couplings $g_i$ and the number of bosonic and fermionic degrees of
freedom. For example, for gauge interactions $f(g)=const \times
g^2$, whereas for the scalar $\lambda \phi^4$ theory one has that
$f(\lambda)=const \times \lambda$.

Now, since we are interested in the moduli-dependence of the
finite temperature corrections to the scalar potential, we can
drop the first term on the RHS of (\ref{1-loop}) and focus only on
the $T^2$ term, which indeed inherits moduli-dependence from
the bosonic and fermionic mass matrices. However, notice that
in string theory the various couplings are generically functions of
the moduli. Thus, also the first term on the RHS of (\ref{2-loop})
depends on the moduli and, even though it is a 2-loop effect, it could
compete with the second term on the RHS of (\ref{1-loop}), because
it scales as $T^4$ whereas the latter one scales only as $T^2$.
This issue has to be addressed on a case by case basis, by studying
carefully what particles form the thermal bath.

\subsection{Thermal equilibrium}
\label{Sec:ThermEq}

In an expanding Universe, a particle species is in equilibrium with
the thermal bath if its interaction rate, $\Gamma$, with the particles
in that bath is larger than the expansion rate of the Universe. The latter
is given by $H\sim g_{*}^{1/2}T^2/M_P$, during the radiation dominated
epoch, with $g_{*}$ being the total number of degrees of freedom. Thermal
equilibrium can be established and maintained by $2\leftrightarrow
2$ interactions, like scattering or annihilation and the inverse
pair production processes, and also by $1\leftrightarrow 2$
processes, like decays and inverse decays (single particle productions).
Let us now consider each of these two cases in detail.

\subsubsection{$2\leftrightarrow 2$ interactions}
\label{Par:2to2}

In this case the thermally averaged interaction rate can be
inferred on dimensional grounds by noticing that:
\begin{equation}
\langle\Gamma\rangle\sim\frac{1}{\langle t_c \rangle},
\end{equation}
where $\langle t_c \rangle$ is the mean time between two
collisions (interactions). Moreover
\begin{equation}
t_c \sim \frac{1}{n\sigma v},
\end{equation}
where $n$ is the number density of the species, $\sigma$ is the
effective cross section and $v$ is the relative velocity between
the particles. Thus $\langle\Gamma\rangle\sim n\la\sigma
v\rangle$. For relativistic particles, one has that $\langle
v\rangle\sim c$ ($\equiv 1$ in our units) and also $n \sim T^3$.
Therefore
\begin{equation}
\langle\Gamma\rangle \sim \la \sigma \ra T^3 \, .
\end{equation}
%Since we are focusing on energies $T>M_{EW}$, all particles are
%ultra-relativistic and so their number density behaves as $n\simT^3$. On the other hand,
The cross-section $\sigma$ has dimension of $(length)^2$ and for
$2\leftrightarrow 2$ processes its thermal average scales with the temperature as:
\begin{enumerate}
\item{For renormalisable interactions:}
\begin{equation}
\la \sigma \ra \sim\alpha^2 \frac{T^2}{\left(T^2+M^2\right)^{2}},
\label{2to2}
\end{equation}
where $\alpha=g^2/(4\pi)$ ($g$ is the gauge coupling) and $M$ is
the mass of the particle mediating the interactions under consideration.
\begin{itemize}
\item[a)] For long-range interactions $M=0$ and
(\ref{2to2}) reduces to:
\begin{equation}
\la \sigma \ra \sim\alpha^2 T^{-2}\text{ \ }\Rightarrow\text{ \
}\langle\Gamma\rangle\sim\alpha^2 T. \label{UNOs}
\end{equation}
This is also the form that (\ref{2to2}) takes for short-range
interactions at energies $E>\!\!>M$.
\item[b)] For short-range interactions at scales lower than the
mass of the mediator, the coupling constant becomes dimensionful
and (\ref{2to2}) looks like:
\begin{equation}
\la \sigma \ra \sim\alpha^2\frac{T^{2}}{M^{4}}\text{ \ }\Rightarrow\text{ \
}\langle\Gamma\rangle\sim\alpha^2\frac{T^{5}}{M^4}. \label{3ef}
\end{equation}
\end{itemize}

\item{For processes including gravity:}
\begin{itemize}
\item[a)] Processes with two gravitational vertices:
\begin{equation}
\la\sigma\ra\sim d\frac{T^{2}}{M_P^4}\text{ \ }\Rightarrow\text{ \
}\langle\Gamma\rangle\sim d\frac{T^{5}}{M_P^4}, \label{gravit}
\end{equation}
where $d$ is a dimensionless moduli-dependent constant.
\item[b)]{Processes with one renormalisable and one gravitational vertex:}
\begin{equation}
\la\sigma\ra\sim \sqrt{d}\frac{g^{2}}{M_P^2}\text{ \ }\Rightarrow\text{
\ }\langle\Gamma\rangle\sim \sqrt{d}\frac{g^2 T^{3}}{M_P^2},
\label{renorm-gravit}
\end{equation}
where $d$ is the same moduli-dependent constant as before.
\end{itemize}
\end{enumerate}

\begin{figure}
\begin{center}
%%\FIGURE{
  \begin{picture}(200,80) (0,0)
     \SetColor{Black}
    \DashLine(0,0)(50,50){6}
    \Gluon(0,100)(50,50){7.5}{4}
    \Vertex(50,50){2.83}
    \Gluon(50,50)(120,50){7.5}{4}
    \Vertex(120,50){2.83}
    \Gluon(120,50)(170,100){7.5}{4}
    \Gluon(120,50)(170,0){7.5}{4}
    \Text(25, 10)[lb]{\Large{\Black{$\Phi$}}}
    \Text(34, 80)[lb]{\Large{\Black{$g$}}}
     \Text(80, 30)[lb]{\Large{\Black{$g^*$}}}
     \Text(130, 10)[lb]{\Large{\Black{$g$}}}
      \Text(130, 80)[lb]{\Large{\Black{$g$}}}
   \end{picture}
   \end{center}
\caption{Scattering process $\Phi g\rightarrow g g$ through which the modulus $\Phi$ and gluons can reach thermal equilibrium.}
\label{ren-grav}
%}
%%\end{center}
\end{figure}
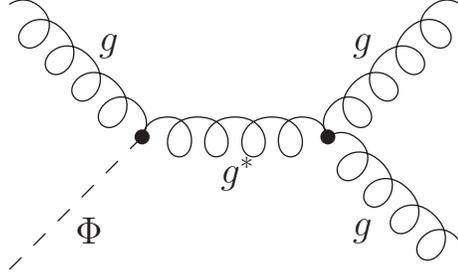

Let us now compare these interaction rates with the expansion rate of
the Universe, $H\sim g_{*}^{1/2}T^2/M_P$, in order to determine at what
temperatures various particle species reach or drop out of thermal
equilibrium, depending on the degree of efficiency of the relevant
interactions.
\begin{enumerate}
\item[1.a)]{Renormalisable interactions with massless
mediators:}
\begin{equation}
\langle\Gamma\rangle>H \text{ \ }\Leftrightarrow\text{ \ }\alpha^2
T>g_{*}^{1/2}T^2 M_P^{-1}\text{ \ }\Rightarrow\text{ \
}T<\alpha^2g_{*}^{-1/2} M_P.
\end{equation}
QCD processes, like the ones shown in Figure \ref{QCD}, are the main
examples of this kind of interactions. The same behaviour of
$\sigma$ is expected also for the other MSSM gauge groups for
energies above the EW symmetry breaking scale. Therefore, MSSM
particles form a thermal bath via strong interactions for
temperatures $T<\alpha_s^2g_{*}^{-1/2}M_P\sim 10^{15}$ GeV \cite{en}.

%\begin{figure}[ht]
%\begin{center}
%\epsfig{file=QCD.eps, height=50mm,width=70mm} \caption{QCD
%scattering process $q\bar{q}\rightarrow g g$ through which quarks
%and gluons reach thermal equilibrium.}
%\end{center}
%\end{figure}
%
\begin{figure}
\begin{center}
%\FIGURE[ht]{
  \begin{picture}(200,80) (0,0)
     \SetColor{Black}

    \ArrowLine(50,50)(0,0)
    \ArrowLine(0,100)(50,50)
    \Vertex(50,50){2.83}
    \Gluon(50,50)(120,50){7.5}{4}
    \Vertex(120,50){2.83}
    \Gluon(120,50)(170,100){7.5}{4}
    \Gluon(120,50)(170,0){7.5}{4}
    \Text(25, 10)[lb]{\Large{\Black{$\bar{q}$}}}
    \Text(25, 80)[lb]{\Large{\Black{$q$}}}
    \Text(80, 30)[lb]{\Large{\Black{$g^*$}}}
    \Text(130, 10)[lb]{\Large{\Black{$g$}}}
    \Text(130, 80)[lb]{\Large{\Black{$g$}}}
 \end{picture}
 \end{center}
 \caption{QCD scattering process $q\bar{q}\rightarrow g g$ through
which quarks and gluons reach thermal equilibrium.} \label{QCD}
\vspace{1.2cm}
\end{figure}
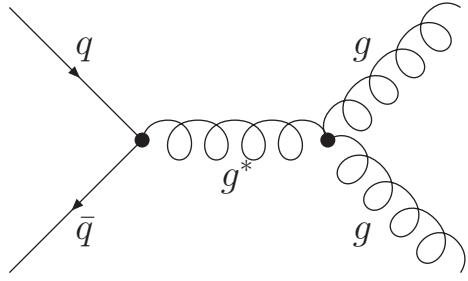

\item[1.b)]{Renormalisable interactions with massive mediators:}
\begin{equation}
\langle\Gamma\rangle>H \text{ \ }\Leftrightarrow\text{ \ }\alpha^2
\frac{T^5}{M^{4}}>g_{*}^{1/2}\frac{T^2}{M_P}\text{ \
}\Rightarrow\text{ \ }\left(\frac{g_{*}^{1/2}M^4}{\alpha^2
M_P}\right)^{1/3}<T<M.
\end{equation}
Examples of interactions with effective dimensionful couplings are
weak interactions below $M_{EW}$. In this case, the theory is well
described by the Fermi Lagrangian. An interaction between
electrons and neutrinos, like the one shown in Figure \ref{WEAK},
gives rise to a cross-section of the form of (\ref{3ef}):
\begin{equation}
\la\sigma_w\ra\sim\frac{\alpha_w^2}{M_Z^4}\la p^2 \ra \sim\frac{\alpha_w^2}{M_Z^4}T^2,
\end{equation}
where $\alpha_w$ is the weak fine structure constant and $p\sim
T$. Thus, neutrinos are coupled to the thermal bath if and only if
\begin{equation}
T>\left(\frac{g_{*}^{1/2}M_Z^4}{\alpha_w^2 M_P}\right)^{1/3}\sim
1\text{ \ MeV}.
\end{equation}

\begin{figure}
\begin{center}
%\FIGURE{
  \begin{picture}(200,80) (0,0)
     \SetColor{Black}
    \ArrowLine(50,50)(0,0)
    \ArrowLine(0,100)(50,50)
    \Vertex(50,50){2.83}
    \DashLine(50,50)(120,50){7}
    \Vertex(120,50){2.83}
    \ArrowLine(170,100)(120,50)
    \ArrowLine(120,50)(170,0)
    \Text(25, 10)[lb]{\Large{\Black{$e^+$}}}
    \Text(25, 80)[lb]{\Large{\Black{$e^-$}}}
    \Text(80, 30)[lb]{\Large{\Black{$Z_0$}}}
    \Text(130, 10)[lb]{\Large{\Black{$\nu_e$}}}
    \Text(130, 80)[lb]{\Large{\Black{$\bar{\nu}_e$}}}
 \end{picture}
 \end{center}
\caption{Weak interaction between electrons and neutrinos through
which they reach thermal equilibrium.} \label{WEAK}
%}
%\end{center}
\end{figure}
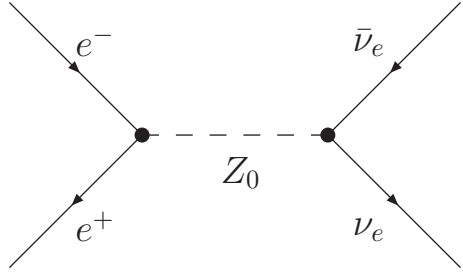
%
%\begin{figure}[ht]
%\begin{center}
%\epsfig{file=Neutrino.eps, height=50mm,width=70mm} \caption{Weak
%interaction between electrons and neutrinos through which they
%reach thermal equilibrium.}
%\end{center}
%\end{figure}

\item[2.]{Gravitational interactions:}
\begin{equation}
\hspace*{-1cm}\text{a) \ } \hspace*{1cm}\langle\Gamma\rangle>H \text{ \ }\Leftrightarrow\text{
\ }d \frac{T^5}{M_P^{4}}>g_{*}^{1/2}\frac{T^2}{M_P}\text{ \
}\Rightarrow\text{ \ }T>g_{*}^{1/6}\frac{M_P}{d^{1/3}}.
\label{casea}
\end{equation}
\begin{equation}
\hspace*{-0.6cm}\text{b) \ } \hspace*{1cm}\langle\Gamma\rangle>H \text{ \ }\Leftrightarrow\text{
\ }\sqrt{d} \frac{g^2
T^3}{M_P^{2}}>g_{*}^{1/2}\frac{T^2}{M_P}\text{ \
}\Rightarrow\text{ \ }T>\frac{g_{*}^{1/2}M_P}{g^2 \sqrt{d}}.
\label{caseb}
\end{equation}
As before, case (a) refers to $2\leftrightarrow 2$ processes with
two gravitational vertices, whereas in case (b) one vertex is
gravitational and the other one is a renormalisable interaction. A
typical K\"{a}hler modulus of string compactifications generically
couples to the gauge bosons of the field theory, that lives on the
stack of branes wrapping the cycle whose volume is given by that
modulus. Scattering processes, annihilation and pair production
reactions, that arise due to that coupling, all have
cross-sections of the form (\ref{gravit}) and
(\ref{renorm-gravit}). For all the K\"{a}hler moduli in KKLT
constructions $d\sim\mathcal{O}(1)$ and so $\langle\Gamma\rangle$
is never greater than $H$ for temperatures below the Planck scale,
for both cases (a) and (b). Therefore, those moduli will never
thermalise through $2\leftrightarrow 2$ processes. However, we
shall see in Section \ref{Sec:ModuliTherm} that the situation is
different for the small modulus in LVS, since in that case $d\sim
\mathcal{V}^2\gg 1$. A typical $2\leftrightarrow 2$ process of
type (b), with a modulus $\Phi$ and a non-abelian gauge boson $g$
going to two $g$'s, is shown in Figure \ref{ren-grav}. Here $\Phi$
denotes the canonically normalized field, which at leading order
in the large-volume expansion corresponds to the small modulus. We
will give the precise definition of $\Phi$ in Section \ref{Mmc}.
\end{enumerate}

\subsubsection{$1\leftrightarrow2$ interactions}
\label{Par:1to2}

In order to work out the temperature dependence of the interaction
rate for decay and inverse decay processes, recall that the
rest frame decay rate $\Gamma_{D}^{(R)}$ does not depend on the
temperature.
%but will take the general form:
%\begin{equation}
%\Gamma_{D}^{(R)}\sim g^2 m, \label{decadi}
%\end{equation}
%where $m$ is the mass of the decaying particle and $g$ is either a
%gauge or a Yukawa coupling.
For renormalisable interactions with massless mediators or
mediated by particles with mass $M$ at temperatures $T>M$,
%(\ref{decadi}) reads
it takes the form:
\begin{equation}
\Gamma_{D}^{(R)}\sim \alpha m, \label{decadim}
\end{equation}
where $m$ is the mass of the decaying particle and $\alpha\sim
g^2$, with $g$ being either a gauge or a Yukawa coupling. On the
other hand, for gravitational interactions or for renormalisable
interactions mediated by particles with mass $M$ at temperatures
$T<M$, we have ($M\equiv M_P$ in the case of gravity):
\begin{equation}
\Gamma_{D}^{(R)}\sim D\frac{m^3}{M^2}, \label{decadime}
\end{equation}
with $D$ a dimensionless constant (note that in the case of
gravity $D=\sqrt{d}$, where $d$ is the same moduli-dependent
constant as in Subsection \ref{Par:2to2}).

Now, the decay rate that has to be compared with $H$ is not
$\Gamma_{D}^{(R)}$, but its thermal average
$\langle\Gamma_{D}\rangle$. In order to evaluate this quantity, we
need to switch to the `laboratory frame' where:
\begin{equation}
\Gamma_{D}=\Gamma_{D}^{(R)}\sqrt{1-v^2}=\Gamma_{D}^{(R)}\frac{m}{E} \, ,
\end{equation}
and then take the thermal average:
\begin{equation}
\langle\Gamma_{D}\rangle=\Gamma_{D}^{(R)}\frac{m}{\langle E\rangle} \, .
\end{equation}
In the relativistic regime, $T\gtrsim m$, the Lorentz factor
$\gamma=\langle E\rangle/m\sim T/m$, whereas in the
non-relativistic regime, $T\lesssim m$, $\gamma=\langle
E\rangle/m\sim 1$.

Notice that, by definition, in a thermal bath the decay rate of
the direct process is equal to the decay rate of the inverse
process. However, for $T<m$ the energy of the final states of the
decay process is of order $T$, which means that the final states
do not have enough energy to re-create the decaying particle. So
the rate for the inverse decay, $\Gamma_{ID}$, is
Boltzmann-suppressed: $\Gamma_{ID}\sim e^{-m/T}$. Hence, the
conclusion is that, for $T<m$, one can never have
$\Gamma_{D}=\Gamma_{ID}$ and thermal equilibrium will not be
attained. Let us now summarize the various decay and inverse decay
rates:

\begin{enumerate}
\item{Renormalisable interactions with massless mediators or mediated by particles with mass $M$ at $T>M$:}
\begin{equation}
\langle \Gamma _{D}\rangle \simeq \left\{
\begin{array}{c}
g^{2}\frac{m^{2}}{T}\text{, \ \ for }T\gtrsim m \\
g^{2}m\text{, \ \ for }T\lesssim m \,\, ,
\end{array}
\right.
\end{equation}
\begin{equation}
\langle \Gamma _{ID}\rangle \simeq \left\{
\begin{array}{c}
g^{2}\frac{m^{2}}{T}\text{, \ \ for }T\gtrsim m \\
g^{2}m\left( \frac{m}{T}\right) ^{3/2}e^{-m/T}\text{, \ \ for
}T\lesssim m \,\, .
\end{array}
\right.
\end{equation}
Therefore, particles will reach thermal equilibrium via decay and
inverse decay processes if and only if
\begin{equation}
\langle\Gamma\rangle>H \text{ \ }\Leftrightarrow\text{ \ }g^2
\frac{m^2}{T}>g_{*}^{1/2}\frac{T^2}{M_P}\text{ \
}\Rightarrow\text{ \ }m<T<\left(\frac{g^2 m^2
M_P}{g_{*}^{1/2}}\right)^{1/3}.
\end{equation}

\item{Gravity or renormalisable interactions mediated by particles with mass $M$ at $T<M$:}
\begin{equation}
\langle \Gamma _{D}\rangle \simeq \left\{
\begin{array}{c}
D\frac{m^{4}}{M^2 T}\text{, \ \ for }T\gtrsim m \\
D\frac{m^3}{M^2}\text{, \ \ for }T\lesssim m \,\, ,
\end{array}
\right.
\end{equation}
\begin{equation}
\langle \Gamma _{ID}\rangle \simeq \left\{
\begin{array}{c}
D\frac{m^{4}}{M^2 T}\text{, \ \ for }T\gtrsim m \\
D\frac{m^3}{M^2}\left( \frac{m}{T}\right) ^{3/2}e^{-m/T}\text{, \
\ for }T\lesssim m
\end{array}
\right.
\end{equation}
with $M\equiv M_P$ in the case of gravity. Therefore, particles
will reach thermal equilibrium via decay and inverse decay
processes if and only if
\begin{equation}
\langle\Gamma\rangle>H \text{ \ }\Leftrightarrow\text{ \
}D\frac{m^{4}}{M^2 T}>g_{*}^{1/2}\frac{T^2}{M_P}\text{ \
}\Rightarrow\text{ \ }1<\frac{T}{m}<\left(D\frac{m
M_P}{g_{*}^{1/2}M^2}\right)^{1/3}. \label{izz}
\end{equation}
In the case of gravitational interactions, (\ref{izz}) becomes
\begin{equation}
1<\frac{T}{m}<\left(D\frac{m}{g_{*}^{1/2}M_P}\right)^{1/3}.
\label{uzz}
\end{equation}
In KKLT constructions, $D\sim\mathcal{O}(1)$ and $m\sim m_{3/2}$.
So (\ref{uzz}) can never be satisfied and hence moduli cannot
reach thermal equilibrium via decay and inverse decay processes.
However, we shall see in Section \ref{Sec:ModuliTherm} that in LVS
one has $D\sim\mathcal{V}\gg 1$ and so $1\leftrightarrow 2$
processes could, in principle, play a role in maintaining
thermal equilibrium between moduli and ordinary MSSM particles.
\end{enumerate}

\section{Moduli masses and couplings}
\label{Mmc}

As we have seen in the previous section, the temperature, at which
a thermal bath is established or some particles drop out of
thermal equilibrium, depends on the masses and couplings of the
particles. To determine the latter, one needs to use canonically
normalised fields. In this Section, we study the canonical normalisation
of the K\"{a}hler moduli kinetic terms and use the results to compute the
masses of those moduli and their couplings to visible sector particles.

\subsection{Single-hole Swiss-cheese}
\label{Sec:ModCoupl}

We start by focusing on the simplest Calabi-Yau realisation of
LVS, the `single-hole Swiss-cheese' case described in Subsection
\ref{Sec:Swiss-CheeseCY} (i.e., the degree 18 hypersurface
embedded in $\mathbb{C}P^4_{[1,1,1,6,9]}$). First of all, we shall
review the canonical normalisation derived in \cite{CQ}. In order
to obtain the Lagrangian in the vicinity of the zero temperature
vacuum, one expands the moduli fields around the $T=0$ minimum:
\bea
\tau_b &=& \langle\tau_b\rangle +\delta \tau_b\, , \\
\tau_s &=& \langle\tau_s\rangle  +\delta \tau_s \, . \eea where
$\langle\tau_b\rangle$ and $\langle\tau_s\rangle$ denote the VEV
of $\tau_b$ and $\tau_s$. One then finds:
\be
{\cal L} = K_{i \bar{j} } \partial_{\mu} (\delta \tau_i)
\partial^{\mu} (\delta \tau_j) -\langle V_0 \rangle - \frac{1}{2}
V_{i\bar{j}} \delta \tau_i \delta \tau_j + {\cal O}(\delta \tau^3)
\, , \label{Lago}
\ee
where $i=b,s$ and $\langle V_0 \rangle$ denotes the value of the
zero temperature potential at the minimum. To find the canonically
normalized fields $\Phi$ and $\chi$, let us write $\delta\tau_b$
and $\delta\tau_s$ as:
\be \delta \tau_i=\frac{1}{\sqrt{2}}\left[(\vec{v}_{\Phi})_i\Phi+
(\vec{v}_{\chi})_i\chi\right]. \label{oiu} \ee
Then the conditions for the Lagrangian (\ref{Lago}) to take the
canonical form:
\be {\cal L}= \half \partial_{\mu} {\Phi}
\partial^{\mu} {\Phi}+\half
\partial_{\mu} {\chi} \partial^{\mu} {\chi}- \langle V_0 \rangle -
\half m_{\Phi}^2 \Phi^2 - \half m_{\chi}^2 \chi^2
\ee
are the following:
\begin{equation}
K_{i\bar{j}}(\vec{v}_{\alpha})_{i}(\vec{v}_{\beta})_j=
\delta_{\alpha\beta}\text{ \ \ and \ \
}\frac{1}{2}V_{i\bar{j}}(\vec{v}_{\alpha})_{i}(\vec{v}_{\beta})_j
=m_{\alpha}^2\delta_{\alpha\beta} \, . \label{gg}
\end{equation}
These relations are satisfied when $\vec{v}_{\Phi}$,
$\vec{v}_{\chi}$ (properly normalised according to the first of
(\ref{gg})) and $m_{\Phi}^2$, $m_{\chi}^2$ are, respectively, the
eigenvectors and the eigenvalues of the mass-squared matrix
$\left(M^2\right)_{ij}\equiv\frac{1}{2}\left(K^{-1}\right)_{i\bar{k}}V_{\bar{k}j}$.

Substituting the results of \cite{CQ} for $\vec{v}_{\Phi}$ and
$\vec{v}_{\chi}$ in (\ref{oiu}), we can write the original
K\"{a}hler moduli $\delta\tau_i$ as (for $a_s\tau_s\gg 1$):
\bea
\delta\tau_b &=&
\left(\sqrt{6}\langle\tau_b\rangle^{1/4}\langle\tau_s\rangle^{3/4}\right)
\frac{\Phi}{\sqrt{2}}
+\left(\sqrt{\frac{4}{3}}\langle\tau_b\rangle\right)\frac{\chi}{\sqrt{2}}
\sim\mathcal{O}\left(\mathcal{V}^{1/6}\right)\Phi
+\mathcal{O}\left(\mathcal{V}^{2/3}\right)\chi \, , \label{big}\\
\delta\tau_s &=&
\left(\frac{2\sqrt{6}}{3}\langle\tau_b\rangle^{3/4}\langle\tau_s\rangle^{1/4}\right)
\frac{\Phi}{\sqrt{2}}
+\left(\frac{\sqrt{3}}{a_s}\right)\frac{\chi}{\sqrt{2}}
\sim\mathcal{O}\left(\mathcal{V}^{1/2}\right)\Phi+\mathcal{O}\left(1\right)\chi
\, . \label{small} \eea
As expected, these relations show that there is a mixing of the
original fields. Nevertheless, $\delta\tau_b$ is mostly $\chi$ and
$\delta\tau_s$ is mostly $\Phi$. On the other hand, the
mass-squareds are \cite{CQ}:
\bea m^2_{\Phi} &\simeq& Tr\left(M^{2}\right)
\simeq\left(\frac{g_s e^{K_{cs}}}{8\pi}\right)\frac{24\sqrt{2}\nu
a_s^2 \langle\tau_s\rangle^{1/2}}{\mathcal{V}^2}M_P^2
\sim \left(\frac{\ln{\mathcal{V}}}{\mathcal{V}}\right)^2 \!M_P^2 \label{canmass1} \\
m^2_{\chi} &\simeq&
\frac{Det\left(M^2\right)}{Tr\left(M^2\right)}\simeq
\left(\frac{g_s e^{K_{cs}}}{8\pi}\right)\frac{27 \nu}{4 a_s
\langle\tau_s\rangle
\mathcal{V}^3}M_P^2\sim\frac{M_P^2}{\mathcal{V}^3\ln{\mathcal{V}}}.
\label{canmass2} \eea
We can see that there is a large hierarchy of masses among the two
particles, with $\Phi$ being heavier than the gravitino mass
(recall that $m_{3/2} \sim M_P/ {\cal V}$)
and $\chi$ lighter by a factor of $\sqrt{\V}$.\\

Using the above results and assuming that the MSSM is built via
magnetised D7 branes wrapped around the small cycle, we can
compute the couplings of the K\"{a}hler moduli fields of the
$\mathbb{C} P^{4}_{[1,1,1,6,9]}$ model to visible gauge and
matter fields. This is achieved by expanding the kinetic and mass
terms of the MSSM particles around the moduli VEVs. The details
are provided in Appendix \ref{App:ModuliCouplings}, where we focus
on $T>M_{EW}$ since we are interested in thermal corrections at
high temperatures. This, in particular, means that all fermions
and gauge bosons are massless and the mixing of the Higgsinos with
the EW gauginos, that gives neutralinos and charginos, is not
present. We summarise the results for the moduli couplings in
Tables \ref{table:Cp11169couplings} and \ref{table:scCp11169}.

\TABULAR[h]{c||c|c|c|c}
 { & Gauge bosons $(F_{\mu \nu} F^{\mu \nu})$ & Gauginos
 $(\bar{\lambda} \lambda)$& Matter fermions $(\bar{\psi} \psi )$
 & Higgsinos $(\bar{\tilde{H}}\tilde{H})$ \\
  \hline\hline
  \\ & & & & \vspace{-0.9cm}\\
  $\chi$ & $\frac{1}{M_p \, \rm{ln} \V}$ & $\frac{1}{\V \rm{ln}\V}$
  & No coupling & $\frac{1}{\mathcal{V}\ln\mathcal{V}}$
  \\ & & & & \vspace{-0.3cm}\\
  $\Phi$ & $ \frac{\sqrt{\V} } {M_p} $ & $\frac{1}{\V^{3/2} \rm{ln}\V} $
  & No coupling & $\frac{1}{\sqrt{\mathcal{V}}\ln\mathcal{V}}$
  \label{table:Cp11169couplings} } {$\mathbb{C} P^{4}_{[1,1,1,6,9]}$ case: moduli couplings
  to spin $1$ and $1/2$ MSSM particles for $T>M_{EW}$.}

 \TABULAR[h]{c|| c | c | c | c | c }
 {
%$\phantom{123}$
 & Higgs $(\bar{H}H)$ & Higgs-Fermions $(H\bar{\psi}\psi)$ & SUSY scalars $(\bar{\varphi}
 \varphi )$ & $\chi^2$ & $\Phi^2$ \\
  \hline\hline
  \\ & & & & \vspace{-0.9cm}\\
  $\chi$ & $\frac{M_P}{\mathcal{V}^2(\ln\mathcal{V})^2}$ & $\frac{1}{M_P\mathcal{V}^{1/3}}$
  & $\frac{M_P}{\mathcal{V}^2(\ln\mathcal{V})^2}$ & $\frac{M_P}{\V^3}$
  & $ \frac{M_P}{\V^2} $
  \\ & & & & \vspace{-0.3cm}\\
  $\Phi$ & $\frac{M_P}{\mathcal{V}^{5/2}(\ln\mathcal{V})^2}$ & $\frac{1}{M_P\mathcal{V}^{5/6}}$
  & $\frac{M_P}{\mathcal{V}^{5/2}(\ln\mathcal{V})^2}$ & $\frac{M_P}
  { \V^{5/2} }$  & $ \frac{M_P} {\V^{3/2} } $
  \label{table:scCp11169}
} {$\mathbb{C} P^{4}_{[1,1,1,6,9]}$ case: moduli couplings to
spin $0$ and $1/2$ MSSM particles and cubic self-couplings for
$T>M_{EW}$.}

\subsection{Multiple-hole Swiss-cheese}
\label{Sec:sottosezione}

Let us now consider the more general Swiss-cheese CY three-folds
with more than one small modulus and with volume given by
(\ref{SCmsm}). In this case we find:
\bea \label{Lagrangiankin} {\cal L}_{kin} &=& \frac{3}{4 \la
\tau_b \ra^2} \pd_{\mu} (\delta \tau_b) \pd^{\mu} (\delta \tau_b)
+ \frac{3}{8} \sum_i \frac{\lambda_i \epsilon_i}
{\la \tau_b \ra \la \tau_i \ra} \pd_{\mu} (\delta \tau_i) \pd^{\mu} (\delta \tau_i) \nn \\
&-& \frac{9}{4} \sum_i \frac{\lambda_i \epsilon_i}{\la \tau_b
\ra^2} \pd_{\mu} (\delta \tau_b) \pd^{\mu} (\delta \tau_i) +
\frac{9}{4} \sum_{i<j} \frac{\lambda_i \lambda_j \epsilon_i
\epsilon_j}{\la \tau_b \ra^2} \pd_{\mu} (\delta \tau_i) \pd^{\mu}
(\delta \tau_j) \,\, , \eea
where $\epsilon_i \equiv \sqrt{\frac{\tau_i}{\tau_b}} <\!\!< 1$
and also we have kept only the leading (in the limit $\tau_b
>\!\!> \tau_i \,\, \forall i$) contribution in each term. Notice
that the mixed terms are subleading compared to the diagonal ones.
So, to start with, one can keep only the first line in
(\ref{Lagrangiankin}). Then at leading order the canonically
normalized fields $\chi$ and $\Phi_i$, $i=1,...,N_{small}$, are
defined via:
\be
\delta \tau_b = \sqrt{\frac{2}{3}} \la \tau_b \ra \chi \sim
{\cal O} \left( {\cal V}^{2/3} \right) \chi \, , \qquad \delta
\tau_i = \frac{2}{\sqrt{3 \lambda_i}} \la \tau_b \ra^{3/4} \la
\tau_i \ra^{1/4} \Phi_i \sim {\cal O} \left( {\cal V}^{1/2}
\right) \Phi_i \, . \ee
As was to be expected, this scaling with the volume agrees with
the behaviour of $\delta \tau_b$ and $\delta \tau_s$ in
(\ref{big}), (\ref{small}). Now, let us work out the volume
scaling of the subdominant mixing terms since it is important for
the computation of the various moduli couplings. Proceeding order
by order in a large-$\mathcal{V}$ expansion, we end up with:
\begin{gather}
\delta \tau _{b} \sim \mathcal{O}\left( \mathcal{V}^{2/3}\right)
\chi+\sum_i\mathcal{O}\left( \mathcal{V}
^{1/6}\right) \Phi_i \, ,  \label{SCbig11} \\
\delta \tau _{i} \sim \mathcal{O} \left( \mathcal{V}^{1/2}\right)
\Phi_i+\mathcal{O }\left( 1\right) \chi +\sum_{j\neq i
}\mathcal{O}\left( \mathcal{V}^{-1/2}\right) \Phi_{j} \, .
\label{SCSmall1}
\end{gather}
This shows that the mixing between the small moduli is strongly
suppressed by inverse powers of the overall volume, in accord with
the subleading behaviour of the last term in
(\ref{Lagrangiankin}). Furthermore, the fact that the leading
order volume-scaling of (\ref{SCbig11})-(\ref{SCSmall1}) is the
same as (\ref{big})-(\ref{small}), implies that all small moduli
behave in the same way as the only small modulus of the
$\mathbb{C} P^{4}_{[1,1,1,6,9]}$ model. Hence, if all the
small moduli are stabilised by non-perturbative effects, the
moduli mass spectrum in the general case will look like
(\ref{canmass1})-(\ref{canmass2}), with (\ref{canmass1}) valid for
all the small moduli. In addition, if we assume that all the
4-cycles corresponding to small moduli are wrapped by MSSM D7
branes, the moduli couplings to matter fields are again given by
Tables \ref{table:Cp11169couplings} and \ref{table:scCp11169},
where now $\Phi$ stands for any small modulus $\Phi_i$.

However, in general the situation may be more complicated. In
fact, the authors of \cite{blumenhagen} pointed out that 4-cycles
supporting MSSM chiral matter cannot always get non-perturbative
effects.\footnote{This is because an ED3 wrapped on the same cycle
will have, in general, chiral intersections with the MSSM branes.
Thus the instanton prefactor would be dependent on the VEVs of
MSSM fields which are set to zero for phenomenological reasons. In
the case of gaugino condensation, this non-perturbative effect
would be killed by the arising of chiral matter.} A possible way
to stabilise these 4-cycles is to use $g_s$ corrections as
proposed in \cite{ccq2}. In this case, the leading-order behaviour
of (\ref{canmass1}) should not change: $m_{\Phi_i}^2\sim
\frac{M_P^2}{\mathcal{V}^2}$\,.\footnote{It may be likely that
$m_{\Phi_i}^2$ depends on subleading powers of $(\ln\mathcal{V})$
due to the fact that the loop corrections are subdominant with
respect to the non-perturbative ones (see \cite{ccq2}), but the
main $\mathcal{V}^{-2}$ dependence should persist.} However, the
moduli couplings to MSSM particles depend on the
underlying brane set-up. So let us consider the following main cases:
\begin{enumerate}
\item All the small 4-cycles are wrapped by MSSM D7 branes except
$\tau_{np}$ which is responsible for non-perturbative effects,
being wrapped by an ED3 brane. It follows that the MSSM couplings
of $\Phi_{np}$ are significantly suppressed compared to the MSSM
couplings of the other small cycles (still given by Tables
\ref{table:Cp11169couplings} and \ref{table:scCp11169}). This is
due to the mixing term in (\ref{SCSmall1}) being highly suppressed
by inverse powers of $\mathcal{V}$.

\item All the small 4-cycles are wrapped by MSSM D7 branes except
$\tau_{np}$ which is supporting a pure $SU(N)$ hidden sector that
gives rise to gaugino condensation. This implies that the coupling
of $\Phi_{np}$ to hidden sector gauge bosons will have the same
volume-scaling as the coupling of the other small moduli with
visible sector gauge bosons. However, the coupling of the MSSM
4-cycles with hidden sector gauge bosons will be highly
suppressed.

\item All the small 4-cycles $\tau_i$ support MSSM D7 branes which are also wrapped
around the 4-cycle responsible for non-perturbative effects
$\tau_{np}$, but they have chiral intersections only on the other
small cycles. In this case, the coupling of $\Phi_{np}$ to MSSM
particles would be the same as the other $\Phi_i$. However, if
$\tau_{np}$ supports an hidden sector that undergoes gaugino
condensation, the coupling of the MSSM 4-cycles with the gauge
bosons of this hidden sector would still be highly suppressed.
\end{enumerate}

\subsection{K3 Fibration}
\label{K3CanNorm}

We turn now to the K3 fibration case described in Section
\ref{Sec:K3FibrCY}. We shall consider first the `LV' case, in
which the modulus related to the K3 divisor is fixed at a very
large value, and then the `SV' case, in which the overall volume
is of the order $\mathcal{V}\sim 10^{3}$ and the K3 fiber is
small.

In order to compute the moduli mass spectroscopy and couplings, it
suffices to canonically normalise the fields just in the vicinity
of the vacuum. The non-canonical kinetic terms look like (with
$\varepsilon\equiv\sqrt{\langle\tau_s\rangle/\langle\tau_1\rangle}$):
\begin{align}
\mathcal{L}_{kin}& = \frac{1}{4\langle\tau_{1}\rangle^{2}}\partial
_{\mu }(\delta\tau _{1})\partial ^{\mu }(\delta\tau
_{1})+\frac{1}{2\langle\tau _{2}\rangle^{2}}\partial _{\mu
}(\delta\tau _{2})\partial ^{\mu }(\delta\tau
_{2})-\frac{3\gamma\varepsilon}{4\langle\tau_2\rangle\langle\tau_1\rangle}
\partial_{\mu }(\delta\tau _{1})\partial ^{\mu
}(\delta\tau_{s})\nonumber\\
& -\frac{3\gamma\varepsilon}{2\langle\tau_{2}\rangle^{2}}
\partial _{\mu }(\delta\tau_2)
\partial ^{\mu }(\delta\tau_s) +\frac{ \gamma
\varepsilon^{3}}{2\langle\tau_2\rangle^2}
\partial _{\mu }(\delta\tau_{1})\partial ^{\mu }
(\delta\tau_{2})+\frac{3\gamma\varepsilon
}{8\langle\tau_2\rangle\langle\tau_{s}\rangle}
\partial_{\mu}(\delta\tau_{s})\partial^{\mu}(\delta\tau_{s}). \label{LKinetic}
\end{align}

\medskip\noindent{\em Large K3 fiber}

\medskip\noindent
In the `LV' case where the K3 fiber is stabilised at large value,
$\varepsilon\ll 1$. Therefore at leading order in a large volume
expansion, where
$\langle\tau_2\rangle>\langle\tau_1\rangle\gg\langle\tau_s\rangle$,
all the cross-terms in (\ref{LKinetic}) are subdominant to the
diagonal ones, and so can be neglected:
\begin{equation}
\mathcal{L}_{kin} \simeq
\frac{1}{4\langle\tau_{1}\rangle^{2}}\partial_{\mu}(\delta\tau
_{1})\partial ^{\mu }(\delta\tau _{1})+\frac{1}{2\langle\tau
_{2}\rangle^{2}}\partial _{\mu }(\delta\tau _{2})\partial ^{\mu
}(\delta\tau _{2})+\frac{ 3\gamma\varepsilon
}{8\langle\tau_2\rangle\langle\tau_{s}\rangle}\partial _{\mu
}(\delta\tau _{s})\partial ^{\mu }(\delta\tau _{s}).
\label{LKinetica}
\end{equation}
Therefore, at leading order the canonical normalisation close to
the minimum becomes rather easy and reads:
\begin{eqnarray}
\delta \tau _{1} &=& \sqrt{2}\langle \tau _{1}\rangle \chi _{1}\sim
\mathcal{O}\left( \mathcal{V}^{2/3}\right) \chi _{1}, \label{recup} \\
\delta \tau _{2} &=&\langle \tau
_{2}\rangle \chi _{2}\sim \mathcal{O}\left( \mathcal{V}^{2/3}\right) \chi _{2},  \label{K3big} \\
\delta \tau _{s} &=&\sqrt{\frac{4\langle \tau _{1}\rangle
^{1/2}\langle \tau _{2}\rangle \langle \tau
_{s}\rangle^{1/2}}{3\gamma }}\Phi \sim \mathcal{O}\left(
\mathcal{V}^{1/2}\right) \Phi . \label{K3small}
\end{eqnarray}
However, in order to derive all the moduli couplings, we need also
to work out the leading order volume-scaling of the subdominant
mixing terms in (\ref{K3big}) and (\ref{K3small}). This can be
done order by order in a large-$\mathcal{V}$ expansion and, after
some algebra, we obtain:
\begin{gather}
\delta \tau _{1}=\alpha _{1}\langle \tau _{1}\rangle \chi
_{1}+\alpha _{2} \frac{\sqrt{\langle \tau _{1}\rangle }}{\langle
\tau _{2}\rangle }\langle \tau _{s}\rangle ^{3/2}\chi _{2}+\alpha
_{3}\frac{\langle \tau _{1}\rangle ^{3/4}}{\sqrt{\langle \tau
_{2}\rangle }}\langle \tau _{s}\rangle ^{3/4}\Phi,  \label{K3big1} \\
\delta \tau _{2}=\alpha _{4}\frac{\sqrt{\langle \tau _{1}\rangle
}}{\langle \tau _{2}\rangle }\langle \tau _{s}\rangle ^{3/2}\chi
_{1}+\alpha _{5}\langle \tau _{2}\rangle \chi _{2}+\alpha
_{6}\frac{\sqrt{\langle \tau _{2}\rangle }}{ \langle \tau
_{1}\rangle ^{1/4}}\langle \tau _{s}\rangle
^{3/4}\Phi,  \label{K3big2} \\
\delta \tau _{s}=\alpha _{7}\frac{\langle \tau _{1}\rangle
}{\langle \tau _{2}\rangle }\langle \tau _{s}\rangle \chi
_{1}+\alpha _{8}\langle \tau _{s}\rangle \chi _{2}+\alpha
_{9}\langle \tau _{1}\rangle ^{1/4}\sqrt{ \langle \tau _{2}\rangle
}\langle \tau _{s}\rangle ^{1/4}\Phi, \label{K3Small}
\end{gather}
where the $\alpha_i$, $i=1,...,9$ are $\mathcal{O}(1)$
coefficients. The volume-scalings of (\ref{K3big1}), (\ref{K3big2})
and (\ref{K3Small}) are the following:
\begin{gather}
\delta \tau _{1} \sim \mathcal{O}\left( \mathcal{V}^{2/3}\right)
\chi _{1}+\mathcal{O}\left( \mathcal{V}^{-1/3}\right) \chi
_{2}+\mathcal{O}\left( \mathcal{V}
^{1/6}\right) \Phi ,  \label{K3big11} \\
\delta \tau _{2}\sim \mathcal{O}\left( \mathcal{V}^{-1/3}\right)
\chi _{1}+\mathcal{O}\left( \mathcal{V}^{2/3}\right) \chi_2
+\mathcal{O}\left( \mathcal{V}^{1/6}\right)
\Phi ,  \label{K3big21} \\
\delta \tau _{s} \sim \mathcal{O }\left( 1\right) \chi
_{1}+\mathcal{O}\left( 1\right) \chi _{2}+\mathcal{O} \left(
\mathcal{V}^{1/2}\right) \Phi.  \label{K3Small1}
\end{gather}
This shows that, if we identify each of $\tau_1$ and $\tau_2$ with the
large modulus $\tau_b$ in the Swiss-cheese case, (\ref{K3big11})
and (\ref{K3big21}) have the same volume scaling as (\ref{big}),
as one might have expected. Moreover, the similarity of
(\ref{K3Small1}) and (\ref{small}) shows that also the small
moduli in the two cases behave in the same way. Therefore, we
can conclude that (\ref{canmass1}) is valid also for the
K3 Fibration case under consideration:
\begin{equation}
m_{\Phi}\sim\left(\frac{\ln\mathcal{V}}{\mathcal{V}}\right)M_P.
\label{XX}
\end{equation}
On the other hand, we need to be more careful in the study of the
mass spectrum of the large moduli $\tau_1$ and $\tau_2$. We can
work out this `fine structure', at leading order in a
large-$\mathcal{V}$ expansion, first integrating out $\tau_{s}$
and then computing the eigenvalues of the matrix. The latter are
obtained by multiplying the inverse K\"{a}hler metric by the
Hessian of the potential both evaluated at the minimum. The
leading order behaviour of the determinant of this matrix is:
\begin{equation}
Det\left(K^{-1}d^2V\right)\sim
\frac{\tau_2^{4}\sqrt{\ln\mathcal{V}}}{\mathcal{V}^9},\text{ \ \
with \ \ }\mathcal{V}\sim\sqrt{\tau_1}\tau_2.
\end{equation}
Because $m_{\chi_2}^2\gg m_{\chi_1}^2$, we have at leading order
at large volume:
\bea
m^2_{\chi_2} &\simeq& Tr\left(K^{-1}d^2V\right)
\sim \frac{\sqrt{\ln\mathcal{V}}}{\mathcal{V}^3}M_P^2 \label{Canmass1} \\
m^2_{\chi_1} &\simeq&
\frac{Det\left(K^{-1}d^2V\right)}{Tr\left(K^{-1}d^2V\right)}
\sim\frac{\tau_2^{4}}{\mathcal{V}^6}M_P^2\sim\frac{M_P^2}{\tau_1^3\tau_2^2}.
\label{Canmass2} \eea
Identifying $\tau_1$ with $\tau_2$, (\ref{Canmass2}) simplifies to
$m^2_{\chi_1} \sim \mathcal{V}^{-10/3}$, confirming the
qualitative expectation that the $\tau_1$ direction is
systematically lighter than $\V$ in the large-$\V$ limit.

Using the results of this Section and assuming that the MSSM
branes are wrapped around the small cycle\footnote{We also ignore
the incompatibility between localising non-perturbative effects
and the MSSM on the same 4-cycle.}, it is easy to repeat the
computations of Appendix \ref{App:ModuliCouplings} for the K3
fibration. Due to the fact that the leading order ${\cal
V}$-scaling of (\ref{K3big11})-(\ref{K3Small1}) matches that of
the single-hole Swiss-cheese model, we again find the same
couplings as those given in Tables \ref{table:Cp11169couplings}
and \ref{table:scCp11169}, where now $\chi$ stands for any of
$\chi_1$ and $\chi_2$.

\medskip\noindent{\em Small K3 fiber}

\medskip\noindent
In the `SV' case where the K3 fiber is stabilised at small value,
$\varepsilon\simeq 1$. Therefore at leading order in a large
volume expansion, where
$\langle\tau_2\rangle\gg\langle\tau_1\rangle>\langle\tau_s\rangle$,
the first term in (\ref{LKinetic}) is dominating the whole kinetic
Lagrangian. Hence we conclude that, at leading order, the
canonical normalisation of $\delta\tau_1$ close to the $T=0$
minimum is again given by (\ref{recup}). However, now its volume
scaling reads:
\begin{equation}
\delta \tau _{1}\sim \mathcal{O}\left(1\right) \chi
_{1}+\left(\text{subleading \ mixing \ terms}\right).
\label{K3Big}
\end{equation}
To proceed order by order in a large volume expansion, note that
the third and the sixth term in (\ref{LKinetic}) are suppressed by
just one power of $\langle\tau_2\rangle$, whereas the second,
fourth and fifth term are suppressed by two powers of the large
modulus. Thus, we obtain the following leading order behaviour for
the canonical normalisation of the two remaining moduli:
\begin{gather}
\delta \tau _{2}\sim \mathcal{O}\left( \mathcal{V}\right) \chi
_{1}+\mathcal{O}\left( \mathcal{V}\right) \chi_2
+\mathcal{O}\left( \mathcal{V}\right)
\Phi ,  \label{K3BIG} \\
\delta \tau _{s} \sim \mathcal{O}\left( \mathcal{V}^{1/2}\right)
\chi _{1}+\mathcal{O}\left( \mathcal{V}^{1/2}\right)
\Phi+\text{subleading \ mixing \ terms}. \label{K3SMALL}
\end{gather}
Notice that the canonically normalised field $\chi_1$ corresponds
to the K3 divisor $\tau_1$, whereas $\Phi$ is a mixing of $\tau_1$
and the blow-up mode $\tau_s$. Finally $\chi_2$ is a combination
of all the three states, and so plays the role of the `large'
field. The moduli mass spectrum will still be given by (\ref{XX}),
(\ref{Canmass1}) and (\ref{Canmass2}). However now the volume
scaling of (\ref{Canmass2}) simplifies to $m^2_{\chi_1} \sim
\mathcal{V}^{-2}$, confirming the qualitative expectation that
$\chi_1$ is also a small field with a mass of the same order of
magnitude of $m_{\Phi}$.

The computation of the moduli couplings depends on the
localisation of the MSSM within the compact CY. As we have seen in
Subsection \ref{Sec:K3FibrCY}, the scalar potential receives
non-perturbative corrections in the blow-up mode $\tau_s$.
Therefore, in order for the non-perturbative contributions to be
non-vanishing, the MSSM branes have to wrap either the small K3
fiber $\tau_1$ or the 4-cycle given by the formal sum
$\tau_s+\tau_1$ with chiral intersections on $\tau_1$. In both
cases, we cannot immediately read off the moduli couplings from
the results of Appendix \ref{App:ModuliCouplings}. This is due to
the difference of the leading order volume scaling of the
canonical normalisation between the `SV' case for the K3 fibration
and the Swiss-cheese scenario.\footnote{We stress also that
presently there is no knowledge of the K\"{a}hler metric for
chiral matter localised on deformable cycles.}

However, as we shall see in the next Section, in the Swiss-cheese
case, the relevant interactions through which the small moduli can
thermalise, are with the gauge bosons. As we shall see in Section
\ref{Sec:K3FibrModTherm}, these interactions will also be the ones
that are crucial for moduli thermalisation in the K3 fibration
case. Therefore, here we shall focus on them only. Following the
calculations in Subsection \ref{Sec:ModCouplOrdPart} of Appendix
\ref{App:ModuliCouplings}, we infer that if only $\tau_1$ is
wrapped by MSSM branes, then the coupling of $\chi_1$ with MSSM
gauge bosons is of the order $g\sim 1/M_P$ without any factor of
the overall volume, while the coupling of $\Phi$ with gauge bosons
will be more suppressed by inverse powers of $\mathcal{V}$. On the
other hand, if both $\tau_1$ and $\tau_s$ are wrapped by MSSM
branes, then the couplings of both small moduli with the gauge
bosons are similar to the ones in the Swiss-cheese case:
$g\sim\sqrt{\mathcal{V}}/M_P$. Moreover, if gaugino condensation
is taking place in the pure $SU(N)$ theory supported on $\tau_s$,
then both $\chi_1$ and $\Phi$ couple to the hidden sector gauge
bosons with strength $g\sim\sqrt{\mathcal{V}}/M_P$.

We end this Subsection by commenting on K3 fibrations with more
than one blow-up mode. In such a case, it is possible to localise
the MSSM on one of the small blow-up modes and the situation is
very similar to the one outlined for the multiple-hole
Swiss-cheese. The only difference is the presence of the extra
modulus related to the K3 fiber, which will couple to the MSSM
gauge bosons with the same strength as the small modulus
supporting the MSSM. This is because of the particular form of the
canonical normalisation, which, for example in the case of two
blow-up modes $\tau_{s1}$ and $\tau_{s2}$, looks like
(\ref{K3Big}) and (\ref{K3BIG}) together with:
\begin{gather}
\delta \tau _{s1} \sim \mathcal{O}\left( \mathcal{V}^{1/2}\right)
\chi _{1}+\mathcal{O}\left( \mathcal{V}^{1/2}\right)
\Phi_1+\text{subleading \ mixing \ terms}, \label{K3SMALL2} \\
\delta \tau _{s2} \sim \mathcal{O}\left( \mathcal{V}^{1/2}\right)
\chi _{1}+\mathcal{O}\left( \mathcal{V}^{1/2}\right)
\Phi_2+\text{subleading \ mixing \ terms}. \label{K3SMALL3}
\end{gather}

\subsection{Modulini}
\label{Sec:Modulini}

In this Subsection we shall concentrate on the supersymmetric
partners of the moduli, the modulini. More precisely, we will consider
the fermionic components of the chiral superfields, whose scalar components
are the K\"{a}hler moduli. The kinetic Lagrangian for these
modulini reads:
\begin{equation}
\mathcal{L}_{kin} = \frac{i}{4}\frac{\partial^2 K}{\partial
\tau_i\partial\tau_j} \delta
\bar{\tilde{\tau}}_j\gamma^{\mu}
\partial_{\mu} (\delta \tilde{\tau}_i) \, , \label{LagModulini}
\end{equation}
where the K\"{a}hler metric is the same as the one that appears in the
kinetic terms of the K\"{a}hler moduli. Therefore, the canonical
normalisation of the modulini takes exactly the same form as the
canonical normalisation of the corresponding moduli. For example, in the
single-hole Swiss-cheese case, we have:
\bea \delta\tilde{\tau}_b &=&
\left(\sqrt{6}\langle\tau_b\rangle^{1/4}\langle\tau_s\rangle^{3/4}\right)
\frac{\tilde{\Phi}}{\sqrt{2}}
+\left(\sqrt{\frac{4}{3}}\langle\tau_b\rangle\right)\frac{\tilde{\chi}}{\sqrt{2}}
\sim\mathcal{O}\left(\mathcal{V}^{1/6}\right)\tilde{\Phi}
+\mathcal{O}\left(\mathcal{V}^{2/3}\right)\tilde{\chi} \, , \label{Bbig}\\
\delta\tilde{\tau}_s &=&
\left(\frac{2\sqrt{6}}{3}\langle\tau_b\rangle^{3/4}\langle\tau_s\rangle^{1/4}\right)
\frac{\tilde{\Phi}}{\sqrt{2}}
+\left(\frac{\sqrt{3}}{a_s}\right)\frac{\tilde{\chi}}{\sqrt{2}}
\sim\mathcal{O}\left(\mathcal{V}^{1/2}\right)\tilde{\Phi}+\mathcal{O}\left(1\right)\tilde{\chi}.
\label{Ssmall} \eea
We focus now on the modulini mass spectrum. We recall that in LVS
the minimum is non-supersymmetric, and so the Goldstino is
eaten by the gravitino via the super-Higgs effect. The Goldstino
is the supersymmetric partner of the scalar field, which is
responsible for SUSY breaking. In our case this is the modulus
related to the overall volume of the Calabi-Yau, as can be checked
by studying the order of magnitude of the various F-terms.
Therefore, the volume modulino is the Goldstino.
More precisely, in the $\mathbb{C}P^4_{[1,1,1,6,9]}$ case,
$\tilde{\chi}$ is eaten by the gravitino, whereas the mass of
$\tilde{\Phi}$ can be derived as follows:
\be \label{TrMfer} m_{\tilde{\Phi}}^2={\rm Tr} M_f^2 = \la e^G
K^{i \bar{j}} K^{l \bar{m}} (\nabla_i G_l +\frac{G_i G_l}{3} )
(\nabla_{\bar{j}} G_{\bar{m}} + \frac{G_{\bar{j}} G_{\bar{m}}}{3})
\ra, \ee
where the function $G = K + \ln |W|^2$ is the supergravity
K\"{a}hler invariant potential, and $\nabla_i G_j = G_{ij} -
\Gamma^{l}_{ij}G_{l}$, with the connection $\Gamma^{l}_{ij} = K^{l
\bar{m}} \partial_i K_{j \bar{m}}$. Equation (\ref{TrMfer}) at
leading order in a large volume expansion, can be approximated as
\be m_{\tilde{\Phi}}^2 \simeq \langle e^G | (K^{s \bar{s}}
(\nabla_s G_s + \frac{G_s G_s}{3}) |^2\rangle \label{TrMfer11169}
\ee
where $\nabla_s G_s \simeq G_{ss} - \Gamma^{s}_{ss} G_s $ and
$\Gamma^{s}_{ss} \simeq K^{s\bar{s}}
\partial_{s} K_{s\bar{s}}$. In the single-hole Swiss-cheese case, for $a_s\tau_s\gg 1$, we obtain:
\begin{equation}
m_{\tilde{\Phi}}^{2}\simeq\langle \frac{g_s
e^{K_{cs}}M_P^2}{\pi}\left(36 a_{s}^{4}A_{s}^{2}\tau
_{s}e^{-2a_{s}\tau
_{s}}-\frac{6\sqrt{2}a_{s}^{2}A_{s}W_{0}}{\mathcal{V}} \sqrt{\tau
_{s}}e^{-a_{s}\tau _{s}}+\frac{
W_{0}^{2}}{2\mathcal{V}^{2}}\right) \rangle. \label{ModulinoMass}
\end{equation}
Evaluating (\ref{ModulinoMass}) at the minimum, we find that the
mass of the modulino $\tilde{\Phi}$ is of the same order of
magnitude as the mass of its supersymmetric partner $\Phi$:
\begin{equation}
m_{\tilde{\Phi}}^{2} \simeq \frac{a_{s}^{2}\langle \tau
_{s}\rangle ^{2}W_{0}^{2}}{ \mathcal{V}
^{2}}M_P^2\sim \left( \frac{\ln \mathcal{V}}{
\mathcal{V} }\right) ^{2}M_{P}^{2}\sim m_{\Phi
}^{2}. \label{mtilde}
\end{equation}
Similarly, it can be checked that, in the general case of multiple-hole
Swiss-cheese Calabi-Yaus and K3 fibrations, the masses of the modulini
also keep being of the same order of magnitude as the masses of the
corresponding supersymmetric partners.

We now turn to the computation of the modulini couplings. In fact,
we are interested only in the modulino-gaugino-gauge boson
coupling since, as we shall see in Section \ref{Sec:ModuliTherm},
this is the relevant interaction through which the modulini reach
thermal equilibrium with the MSSM thermal bath. This coupling can
be worked out by recalling that the small modulus $\tau_s$ couples
to gauge bosons $X$ as (see appendix \ref{Sec:ModCouplOrdPart}):
\begin{equation}
\mathcal{L}_{gauge}\sim\frac{\tau_s}{M_P}F_{\mu\nu}F^{\mu\nu} \, .
\label{tauFF}
\end{equation}
The supersymmetric completion of this interaction term contains the
following modulino-gaugino-gauge boson coupling:
\begin{equation}
\mathcal{L}\sim\frac{\tilde{\tau}_s}{M_P}\sigma^{\mu\nu}\lambda'
F_{\mu\nu} \, . \label{tildetaulambdaF}
\end{equation}
Now, expanding $\tilde{\tau}_s$ around its minimum and going to the
canonically normalised fields $G_{\mu\nu}$ and $\lambda$ defined
as (see appendices \ref{Sec:ModCouplOrdPart} and
\ref{Sec:ModCouplSUSY}):
\begin{equation}
G_{\mu\nu}=\sqrt{\langle\tau_s\rangle}F_{\mu\nu} \, ,\text{ \ \
}\lambda=\sqrt{\langle\tau_s\rangle}\lambda' \, , \label{Redef}
\end{equation}
we obtain:
\begin{equation}
\mathcal{L}\sim \frac{\delta\tilde{\tau}_s}
{M_P\langle\tau_s\rangle}\sigma^{\mu\nu}\lambda G_{\mu\nu} \, .
\end{equation}
Hence, by means of (\ref{Ssmall}), we end up with the following
\textit{dimensionful} couplings:
\begin{eqnarray}
\mathcal{L}_{\tilde{\chi}\tilde{X}X}&\sim&\left(\frac{1}
{M_P\ln{\mathcal{V}}}\right)\tilde{\chi} \sigma^{\mu\nu}\lambda G_{\mu\nu} \, , \\
\mathcal{L}_{\tilde{\Phi}\tilde{X}X}&\sim&\left(\frac{\sqrt{\mathcal{V}}}
{M_P}\right)\tilde{\Phi} \sigma^{\mu\nu}\lambda G_{\mu\nu} \, .
\label{ImpModuliniCoupl}
\end{eqnarray}

\section{Study of moduli thermalisation}
\label{Sec:ModuliTherm}

Using the general discussion of Section \ref{Sec:ThermEq} and the
explicit expressions for the moduli masses and couplings of
Section \ref{Mmc}, we can now study in detail which particles form
the thermal bath. Consequently, we will be able to write down the
general form that the finite temperature corrections of Section
\ref{Sec:GenTemp} take in the LVS.

We shall start by focusing on the simple geometry
$\mathbb{C}P^4_{[1,1,1,6,9]}$, and then extend our analysis to
more general Swiss-cheese and fibred CY manifolds. We will show
below that, unlike previous expectations in the literature,
the moduli corresponding to small cycles that support chiral matter
can reach thermal equilibrium with the matter fields.

\subsection{Single-hole Swiss-cheese}
\label{Sec:11169ModTherm}

As we have seen in Section \ref{Sec:ThermEq}, both
$2\leftrightarrow 2$ and $1\leftrightarrow 2$ processes can
establish and maintain thermal equilibrium. Let us now apply the
general conditions of Sections \ref{Par:2to2} and \ref{Par:1to2}
to our case.

As we have already pointed out, scattering and annihilation
processes involving strong interactions will establish thermal
equilibrium between MSSM particles for temperatures $T<\alpha_s^2
g_{*}^{-1/2}M_P\sim 10^{15}$ GeV. Let us now concentrate on the
moduli.

\medskip\noindent{\em Small modulus $\Phi$}

\medskip\noindent
From Section \ref{Sec:ModCoupl}, we know that the largest coupling
of the small canonical modulus $\Phi$ is with the non-abelian
gauge bosons denoted by $X$:
\begin{equation}
\mathcal{L}_{\Phi XX}=g_{\Phi XX}\Phi F_{\mu\nu}F^{\mu\nu},\text{
\ \ }g_{\Phi XX}\sim\frac{\sqrt{\cal V}}{M_P}\sim\frac{1}{M_s}.
\label{1}
\end{equation}
Therefore according to (\ref{casea}), scattering or annihilation
and pair production processes with two gravitational vertices like
$X+X\leftrightarrow\Phi+\Phi$, $X+\Phi\leftrightarrow X+\Phi$, or
$X+X\leftrightarrow X+X$, can establish thermal equilibrium
between $\Phi$ and $X$ for temperatures:
\begin{equation}
T>T_f^{(1)}\equiv g_{*}^{1/6}\frac{M_P}{\mathcal{V}^{2/3}},
\label{caseA}
\end{equation}
where $T_f^{(1)}$ denotes the freeze-out temperature of the
modulus. Taking the number of degrees of freedom $g_*$ to be
${\cal O}(100)$, as in the MSSM, we find that (\ref{caseA})
implies $T>5 \times 10^{8}$ GeV for ${\cal V} \sim 10^{15}$,
whereas $T>10^{16}$ GeV for ${\cal V} \sim 10^4$.\footnote{Recall
that $M_P$ here is the reduced Planck mass, which equals $(8\pi
G_N)^{-1/2} = 2.4\times 10^{18}$ GeV.} In fact, for a typically
large volume (${\cal V}>10^{10}$) a more efficient $2
\leftrightarrow 2$ process is $X+X\leftrightarrow X+\Phi$ with one
gravitational and one renormalisable vertex with coupling constant
$g$. Indeed, according to (\ref{caseb}), such scattering processes
maintain thermal equilibrium for temperatures:
\begin{equation}
T>T_f^{(2)}\equiv\frac{g_{*}^{1/2}M_P}{g^2 \mathcal{V}}\sim
10^3\frac{M_P}{\mathcal{V}}\text{ \ for \ }g_{*}\sim 100\text{ \
and \ }g\sim 0.1 \, , \label{caseB}
\end{equation}
which for ${\cal V} \sim 10^{15}$ gives $T>10^6$ GeV while for
${\cal V} \sim 10^{4}$ it gives $T>10^{17}$ GeV.

Finally, let us investigate the r\^{o}le played by decay and
inverse decay processes of the form $\Phi\leftrightarrow X+X$. We
recall that such processes can, in principle, maintain thermal
equilibrium only for temperatures:
\begin{equation}
T>m_{\Phi}\sim \frac{\ln{\mathcal{V}}}{\mathcal{V}}M_P,
\label{mass}
\end{equation}
because the energy of the gauge bosons is given by $E_{X}\sim T$
and hence for $T<m_{\Phi}$ it is insufficient for the inverse
decay process to occur. However, for $T>m_{\Phi}$ the process $X +
X \to \Phi$ does take place and so one only needs to know the rate
of the decay $\Phi \to X + X$ in order to find out whether thermal
equilibrium is achieved. According to (\ref{uzz}) with $D\sim
g_{\Phi XX}^2/4\pi\sim\mathcal{V}/4\pi$, where we have also used
(\ref{1}), the condition for equilibrium is that:
\begin{equation}
T<T_{eq}\equiv\left(\frac{\mathcal{V}m_{\Phi}}{4\pi
g_{*}^{1/2}M_P}\right)^{1/3}m_{\Phi}\,\sim
\,\left(\frac{\ln{\mathcal{V}}}{4\pi
g_{*}^{1/2}}\right)^{1/3}m_{\Phi}\equiv \kappa m_{\Phi} \, .
\label{Uzz}
\end{equation}
Hence thermal equilibrium between $\Phi$ and $X$ can be maintained
by $1\leftrightarrow 2$ processes only if $\kappa>1$ \footnote{The
exact value of $\kappa$ can be worked out via a more detailed
calculation, very similar to the one that we will carry out in
Section \ref{Sec:VolBound}. It turns out that this value differs
from the `$\sim$' estimate in (\ref{Uzz}) just by a multiplicative
factor $c^{1/3}$ of $\mathcal{O}(1)$. More precisely, $c=18
(\pi\langle\tau_s\rangle)^{-3/2}e^{K_{\textrm{cs}}/2}W_0\sqrt{10g_s}$
and so, for natural values of all the parameters: $W_0=1$,
$g_s=0.1$, $\langle\tau_s\rangle=5$, $K_{\textrm{cs}}=3$, we
obtain $c^{1/3}=1.09$.}. However, estimating the total number of
degrees of freedom as $g_{*}\sim\mathcal{O}(100)$, and writing the
volume as $\mathcal{V}\sim 10^{x}$, we obtain that $\kappa>1$
$\Leftrightarrow$ $x>55$. Such a large value is unacceptable, as
it makes the string scale too small to be compatible with
observations. Therefore, we conclude that in LVS the small modulus
$\Phi$ never thermalises via decay and inverse decay processes.

The final picture is the following:
\begin{itemize}
\item For ${\cal V}$ of order $10^{15}$ ($10^{10}$), as in typical LVS,
from (\ref{caseB}) we deduce that the modulus $\Phi$ is in thermal
equilibrium with MSSM particles for temperatures $T
>T_f^{(2)}\simeq 10^6$ GeV ($T > T_f^{(2)}\simeq10^{11}$ GeV) due to
$X+X\leftrightarrow\Phi+X$ processes.

\item On the other hand,
for ${\cal V}<10^{10}$, as for LVS that allow gauge coupling
unification, the main processes that maintain thermal equilibrium
of the modulus $\Phi$ with MSSM particles are purely
gravitational: $X+X\leftrightarrow\Phi+\Phi$,
$\Phi+X\leftrightarrow\Phi+X$ or $X+X\leftrightarrow X+X$ and the
freeze-out temperature is given by (\ref{caseA}). For example for
${\cal V}\sim 10^{4}$ ($\Leftrightarrow$ $M_{s} \sim 10^{16}$
GeV), $\Phi$ is in thermal equilibrium for temperatures $T
>T_f^{(1)}\simeq 5\times 10^{15}$ GeV.
\end{itemize}
We stress that this is the first example in the literature of a
modulus that reaches thermal equilibrium with ordinary particles
for temperatures significantly less than $M_P$, and so completely
within the validity of the low energy effective theory.
Note that we did not focus on the interactions of $\Phi$ with other
ordinary and supersymmetric particles, since the
corresponding couplings, derived in Appendix
\ref{App:ModuliCouplings}, are not large enough to establish
thermal equilibrium.

Finally, let us also note that, once the modulus $\Phi$ drops out
of thermal equilibrium, it will decay before its energy density
can begin to dominate the energy density of the Universe, unlike
traditional expectations in the literature. We will show this in
more detail in Subsection \ref{Sec:MaxTemp}.

\medskip\noindent{\em Large modulus $\chi$}

\medskip\noindent
As summarised in Section \ref{Sec:ModCoupl}, the coupling of the
large modulus $\chi$ with gauge bosons is given by
\begin{equation}
\mathcal{L}_{\chi XX}=g_{\chi XX}\chi F_{\mu\nu}F^{\mu\nu},\text{
\ \ }g_{\chi XX }\sim\frac{1}{M_P\ln\mathcal{V}}. \label{10}
\end{equation}
Consequently purely gravitational $2\leftrightarrow 2$ processes
like $X+X\leftrightarrow\chi+\chi$, $X+\chi\leftrightarrow
X+\chi$, or $X+X \leftrightarrow X+X$, could establish thermal
equilibrium between $\chi$ and $X$ for temperatures:
\begin{equation}
T>T_{f}^{(1)}\equiv
g_{*}^{1/6}M_P\left(\ln\mathcal{V}\right)^{4/3}. \label{CaseA}
\end{equation}
On the other hand, scattering processes like $X+X\leftrightarrow
X+\chi$ with one gravitational and one renormalisable vertex with
coupling constant $g$, could maintain thermal equilibrium for
temperatures:
\begin{equation}
T>T_{f}^{(2)}\equiv\frac{g_{*}^{1/2}M_P}{g^2}\left(\ln\mathcal{V}\right)^2\sim
10^3M_P\left(\ln\mathcal{V}\right)^2,\text{ \ for \ }g_{*}\sim
100\text{ \ and \ }g\sim 0.1 \, . \label{CaseB}
\end{equation}
Clearly, both $T_{f}^{(1)}$ and $T_{f}^{(2)}$ are greater than
$M_P$ and so we conclude that $\chi$ can never thermalise via
$2\leftrightarrow 2$ processes. It is also immediate to notice
that thermal equilibrium cannot be maintained by $1\leftrightarrow
2$ processes, like $\chi\leftrightarrow X+X$, either. The reason
is that, as derived in \cite{CQ}, for typical LARGE values of the
volume $\mathcal{V}\sim 10^{10} - 10^{15}$, the lifetime of the
large modulus $\chi$ is greater than the age of the Universe.
Hence this modulus could contribute to dark matter and its decay
to photons or electrons could be one of the smoking-gun signal of
LVS.

Furthermore, as can be seen from Section \ref{Sec:ModCoupl}, the
couplings of $\chi$ to other MSSM particles are even weaker than
its coupling to gauge bosons. So $\chi$ cannot thermalise via any
other kind of interaction. Finally, one can also verify that
thermal equilibrium between $\chi$ and $\Phi$ can never be
maintained via $1\leftrightarrow 2$ and $2\leftrightarrow 2$
processes involving only the moduli, which processes arise due to
the moduli triple self-couplings computed in Appendix \ref{SELF}.
Therefore, $\chi$ behaves as a typical modulus studied in the
literature.

\subsection{Multiple-hole Swiss-cheese}
\label{Sec:MultHoleModTherm}

We shall now extend the results of Section \ref{Sec:11169ModTherm}
to the more general case of CY three-folds with one large cycle
and several small ones. We shall not focus on explicit models
since this is beyond the scope of our paper, but we will try to
discuss qualitatively the generic behaviour of small moduli in the
case of `multiple-hole Swiss-cheese' CY manifolds.

As we have seen in Section \ref{Sec:ModCoupl}, the couplings with
MSSM particles of all the small cycles wrapped by MSSM branes have
the same volume scaling as the corresponding couplings of the
single small modulus in the $\mathbb{C}P^4_{[1,1,1,6,9]}$ case.
Moreover, in Section \ref{Sec:11169ModTherm} we have learned that
$\Phi$ can thermalise via its interaction with gauge bosons.
Hence, we conclude that the same arguments as in Section
\ref{Sec:11169ModTherm} can be applied for $h_{1,1}>2$ and so all
small cycles, that support MSSM chiral matter, reach thermal
equilibrium with the gauge bosons.

Note however that, as we already pointed out in Section
\ref{Sec:ModCoupl}, the situation may be more complicated in
concrete phenomenological models due to the possibility that
non-perturbative effects may be incompatible with MSSM branes,
which are localized on the same 4-cycle \cite{blumenhagen}.
Whether or not such an incompatibility arises depends on the
particular features of the model one considers, including the
presence or absence of charged matter fields with non-vanishing
VEVs. As a consequence of these subtleties, the issue of moduli
thermalisation is highly dependent on the possible underlying
brane set-ups. To gain familiarity with the outcome, let us
explore in more detail several brane set-ups in the case of only
two small moduli. At the end we will comment on the generalization
of these results to the case of arbitrary $h_{1,1}$.

We will focus on the case $h_{1,1}=3$ with two small moduli $\tau_1$
and $\tau_2$, that give the volumes of the two rigid divisors
$\Gamma_1$ and $\Gamma_2$. The results of Subsection \ref{Sec:11169ModTherm}
imply the following for the different brane set-ups below:

\begin{enumerate}
\item If $\Gamma_1$ is wrapped by an ED3 instanton and $\Gamma_2$ is wrapped by MSSM branes:

\begin{itemize}
\item $\tau_1$ couples to MSSM gauge bosons with strength
$g\sim1/(\sqrt{\mathcal{V}}M_P)$ $\Rightarrow$ $\tau_1$ does not
thermalise.\footnote{The coupling $g\sim1/(\sqrt{\mathcal{V}}M_P)$
can be worked out by substituting the expression (\ref{SCSmall1})
in (\ref{FmunuFmunu}). As pointed out in point 1 at the end of
Subsection \ref{Sec:sottosezione}, the weakness of this coupling
is due to the mixing term in (\ref{SCSmall1}) being highly
suppressed by inverse powers of $\mathcal{V}$.}
\item $\tau_2$ couples to MSSM gauge bosons with strength
$g\sim\sqrt{\mathcal{V}}/M_P$ $\Rightarrow$ $\tau_2$ thermalises.
\end{itemize}

\item If $\Gamma_1$ is wrapped by an ED3 instanton and $\Gamma_1+\Gamma_2$ is wrapped by MSSM
branes with chiral intersections on $\Gamma_2$\footnote{We assume
that a single D7 brane is wrapping $\Gamma_2$ in order to get
chirality from the intersection with the MSSM branes. The same
assumption applies throughout the paper everywhere we use the
expression `chiral intersections on some divisor'.}:

\begin{itemize}
\item $\tau_1$ couples to MSSM gauge bosons with strength
$g\sim\sqrt{\mathcal{V}}/M_P$ $\Rightarrow$ $\tau_1$ thermalises.
\item $\tau_2$ couples to MSSM gauge bosons with strength
$g\sim\sqrt{\mathcal{V}}/M_P$ $\Rightarrow$ $\tau_2$ thermalises.
\end{itemize}

\item If $\Gamma_1$ is supporting a pure $SU(N)$ theory, that undergoes gaugino condensation,
and $\Gamma_2$ is wrapped by MSSM branes:

\begin{itemize}
\item $\tau_1$ couples to MSSM gauge bosons with strength
$g\sim 1/(\sqrt{\mathcal{V}}M_P)$ and to hidden sector gauge
bosons with strength $g\sim\sqrt{\mathcal{V}}/M_P$ $\Rightarrow$
$\tau_1$ thermalises via its interaction with hidden sector gauge
bosons.
\item $\tau_2$ couples to MSSM gauge bosons with strength
$g\sim\sqrt{\mathcal{V}}/M_P$ and to hidden sector gauge bosons
with strength $g\sim1/(\sqrt{\mathcal{V}}M_P)$ $\Rightarrow$
$\tau_2$ thermalises via its interaction with MSSM gauge bosons.
\end{itemize}
Hence in this case there are two separate thermal baths: one
contains $\tau_1$ and the hidden sector gauge bosons at
temperature $T_1$, whereas the other one is formed by $\tau_2$ and
the MSSM particles at temperature $T_2$. Generically, we would
expect that $T_1\neq T_2$ since the two thermal baths are not in
contact with each other.

\item If $\Gamma_1$ is supporting a pure $SU(N)$ theory, that undergoes gaugino condensation,
and $\Gamma_1+\Gamma_2$ is wrapped by MSSM branes with chiral
intersections on $\Gamma_2$:

\begin{itemize}
\item $\tau_1$ couples both to MSSM and hidden sector gauge bosons
with strength $g\sim\sqrt{\mathcal{V}}/M_P$ $\Rightarrow$ $\tau_1$
thermalises.
\item $\tau_2$ couples to MSSM gauge bosons with strength
$g\sim\sqrt{\mathcal{V}}/M_P$ and to hidden sector gauge bosons
with strength $g\sim 1/(\sqrt{\mathcal{V}}M_P)$ $\Rightarrow$
$\tau_2$ thermalises via its interaction with MSSM gauge bosons.
\end{itemize}
Unlike the previous case, now there is only one thermal bath, which contains
both $\tau_1$ and $\tau_2$ together with the MSSM particles and the hidden
sector gauge bosons, since in the present case $\tau_1$ interacts strongly
enough with the MSSM gauge bosons.
\end{enumerate}
We can now extend these results to the general case with
$h_{1,1}>3$ by noticing that a small 4-cycle wrapped by MSSM
branes will always thermalise via its interaction with MSSM gauge
bosons. On the other hand, for a 4-cycle that is not wrapped by
MSSM branes there are the following two options. If it is wrapped
by an ED3 instanton, it will not thermalise. If instead it is
supporting gaugino condensation, it will reach thermal equilibrium
with the hidden sector gauge bosons.

\subsection{K3 Fibration}
\label{Sec:K3FibrModTherm}

Let us now turn to the issue of moduli thermalisation for K3
fibrations. As we have seen in Subsection \ref{K3CanNorm}, there
is an essential difference between the cases when the K3 fiber is
stabilized at a large and at a small value. Let us consider
separately each of these two situations.

\medskip\noindent{\em Large K3 fiber}

\medskip\noindent

As we have already stressed in Subsection \ref{K3CanNorm}, in the
case `LV' where the K3 divisor is stabilised large, the small
modulus $\Phi$ plays exactly the same role as the small
modulus of the single-hole Swiss-cheese case, whereas both
$\chi_1$ and $\chi_2$ behave as the single large modulus. Hence we
can repeat the same analysis as in Subsection
\ref{Sec:11169ModTherm} and conclude that only $\Phi$ will reach
thermal equilibrium with the MSSM particles via its interaction
with the gauge bosons.

\medskip\noindent{\em Small K3 fiber}

\medskip\noindent

The study of moduli thermalisation in the case of small K3 fiber
is more complicated. We shall first focus on CY three-folds with
just one blow-up mode and later on will infer the general features
of the situation with several blow-ups.

K3 fibrations with $h_{1,1}=3$ are characterised by two small
moduli: $\tau_1$ that gives the volume of the K3 divisor
$\Gamma_1$, and $\tau_s$ which is the volume of the rigid divisor
$\Gamma_s$. The canonically normalised fields $\chi_1$ and $\Phi$
are defined by (\ref{K3Big}) and (\ref{K3SMALL}). We recall that one has
to be careful about the possible incompatibility of
MSSM branes on $\Gamma_s$ with the non-perturbative effects that this
cycle supports. Hence, to avoid dealing with such subtleties, below we will
assume that the MSSM branes are not wrapping $\Gamma_s$. Again, using the
results of Subsection \ref{Sec:11169ModTherm}, we infer the following for
the different brane set-ups below:

\begin{enumerate}
\item If $\Gamma_s$ is wrapped by an ED3 instanton and $\Gamma_1$ is wrapped by MSSM branes:

\begin{itemize}
\item $\chi_1$ couples to MSSM gauge bosons with strength
$g\sim 1/M_P$ $\Rightarrow$ $\chi_1$ does not thermalise.
\item $\Phi$ couples to MSSM gauge bosons more weakly than $\chi_1$
$\Rightarrow$ $\Phi$ does not thermalise.
\end{itemize}

\item If $\Gamma_s$ is wrapped by an ED3 instanton and $\Gamma_s+\Gamma_1$ is wrapped by MSSM
branes with chiral intersections on $\Gamma_1$:

\begin{itemize}
\item $\chi_1$ couples to MSSM gauge bosons with strength
$g\sim\sqrt{\mathcal{V}}/M_P$ $\Rightarrow$ $\chi_1$ thermalises.
\item $\Phi$ couples to MSSM gauge bosons with strength
$g\sim\sqrt{\mathcal{V}}/M_P$ $\Rightarrow$ $\Phi$ thermalises.
\end{itemize}

\item If $\Gamma_s$ is supporting a pure $SU(N)$ theory, that undergoes gaugino condensation,
and $\Gamma_1$ is wrapped by MSSM branes:

\begin{itemize}
\item $\chi_1$ couples to MSSM gauge bosons with strength
$g\sim 1/M_P$ and to hidden sector gauge boson with strength
$g\sim\sqrt{\mathcal{V}}/M_P$ $\Rightarrow$ $\chi_1$ thermalises
via its interaction with hidden sector gauge bosons.
\item $\Phi$ couples to MSSM gauge bosons more weakly than $\chi_1$
and to hidden sector gauge bosons with strength
$g\sim\sqrt{\mathcal{V}}/M_P$ $\Rightarrow$ $\Phi$ thermalises via
its interaction with hidden sector gauge bosons.
\end{itemize}
In this case, two separate thermal baths are established: one
contains $\chi_1$, $\Phi$ and the hidden sector gauge bosons at
temperature $T_1$, whereas the other one is formed by the MSSM
particles at temperature $T_2$. Generically, we expect that
$T_1\neq T_2$ since the two thermal baths are not in contact with
each other.

\item If $\Gamma_s$ is supporting a pure $SU(N)$ theory, that undergoes gaugino condensation,
and $\Gamma_s+\Gamma_1$ is wrapped by MSSM branes with chiral
intersections on $\Gamma_1$:

\begin{itemize}
\item $\chi_1$ couples both to MSSM and hidden sector gauge bosons
with strength $g\sim\sqrt{\mathcal{V}}/M_P$ $\Rightarrow$ $\chi_1$
thermalises.
\item $\Phi$ couples both to MSSM and hidden sector gauge bosons
with strength $g\sim\sqrt{\mathcal{V}}/M_P$ $\Rightarrow$ $\Phi$
thermalises.
\end{itemize}
Now only one thermal bath is established containing
$\chi_1$, $\Phi$, the hidden sector gauge bosons and the MSSM
particles, since both moduli interact with equal strength with the
gauge bosons of the MSSM and of the hidden sector.
\end{enumerate}
It is interesting to notice that both moduli $\chi_1$ and $\Phi$
thermalise in all situations, except when the blow-up mode is
wrapped by an ED3 instanton only. In this particular case, no
modulus thermalises. It is trivial to generalise these conclusions
for more than one blow-up mode and the MSSM still localised on the
K3 fiber.

On the other hand, if the MSSM is localised on one of the rigid divisors,
then for the case of more than one blow-up mode one can repeat the same
general conclusions as at the end of Subsection \ref{Sec:MultHoleModTherm},
with in addition the fact that $\chi_1$ will always thermalise as soon as
one of the blow-up modes thermalises. This is due to the leading order
mixing between $\Phi$ and any other small modulus, as can be seen explicitly
in (\ref{K3SMALL2}) and (\ref{K3SMALL3}).

\subsection{Modulini thermalisation}
\label{Sec:ModuliniTherm}

The study of modulini thermalisation is straightforward since, as
we have seen in Subsection \ref{Sec:Modulini}, the canonical
normalisation for the modulini takes exactly the same form as the
canonical normalisation for the moduli. This implies that, after
supersymmetrisation, the small modulino-gaugino-gauge boson
coupling has the same strength as the small modulus-gauge
boson-gauge boson coupling. Given that this is the relevant
interaction for moduli thermalisation, we can repeat the
same considerations as those in Subsections
\ref{Sec:11169ModTherm}-\ref{Sec:K3FibrModTherm}
%\ref{Sec:MultHoleModTherm}
and conclude that the modulini thermalise every time, when their
supersymmetric partners reach thermal equilibrium with the MSSM
thermal bath. Note however that, if for the moduli the relevant processes
are $2\leftrightarrow 2$ interactions with gauge bosons, the crucial
$2\leftrightarrow 2$ processes for the modulini are:
\begin{itemize}
\item $2\leftrightarrow 2$ processes with two gravitational vertices dominant for $\mathcal{V}<10^{10}$:
$\tilde{X}+\tilde{X}\leftrightarrow \tilde{\Phi}+\tilde{\Phi}$,
$X+X\leftrightarrow \tilde{\Phi}+\tilde{\Phi}$,
$\tilde{X}+\tilde{\Phi}\leftrightarrow \tilde{X}+\tilde{\Phi}$,
$X+\tilde{\Phi}\leftrightarrow X+\tilde{\Phi}$,
$\tilde{X}+\tilde{X}\leftrightarrow X+X$,
$\tilde{X}+X\leftrightarrow \tilde{X}+X$.
\item $2\leftrightarrow 2$ processes with one gravitational and one renormalisable
vertex dominant for $\mathcal{V}>10^{10}$:
$X+\tilde{\Phi}\leftrightarrow \tilde{X}+\tilde{X}$,
$\tilde{X}+\tilde{\Phi}\leftrightarrow X+X$.
\end{itemize}

\section{Finite temperature corrections in LVS}
\label{FTCLVS}

In this Section we study the finite temperature effective
potential in LVS. We show that it has runaway behaviour at high
$T$ and compute the decompactification temperature $T_{max}$. We
also investigate the cosmological implications of the small
modulus decay. By imposing that the temperature just after its
decay (regardless of whether or not that decay leads to reheating)
be less than $T_{max}$, in order to avoid decompactification of
the internal space, we find important restrictions on the range of
values of the CY volume.

\subsection{Effective potential}
\label{Sec:ExplicitTPot}

We shall now derive the explicit form of the finite temperature
effective potential for LVS, following the analysis of moduli
thermalisation performed in Section \ref{Sec:ModuliTherm}. We
will study in detail the behaviour of thermal corrections to the
$T=0$ potential of the simple $\mathbb{C}P^4_{[1,1,1,6,9]}$
model, and then realise that the single-hole Swiss-cheese case
already incorporates all the key properties of the general LVS.

\medskip\noindent{\em Single-hole Swiss-cheese}

\medskip\noindent

As we have seen in Section \ref{Sec:11169ModTherm}, not only
ordinary MSSM particles thermalise via Yang-Mills interactions but
also the small modulus and modulino reach thermal equilibrium with
matter via their interactions with the gauge bosons. Therefore,
the general expression (\ref{1-loop}) for the 1-loop finite
temperature effective potential, takes the following form:
\begin{equation}
V_T^{1-loop}=-\frac{\pi^2 T^4}{90} \left( g_B+ \frac{7}{8} g_F
\right)+\frac{T^2}{24}\left(m_{\Phi}^2+m_{\tilde{\Phi}}^2+\sum_i
M_{\rm{SOFT},i}^2\right)+...\,\,. \label{ONE-loop}
\end{equation}
We recall that (\ref{ONE-loop}) is a high temperature expansion of
the general 1-loop integral (\ref{klo}), and so it is valid only
for $T\gg m_{\Phi},m_{\tilde{\Phi}},M_{\rm{SOFT},i}$. The general
moduli-dependent expression for the modulino mass-squared
$m_{\tilde{\Phi}}^2$ is given by (\ref{ModulinoMass}) without the
vacuum expectation value. On the other hand, in the limit
$\tau_b\gg\tau_s$, $m_{\Phi}^2$ can be estimated as follows:
\begin{equation}
m_{\Phi}^2\simeq{\rm Tr} M^2_b = \frac{ K^{ij}}{2}
\frac{\partial^2 V_0} {\partial \tau_i\partial\tau_j}\simeq
\frac{K^{ss}}{2}\frac{\partial^2 V_0} {\partial \tau_s^2}.
\label{TraceMb}
\end{equation}
For $a_s\tau_s\gg 1$, the previous expression (\ref{TraceMb}), at
leading order, becomes:
\begin{equation}
m_{\Phi}^{2}\simeq \frac{A_s a_s^3 g_s
e^{K_{cs}}M_P^2}{\pi}\left(72 A_s a_s \tau_s
e^{-2a_{s}\tau_{s}}-\frac{3
W_0\tau_s^{3/2}e^{-a_s\tau_s}}{\sqrt{2}\mathcal{V}} \right).
\label{GenModulusMass}
\end{equation}
It can be shown that the gaugino and scalar masses arising from
gravity mediated SUSY breaking\footnote{The contribution from
anomaly mediation is subleading with respect to gravity mediation
as shown in \cite{SoftSUSY}.} are always parametrically smaller
than $m_{\Phi}$ and $m_{\tilde{\Phi}}$, and so we shall neglect
them. Moreover we shall drop also the $\mathcal{O}(T^4)$ term in
(\ref{ONE-loop}) since it has no moduli dependence. Therefore, the
relevant 1-loop finite-temperature effective potential reads:
\begin{equation}
V_T^{1-loop}=\frac{T^2}{24}\left(m_{\Phi}^2+m_{\tilde{\Phi}}^2\right)+...\,\,,
\label{1LOOP}
\end{equation}
which using (\ref{ModulinoMass}) and (\ref{GenModulusMass}), takes
the form:
\begin{equation}
V_{T}^{1-loop}=\frac{T^{2}}{24}\left( \frac{g_s
e^{K_{cs}}M_{P}^{2}}{ \pi }\right) \left[ \lambda _{1}\tau
_{s}e^{-2a_{s}\tau _{s}}-\lambda _{2}\left(4+a_s\tau_s\right)
\frac{\sqrt{\tau _{s}} e^{-a_{s}\tau
_{s}}}{\mathcal{V}}+\frac{W_{0}^{2}}{2\mathcal{V}^{2}}\right]
+...\,\,, \label{Full1-loop}
\end{equation}
with
\begin{equation}
\lambda _{1}\equiv 108 A_{s}^{2}a_{s}^{4},\text{ \ \ \ \ \ \
}\lambda _{2}\equiv 3 a_{s}^{2}A_{s}W_{0}/\sqrt{2}.
\end{equation}
Given that the leading contribution in (\ref{ONE-loop}), namely
the $\mathcal{O}(T^4)$ term, does not bring in any moduli
dependence, we need to go beyond the ideal gas approximation and
consider the effect of 2-loop thermal corrections, as the latter
could in principle compete with the terms in (\ref{Full1-loop}).
The high temperature expansion of the 2-loop contribution looks
like:
\begin{equation}
V_T^{2-loops}=T^4\left(\kappa_1 g_{MSSM}^2+\kappa_2 g_{\Phi XX}^2
m_{\Phi}^2+\kappa_3 g_{\tilde{\Phi}\tilde{X}X}^2
m_{\tilde{\Phi}}^2+...\right)+...\,\,, \label{TWO-loop}
\end{equation}
where the $\kappa$'s are $\mathcal{O}(1)$ coefficients and:
\begin{itemize}
\item the $\mathcal{O}(g_{MSSM}^2)$ contribution comes from two
loops involving MSSM particles,
\item the $\mathcal{O}(g_{\Phi XX}^2)$ contribution is due to two loop
diagrams with $\Phi$ and two gauge bosons,
\item the $\mathcal{O}(g_{\tilde{\Phi}\tilde{X}X}^2)$ contribution
comes from two loops involving the modulino $\tilde{\Phi}$, the
gaugino $\tilde{X}$ and the gauge boson $X$,
%\item the $\mathcal{O}(g_{\Phi^3}^2)$ term originates from two loops of
%the self-interacting modulus $\Phi$,
\item all the other two loop diagrams give
rise to subdominant contributions, and so they have been
neglected. Such diagrams are the ones with $\Phi$ or
$\tilde{\Phi}$ plus other MSSM particles, the self-interactions of
the moduli and of the modulini, and two loops involving both
$\Phi$ and $\tilde{\Phi}$. For example, the subleading
contribution originating from the two-loop vacuum diagram due to
the $\Phi^3$ self-interaction takes the form: $\delta
V_T^{2-loops}=\kappa_4 T^4 \frac{g_{\Phi^3}^2}{m_{\Phi}^2}\sim T^4
\frac{const}{{\cal V}(\ln {\cal V})^2}$.
\end{itemize}
Note that in (\ref{TWO-loop}) we have neglected the
$\mathcal{O}(T^2)$ term since it is subleading compared to both
the ${\cal O}(T^4)$ 2-loop term and the ${\cal O}(T^2)$ 1-loop
one. Now, the relevant gauge couplings in (\ref{TWO-loop}), have
the following moduli dependence:
\begin{itemize}
\item $g_{MSSM}^2=4\pi/\tau_s$ since we assume that the
MSSM is built via magnetised D7 branes wrapping the small cycle.
In the case of a supersymmetric $SU(N_c)$ gauge theory with $N_f$
matter multiplets, the coefficient $\kappa_1$ reads \cite{FTBook}:
\begin{equation}
\kappa_1=\frac{1}{64}\left(N_c^2-1\right)(N_c+3N_f)>0.
\label{kappa1}
\end{equation}
\item $g_{\Phi XX}^2\sim g_{\tilde{\Phi}\tilde{X}X}^2
\sim \frac{\sqrt{\mathcal{V}}}{M_P}$ as derived in
(\ref{ImpModuliniCoupl}) and (\ref{ImpModuliCoupl}).
\end{itemize}
Adding (\ref{1LOOP}) and (\ref{TWO-loop}) to the $T=0$ potential
$V_0$, we obtain the full finite temperature effective potential:
\begin{equation}
V_{TOT}=V_0+T^4\left(\kappa_1 g_{MSSM}^2+\kappa_2 g_{\Phi XX}^2
m_{\Phi}^2+\kappa_3 g_{\tilde{\Phi}\tilde{X}X}^2
m_{\tilde{\Phi}}^2\right)+\frac{T^2}{24}\left(m_{\Phi}^2
+m_{\tilde{\Phi}}^2\right)+...\,\,. \label{FULL}
\end{equation}
Despite the thermalisation of $\Phi$ and $\tilde{\Phi}$, which in
principle leads to a modification of $V_{TOT}$ compared to
previous expectations in the literature, we shall now show that
the thermal corrections due to $\Phi$ and $\tilde{\Phi}$ are, in
fact, negligible compared to the other contributions in
(\ref{FULL}), everywhere in the moduli space of these models. In
particular, the 2-loop MSSM effects dominate the
temperature-dependent term.\footnote{Note that this is consistent
with the results of \cite{AC} in the context of the O'KKLT model,
where it was also found that the T-dependent contribution of
moduli, that were assumed to be in thermal equilibrium, is
negligible compared to the dominant contribution of the rest of
the effective potential.}

Let us start by arguing that the $\mathcal{O}(T^4)$ corrections
arising from the modulus $\Phi$ and the modulino $\tilde{\Phi}$
are subleading compared to the 1-loop $\mathcal{O}(T^2)$ term.
Indeed, the relevant part of the effective scalar potential
(\ref{FULL}) may be rewritten as:
\begin{equation}
T^{4}\left( \kappa _{2}g_{\Phi XX}^{2}m_{\Phi }^{2}+\kappa
_{3}g_{\tilde{\Phi}\tilde{X}X}^{2}m_{\tilde{\Phi}}^{2}\right) \sim
T^{2}\left( m_{\Phi }^{2}+m_{\tilde{\Phi}}^{2}\right) \underset
{\left( \frac{T}{M_{s}}\right) ^{2}\ll
1}{\underbrace{T^{2}\frac{\mathcal{V} }{M_{P}^{2}}}},
\end{equation}
where the $<\!\!<$ inequality is due to the fact that our
effective field theory treatment makes sense only at energies
lower than the string scale $M_s$. Therefore, we can neglect the
effect of 2-loop thermal corrections involving $\Phi$ and
$\tilde{\Phi}$. So we see that, although the interactions of
$\Phi$ and $\tilde{\Phi}$ with gauge bosons and gauginos are
strong enough to make them thermalise, they are not sufficient to
produce thermal corrections large enough to affect the form of the
total effective potential. Let us also stress that this result is
valid everywhere in moduli space, i.e. for each value of
$m_{\Phi}^2$ and $m_{\tilde{\Phi}}^2$, not just in the region
around the zero-temperature minimum.

We now turn to the study of the general behaviour of the 1-loop
$\mathcal{O}(T^2)$ term arising from $\Phi$ and $\tilde{\Phi}$. We
shall show that it is always subdominant compared to the
zero-temperature potential (\ref{Vsimple}), and so it can be
safely neglected. In fact, the two relevant terms (\ref{Vsimple})
and (\ref{Full1-loop}) can be written as (ignoring the subleading
loop corrections in $V_0$):
\begin{equation}
V_0+\frac{T^2}{24}\left(m_{\Phi}^2
+m_{\tilde{\Phi}}^2\right)=\frac{g_s e^{K_{cs}}M_{P}^{4}}{8\pi
}\left[ p_{1}A_{1}\sqrt{\tau _{s}} \frac{e^{-2a_{s}\tau
_{s}}}{\mathcal{V}}-p_{2}A_{2}\frac{\tau _{s}e^{-a_{s}\tau
_{s}}}{\mathcal{V}^{2}}+p_{3}A_{3}\frac{1}{\mathcal{V}^{3}}
\right] ,  \label{VSimple}
\end{equation}
with
\begin{equation}
p_{1}=36 a_s^4 A_s^2,\text{ \ \ \ \ \ \
}p_{2}=4a_{s}A_{s}W_{0},\text{ \ \ \ \ \ }p_{3}=W_{0}^{2}/6,
\label{Numbers}
\end{equation}
and
\begin{equation}
A_{1}\equiv \frac{2\sqrt{2}}{3 a_s^2}+\underset{\left(
\frac{T}{M_{KK}}\right) ^{2}\ll 1}{
\underbrace{\frac{T^{2}\mathcal{V}\sqrt{\tau
_{s}}}{M_{P}^{2}}}},\text{ \ \ }A_{2}\equiv
1+\frac{a_s^2}{4\sqrt{2}}\underset{\left( \frac{T}{M_{KK}}\right)
^{2}\ll 1}{\underbrace{\frac{T^{2}\mathcal{V}\sqrt{\tau
_{s}}}{M_{P}^{2}}}} \left( 1+\frac{4}{a_{s}\tau _{s}}\right)
,\text{ \ \ }A_{3}\equiv \frac{9\hat{\xi}}{2}+\underset{\left(
\frac{T}{M_{s}}\right) ^{2}\ll 1}{\underbrace{
\frac{T^{2}\mathcal{V}}{M_{P}^{2}}.}} \notag
\end{equation}
where the appearance of the Kaluza-Klein scale comes from the
assumption that the MSSM branes are wrapping the small cycle
$\tau_s$:
\begin{equation}
M_{KK}\sim\frac{M_s}{\tau_s^{1/4}}\simeq\frac{M_P}{\sqrt{\mathcal{V}}\tau_s^{1/4}}
\, . \label{Mkk}
\end{equation}
Therefore, we can see that the 1-loop $\mathcal{O}(T^2)$ thermal
corrections can never compete with $V_0$ for temperatures below
the compactification scale $M_{KK}<M_s$, where our low energy
effective field theory is trustworthy. Once again, we stress that
the previous considerations are valid in all the moduli space
(within our large volume approximations) and not just in the
vicinity of the $T=0$ minimum. We have seen that the only
finite-temperature contribution that can compete with $V_0$ is the
2-loop $T^4 g_{MSSM}^2$ term, and so we can only consider from now
on the following potential:
\begin{equation}
V_{TOT}=V_0+4\pi\kappa_1 \frac{T^4}{\tau_s}=\left(\frac{g_s
e^{K_{cs}}}{8\pi}\right)\left[\frac{\lambda \sqrt{\tau_s} e^{-2
a_s \tau_s}}{{\cal V}} - \frac{\mu \tau_s e^{-a_s \tau_s}}{{\cal
V}^2} + \frac{\nu}{{\cal V}^3}+\frac{4\pi\tilde{\kappa}_1}{\tau_s}
\left(\frac{T}{M_P}\right)^4\right]M_P^4, \label{effPot}
\end{equation}
valid for temperatures $T\gg M_{SOFT}$, and with the constants
given in (\ref{numbers}) and (\ref{kappa1})\footnote{For
convenience, here we have redefined $\tilde{\kappa}_1\equiv8
\pi\kappa_1 g_s^{-1} e^{-K_{cs}}$.}. We realize that the leading
moduli-dependent finite temperature contribution to the effective
potential comes from 2-loops instead of 1-loop. This, however,
does not mean that perturbation theory breaks down, since 1-loop
effects still dominate when one takes into account the moduli
independent $\mathcal{O}(T^4)$ piece that we dropped.

Now, from (\ref{effPot}) it is clear that the thermal correction
cannot induce any new $T$-dependent extremum of the effective
potential. Its presence only leads to destabilization of the $T=0$
minimum at a certain temperature, above which the potential has a
runaway behaviour. Therefore, we are led to the following
qualitative picture. Let us assume that at the end of inflation
the system is sitting at the $T=0$ minimum. Then, after reheating
the MSSM particles thermalise and the thermal correction $T^4
g_{MSSM}^2\sim T^4/\tau_s$ gets switched on. As a result, the
system starts running away along the $\tau_s$ direction only,
since $V_T$ does not depend on ${\cal V}$. However, as soon as
$\tau_s$ becomes significantly larger than its $T=0$ VEV, the two
exponential terms in (\ref{effPot}) become very suppressed with
respect to the $\mathcal{O}(\mathcal{V}^{-3})$ $\alpha'$
correction (the $\nu$ term). Hence, the potential develops a
run-away behaviour also along the $\mathcal{V}$-direction, thus
allowing the K\"{a}hler moduli to remain within the K\"{a}hler
cone.

In Section \ref{Sec:DecTemp}, we shall compute the
decompactification temperature, at which the $T=0$ minimum gets
destabilised. Hence we shall focus on the region in the vicinity
of the zero-temperature minimum, where
the regime of validity of the expression (\ref{effPot}) takes the
form:
\begin{equation}
M_{SOFT}\ll T \ll M_{KK}\textit{ \ \ \ }\Leftrightarrow\textit{ \
\ \
}\frac{1}{\mathcal{V}\ln\mathcal{V}}\ll\frac{T}{M_P}\ll\frac{1}
{\sqrt{\mathcal{V}}\tau_s^{1/4}}. \label{validity}
\end{equation}
In the typical LVS where $\mathcal{V}\sim 10^{14}$ allows low
energy SUSY, we get $M_{SOFT}\sim 10^{3}$ GeV and $M_{KK}\sim
10^{11}$ GeV; thus, in that case, eq. (\ref{effPot}) makes sense
only for energies $10^3$ GeV $\ll T\ll 10^{11}$ GeV. On the other
hand, for LVS that allow GUT string scenarios, $\mathcal{V}\sim
10^{4}$, which implies $M_{SOFT}\sim 10^{13}$ GeV and $M_{KK}\sim
10^{16}$ GeV; thus, in that case, (\ref{validity}) becomes
$10^{13}$ GeV $\ll T\ll 10^{16}$ GeV.

\medskip\noindent{\em General LARGE Volume Scenario}

\medskip\noindent

As we have seen in Section \ref{ELV}, one of the conditions on an
arbitrary Calabi-Yau to obtain LVS, is the presence of a blow-up
mode resolving a point-like singularity (del Pezzo 4-cycle). The
moduli scaling of the scalar potential, at leading order and in
the presence of $N_{small}$ blow-up modes $\tau_{s_{i}}$,
$i=1,...,N_{small}$, is still of the form (\ref{Vsimple})
(neglecting loop corrections):
\begin{equation}
V_0=\left(\frac{g_s e^{K_{cs}}M_P^4}{8
\pi}\right)\left[\sum_{i=1}^{N_{small}}\left(\frac{\lambda
\sqrt{\tau_{s_{i}}} e^{-2 a_{s_{i}} \tau_{s_{i}}}}{{\cal V}} -
\frac{\mu \tau_{s_{i}} e^{-a_{s_{i}} \tau_{s_{i}}}}{{\cal
V}^2}\right) + \frac{\nu}{{\cal V}^3}\right]. \label{GenVsimple}
\end{equation}
All the other moduli which are neither the overall volume nor a
blow-up mode will appear in the scalar potential at subleading
order. Moreover, due to the topological nature of $\tau_{s,i}$,
$K^{-1}_{s_{i}s_{i}}\sim \mathcal{V}\sqrt{\tau_{s_{i}}}$ $\forall
i=1,...,N_{small}$ \cite{ccq2}.

As derived in Section \ref{Sec:ModCoupl}, these blow-up modes
correspond to the heaviest moduli and modulini, which play the
same role as $\Phi$ and $\tilde{\Phi}$ in the single-hole
Swiss-cheese case. Hence the leading order behaviour of the
mass-squareds of the blow-up moduli $\tau_{s_{i}}$ and the
corresponding modulini $\tilde{\tau}_{s_{i}}$ are still given by
(\ref{GenModulusMass}) and (\ref{ModulinoMass}) $\forall
i=1,...,N_{small}$. Therefore we can repeat the same
considerations made in the previous paragraph and conclude that,
for a general LVS, the 1-loop $\mathcal{O}(T^2)$ thermal
corrections are always subdominant with respect to $V_0$ for
temperatures below the compactification scale\footnote{As we have
seen in Section \ref{Sec:MultHoleModTherm}, if all the
$\tau_{s_{i}}$ are wrapped by ED3 instantons then they do not
thermalise. Only the moduli corresponding to 4-cycles wrapped by
MSSM branes would then thermalise but, since they are lighter than
the ED3 moduli, our argument is still valid. The same is true for
all the possible scenarios outlined for the K3 fibration case in
Section \ref{Sec:K3FibrModTherm}.}. The only finite-temperature
contribution that can compete with $V_0$ is again the 2-loop $T^4
g_{MSSM}^2$ term.

\subsection{Decompactification temperature}
\label{Sec:DecTemp}

As we saw in the previous subsection, the finite temperature
corrections destabilize the large volume minimum of a general LVS.
In this subsection we will derive the decompactification
temperature $T_{max}$, that is the temperature above which the full
effective potential has no other minima than the one at infinity.

Before performing a more precise calculation of $T_{max}$, let us
present a qualitative argument that gives a good intuition for
its magnitude. Let us denote by $V_b$ the height
of the potential barrier that separates the
supersymmetric minimum at infinity from the zero temperature SUSY
breaking one. Now, in order for the moduli to overcome the potential
barrier and run away to infinity, one needs to supply energy of at
least the same order of magnitude as $V_b$. In our case, the source
of energy is provided by the finite-temperature effects, which give a
contribution to the scalar potential of the order $V_T\sim T^4$.
Hence a very good estimate for the decompactification temperature
is given by $T_{max}\sim V_b^{1/4}$.

It is instructive to compare the implications of this estimate for
the KKLT and LVS cases. In the simplest KKLT models the potential
reads:
\begin{equation}
V_{KKLT}=\lambda_1\frac{e^{-2a\tau}}{\tau}-\lambda_2 W_0
\frac{e^{-a\tau}}{\tau^2},
\end{equation}
where $\lambda_1$ and $\lambda_2$ are constants of order unity.
The minimum is achieved by fine tuning the flux parameter $W_0\sim
\tau e^{-a\tau}$ and so the height of the barrier is given by
\begin{equation}
V_b\sim \langle V_{KKLT}\rangle\sim
\frac{W_0^2}{\mathcal{V}^2}M_P^4\sim m_{3/2}^2 M_P^2,
\end{equation}
where we have used the fact that $\mathcal{V}=\tau^{3/2}$ and
$m_{3/2}=W_0 M_P/\cal V$. Therefore the decompactification
temperature becomes $T_{max}\sim\sqrt{m_{3/2}M_P}\sim 10^{10}$
GeV, as estimated in \cite{BHLR}.

In the case of LVS, the height of the barrier is lower and so we
expect a lower decompactification temperature $T_{max}$. Indeed,
to leading order the potential is given by
\begin{equation}
V_{LVS}=\lambda_1\sqrt{\tau_s}\frac{e^{-2a_s\tau_s}}{\cal
V}-\lambda_2 W_0
\tau_s\frac{e^{-a_s\tau_s}}{\mathcal{V}^2}+\lambda_3
\frac{W_0^2}{\mathcal{V}^3}
\end{equation}
with $\lambda_1$, $\lambda_2$ and $\lambda_3$ being constants of
order one, as reviewed in Section \ref{Sec:Review}. The minimum is
achieved for natural values of the flux parameter $W_0\sim
\mathcal{O}(1)$ and at exponentially large values of the overall
volume $\mathcal{V}\sim W_0\sqrt{\tau_s}e^{a_s\tau_s}$. Hence the
height of the barrier can be estimated as:
\begin{equation}
V_b\sim \langle V_{LVS}\rangle\sim
\frac{W_0^2}{\mathcal{V}^3}M_P^4\sim m_{3/2}^3 M_P,
\end{equation}
which gives a decompactification temperature of the order:
\begin{equation}
T_{max}\sim\left(m_{3/2}^3
M_P\right)^{1/4}\sim\frac{M_P}{\mathcal{V}^{3/4}}.
\label{KeyResult}
\end{equation}

Let us now turn to a more precise computation. Without loss of
generality, we shall focus here on the effective potential
(\ref{effPot}), valid for the single-hole Swiss-cheese case, and
look for its extrema. Given that the thermal contribution
does not depend on the volume, the derivative of the potential
with respect to $\mathcal{V}$ gives the same result as in the
$T=0$ case:
\begin{equation}
\frac{\partial V_{TOT}}{\partial \mathcal{V}}=0\text{ \ \ \ \ }
\Longrightarrow \text{ \ \ \ \ }\mathcal{V}_*=\frac{\mu}{\lambda
}A(\tau_s)\sqrt{\tau_s}e^{a_s\tau_s}, \label{Vpmc}
\end{equation}
where\footnote{We discard the solution with the positive sign
in front of the square root in (\ref{A}) since, upon its substitution one
finds that the other extremum condition, $\partial V_{TOT}/\partial\tau_s=0$,
does not have any solution.}
\begin{equation}
A(\tau_s)\equiv 1-\sqrt{1-\frac{3}{4}\left(
\frac{\langle\tau_s\rangle}{\tau _{s}}\right) ^{3/2}}, \label{A}
\end{equation}
and $\langle\tau_s\rangle \simeq \left(4 \lambda \nu/\mu^2\right)^{2/3}$
is the $T=0$ VEV of $\tau_s$.
Substituting (\ref{Vpmc}) in the derivative of $V_{TOT}$ with
respect to $\tau_s$ and working in the limit $a_s \tau_s\gg 1$, in
which one can neglect higher order instanton corrections, we obtain:
\begin{equation}
\left. \frac{\partial V_{TOT}}{\partial \tau _{s}}\right\vert
_{\mathcal{V}= \mathcal{V}_*}=0\text{ \ \ }\Longrightarrow \text{
\ \ }4\pi \tilde{\kappa}_1 \frac{ \mu e^{3a_{s}\tau _{s}}}{\lambda
^{2}a_{s}\tau _{s}^{2}}\left( \frac{T}{M_{P}}\right)^{4}\!A(\tau
_{s})^{2}+2A(\tau _{s})-1=0.  \label{tausc}
\end{equation}
Notice that at zero temperature (\ref{tausc}) simplifies to
$A(\tau_s)=1/2$, which from (\ref{A}) correctly implies
$\tau_s=\langle\tau_s\rangle$.
Now, since equation (\ref{tausc}) is transcendental, one cannot write
down an analytical solution, that gives the general relation between the
location of the $\tau_s$ extrema and the temperature. Nevertheless, we will
see shortly that it is actually possible to extract an analytic estimate
for the decompactification temperature. To understand why, let us gain
insight into the behaviour of the function on the LHS of (\ref{tausc})
by plotting it and looking at its intersections with the $\tau_s$-axis.

\begin{center}
\FIGURE[ht]{ \epsfig{file=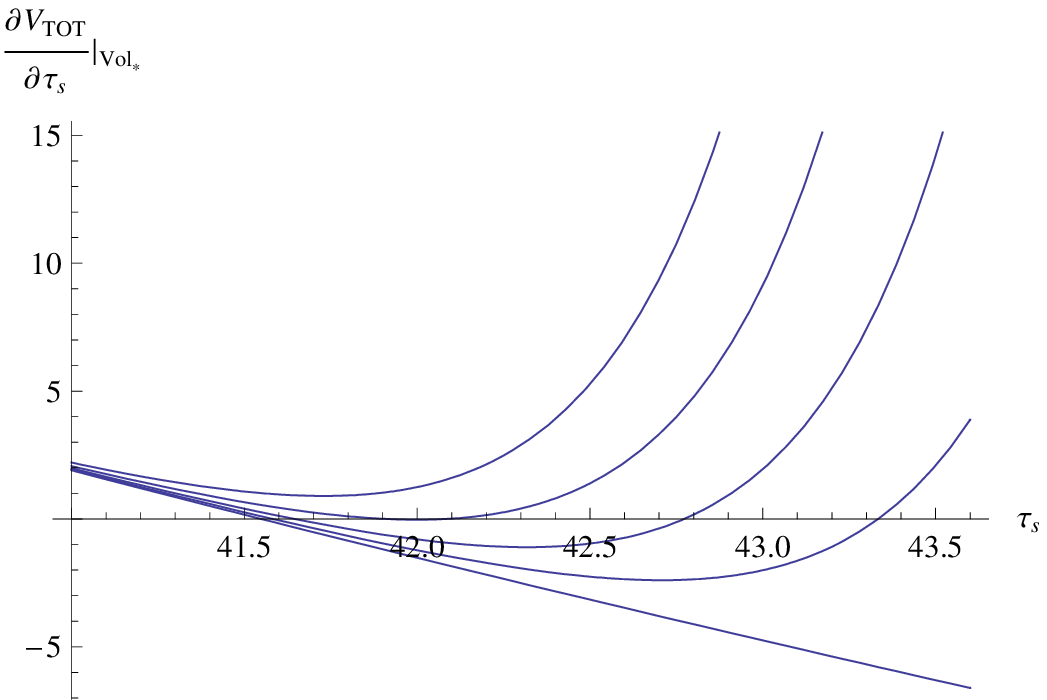, height=70mm,width=100mm}
\caption{The LHS of eq. (\ref{tausc}) is plotted versus $\tau_s$.
The temperature increases from right to left. The straight line
represents the zero temperature case. The other values of the
temperature are $T/M_P= 0.8 \cdot 10^{-10}$, $1.0 \cdot 10^{-10}$,
$1.2 \cdot 10^{-10}$, $1.4 \cdot 10^{-10}$. To obtain the plots we
used the following numerical values: $\xi =1.31$, $A_s=1$,
$W_0=1$, $a_s=\pi/4$, $e^{K_{cs}}= 8\pi /g_s$, $g_s=0.1$, $N_c=5$,
$N_f=7$. With these values one has that
$\langle\tau_{s} \rangle = 41.55$ and $\langle \mathcal{V} \rangle
= 7.02\cdot 10^{13}$, which implies that $T_{max}=1.58 \cdot 10^{-10}
M_P\simeq 3.79 \cdot 10^{8}$ GeV according to (\ref{final}). Note that
the numerically found value of the decompactification temperature is
$T_{max,num} = 1.20 \cdot 10^{-10} M_P$.} \label{Fig:dVdtaus} }
\end{center}
We plot the LHS of equation (\ref{tausc}) on Figure
\ref{Fig:dVdtaus} for several values of the temperature; $T$
increases from right to left. From this figure it is easy to see
that the temperature-dependent correction to $V_{TOT}$ behaves
effectively as an up-lifting term. Namely, the finite-temperature
contribution lifts the potential, giving rise to a local maximum
(the right intersection with the $\tau_s$ axis) in addition to the
$T=0$ minimum (the left intersection). As the temperature
increases, the maximum increases as well and shifts towards
smaller values of $\tau_s$. On the other hand, the minimum remains
very close to the zero-temperature one at all temperatures.
Clearly, the decompactification temperature $T_{max}$ is reached
when the two extrema coincide. The key observation here is that
this happens in a small neighborhood of the $T=0$ minimum, located
at $\la \tau_s \ra \simeq \left(4 \lambda \nu/\mu^2\right)^{2/3}$.

In view of the considerations of the previous paragraph, to find
an analytic estimate for $T_{max}$ we shall utilize the following
strategy. We will Taylor-expand the function $F(\tau_s)$, defined
by the LHS of equation (\ref{tausc}), to second order in a small
neighborhood of the point $\tau_s = \la \tau_s \ra$. Then we will
use the resulting quadratic function $f(\delta)$, where $\delta
\equiv \tau_s - \la \tau_s \ra$, as an approximation of
$F(\tau_s)$ in a larger neighborhood and will look for the zeros
of $f(\delta)$. Requiring that the two roots of $f(\delta)$
coincide, will give us an estimate for the decompactification
temperature. Clearly, this procedure is not exact. In particular,
the function $F(\tau_s)$ is better approximated by keeping higher
orders in the Taylor expansion. In our case, we have checked
numerically that a really good approximation is obtained by going
to at least sixth order. However, in doing so one again ends up
with an equation that cannot be solved analytically. So the key
point is that the systematic error introduced by the quadratic
approximation is rather small (we have checked that the analytical
results obtained by following the above procedure are in very good
agreement with the exact numerical values).
%%that the difference between the value of $T_d$, obtained by
%%following the above procedure, and the numerical solution of
%%(\ref{tausc}), is never greater than $\pm 2$.

Now let us substitute $\tau_s = \langle\tau_s\rangle + \delta$ in
(\ref{tausc}) and read off the terms up to order $\delta^2$. The
result is:
\begin{equation}
a\,\delta^2 + b\,\delta +c =0, \label{eqcri}
\end{equation}
where the corresponding coefficients, in the limit
$a_s\langle\tau_s\rangle\gg 1$, take the form:
\begin{equation}
\left\{
\begin{array}{c}
a\simeq \frac{9}{2}\mathcal{T}a_{s}^{2}+\frac{171}{8}\lambda
^{2}a_{s}, \\
b\simeq 3\mathcal{T}a_{s} -9\lambda
^{2}a_{s}\langle \tau _{s}\rangle,\text{ \ \ \ }c\simeq
\mathcal{T},
\end{array}
\right.
\end{equation}
and we have set
\begin{equation}
{\cal T} \equiv 4 \pi \tilde{\kappa}_1
\left(\frac{T}{M_P}\right)^4 \mu e^{3a_s \langle\tau_s\rangle}.
\end{equation}
Finally, to find the decompactification temperature, we require
that the two solutions $\delta_1$ and $\delta_2$ coincide:
\begin{equation}
\delta_1=\delta_2\text{ \ \ \ }\Longleftrightarrow\text{ \ \ \
}b^2-4a\, c =0,
\end{equation}
which, for $a_s \la \tau_s \ra \gg 1$, gives:
\begin{equation}
{\cal T}_{max} = 3 (\sqrt{2}-1)\lambda ^2 \langle\tau_s\rangle
\text{ \ \ \ }\Longleftrightarrow\text{ \ \ \ }T_{max}^4= \frac{3
(\sqrt{2}-1) \lambda^2 \langle\tau_s\rangle }{4 \pi
\tilde{\kappa}_1 \mu} e^{-3a_s \langle\tau_s\rangle} M_P^4.
\end{equation}
Notice that we can rewrite the decompactification temperature in
terms of $\mathcal{V}$ as:
\begin{equation}
T_{max}^4=\frac{3(\sqrt{2}-1)}{32 \pi}\frac{\mu^2}{\lambda
\tilde{\kappa}_1}\frac{\langle\tau_s\rangle^{5/2}}
{\mathcal{V}^{3}}M_P^4\text{ \ \ \ }\Longrightarrow\text{ \ \ \
}T_{max}\sim\left(m_{3/2}^3
M_P\right)^{1/4}\sim\frac{M_P}{\mathcal{V}^{3/4}}\, ,
\label{final}
\end{equation}
where we have used the relation between the $T=0$ VEV of the
volume and $\langle\tau_s\rangle$, which is given by (\ref{Vpmc})
with $\tau_s=\langle\tau_s\rangle$ and $A=1/2$. It is reassuring
that (\ref{final}) is of the same form as the
result (\ref{KeyResult}), obtained from the
intuitive arguments based on the height of the potential barrier.

\subsection{Small moduli cosmology}
\label{Sec:MaxTemp}

Clearly, the decompactification temperature (\ref{final}) sets an
upper bound on the temperature in the early Universe, in
particular on the reheating temperature, $T_{RH}^0$, at the end of
inflation. We will investigate now how this constraint affects the
moduli thermalisation picture studied in Subsection
\ref{Sec:11169ModTherm}.\footnote{Similar considerations apply for
the more general multiple-hole Swiss-cheese and K3 fibration
cases.}

Recall that there we derived the
following:
\begin{itemize}
\item For small values of
the volume ($\mathcal{V}<10^{10}$), the freeze-out temperature for
the small modulus $\Phi$ is given by (\ref{caseA}): $T_f^{SV}\sim
M_P \mathcal{V}^{-2/3}$.

\item For large values of
the volume ($\mathcal{V}>10^{10}$), the freeze-out temperature for
$\Phi$ is given by (\ref{caseB}): $T_f^{LV}\sim 10^3 M_P
\mathcal{V}^{-1}$.
\end{itemize}
Note also that, in both cases, the condition $T_f<T_{RH}^0<T_{max}$ has
to be satisfied in order for the modulus to reach equilibrium
with the MSSM thermal bath. Now, for
small values of $\mathcal{V}$ we have that:
\begin{equation}
\frac{T_{max}}{T_f^{SV}}\sim\frac{\mathcal{V}^{2/3}}{\mathcal{V}^{3/4}}
=\mathcal{V}^{-1/12}<1,
\end{equation}
which implies that $\Phi$ actually never thermalises.
On the other hand, for large values of
$\mathcal{V}$ we have that (writing $\mathcal{V}\sim 10^x$):
\begin{equation}
\frac{T_{max}}{T_f^{LV}}\sim\frac{\mathcal{V}^{1/4}}{10^3}
=10^{x/4-3}>1\text{ \ \ }\Leftrightarrow\text{ \ \ }x>12.
\end{equation}
Hence, for $\mathcal{V}>10^{12}$, $\Phi$ can reach thermal
equilibrium with the MSSM plasma, as long as $T_{RH}^0$
is such that $T_f^{LV}<T_{RH}^0<T_{max}$. Let us stress,
however, that if $T_{RH}^0<T_f^{LV}$ the modulus will never
thermalise even though $T_f^{LV}<T_{max}$. Note that, since
the temperature $T_{RH}^0$ depends on the concrete realization
of inflation and the details of the initial reheating process,
its determination is beyond the scope of the present paper.
So we will treat it as a free parameter, satisfying only the
constraint $T_{RH}^0<T_{max}$.

We would like now to study the cosmological history of $\Phi$
which, in our case, presents two possibilities:
\begin{enumerate}
\item The modulus $\Phi$ decays at the end of
inflation being the main responsible for initial reheating. We may
envisage two physically different situations where this could
happen: in one case, $\Phi$ is the inflaton and it decays at the
end of inflation. In the other case, $\Phi$ is not the inflaton,
but it starts oscillating around its VEV when the inflaton is
still driving inflation by rolling down its flat potential. In
this case, the decay of $\Phi$ occurs just after the slow-roll
conditions stop being satisfied and the inflaton reaches its VEV.

After $\Phi$ decays, its energy density is converted into
radiation. The decay products thermalise rapidly and re-heat the
Universe to a temperature $T_{RH}=T_{RH}^0$. The latter can be
computed by noticing that the $\Phi$ energy density
$\rho_{\Phi}\sim \Gamma_{\Phi\to XX}^2 M_P^2$ will be converted
into radiation energy density $\rho_{R}\sim g_{*}T^4$. Hence
$T_{RH}^0$ can be obtained by comparing $\Gamma_{\Phi\to XX}$ with
the value of $H$, given by the Friedmann equation for radiation
dominance:
\begin{gather}
\Gamma _{\Phi \rightarrow XX}\sim \frac{\ln \mathcal{V}}{16\pi
}\frac{ m_{\Phi }^{2}}{M_{P}}\simeq H\sim g_{\ast
}^{1/2}\frac{\left(
T_{RH}^{0}\right) ^{2}}{M_{P}}  \notag \\
\Leftrightarrow\text{ \ } T_{RH}^{0}\simeq \left( \frac{\ln
{\mathcal{V}}}{16\pi \sqrt{g_{\ast }}}\right) ^{1/2}m_{\Phi
}=\frac{\left( \ln {\mathcal{V}} \right) ^{3/2}}{4\sqrt{\pi
}g_{\ast }^{1/4}}\frac{M_{P}}{\mathcal{V}}. \label{222}
\end{gather}
In order for this picture to be compatible with the presence of a
decompactification temperature (\ref{final}), that sets the
maximal temperature of the Universe, we need to require that
$T_{RH}^0<T_{max}$. As we shall see in Subsection
\ref{Sec:VolBound}, this requirement can be translated into a
constraint on the values that the internal volume can take.

\item The modulus $\Phi$ is not the main source of initial reheating,
which we suppose to be the inflaton. After the inflaton decays,
the Universe is re-heated to a temperature $T_{RH}^0$ and an epoch
of radiation dominance begins. The modulus $\Phi$ will only
thermalise if $\mathcal{V}>10^{12}$ and $T_f^{LV}<T_{RH}^0$.
However, $T_f^{LV}$ is rather close to $T_{max}$ and so, even when
$\Phi$ thermalises, it will drop out of equilibrium very quickly
at $T_f^{LV}$. Then, for general values of $\mathcal{V}$, the
modulus $\Phi$ will decay out of equilibrium at a temperature
$T_D<T_{RH}^0$. As we shall show below, this decay will occur
during radiation domination, since $T_D>T_{dom}$, with $T_{dom}$
being the temperature at which the modulus energy density would
dominate over the radiation energy density. So the temperature
$T_D$ at which $\Phi$ decays, is still given by (\ref{222}) upon
replacing $T_{RH}^0$ with $T_D$:
\begin{equation}
T_D\simeq
\frac{\left(\ln{\mathcal{V}}\right)^{3/2}}{4\sqrt{\pi}g_{*}^{1/4}}
\frac{M_P}{\mathcal{V}}. \label{DUE}
\end{equation}
Note that the above expression satisfies $T_D<T_f^{SV,LV}$, as
should be the case for consistency. Another important observation
is that (\ref{DUE}) is also the usual expression for the
temperature $T_{RH}$, to which the Universe is re-heated by the
decay of a particle releasing its energy to the thermal bath. In
other words, for us $T_{RH}=T_D$ since the modulus $\Phi$ decays
during radiation domination. On the contrary, if a modulus decays
when its energy density is dominating the energy density of the
Universe, then $T_D < T_{RH}$ and the decay produces an increase
in the entropy density $S$, which is determined by:
\begin{equation}
\Delta \equiv
\frac{S_{fin}}{S_{in}}\sim\left(\frac{T_{RH}}{T_D}\right)^3.
\label{Delta}
\end{equation}
As already mentioned, since for us $T_{RH}=T_D$, the decay of
$\Phi$ does not actually lead to reheating or, equivalently, to an
increase in the entropy density, given that from (\ref{Delta}) we
have $\Delta = 1$. As a consequence, $\Phi$ cannot dilute any
unwanted relics, like for example the large modulus $\chi$ which
suffers from the cosmological moduli problem.\footnote{This kind
of solution of the cosmological moduli problem, i.e. dilution via
saxion or modulus decay, is used both in \cite{Vafa} and in
\cite{Acharya}.}

To recapitulate: in the present case 2, we have the following system
of inequalities:
\begin{eqnarray}
\text{for }\mathcal{V} &<&10^{12}\text{: \ \ \
}T_{dom}<T_{D}<T_{RH}^{0}<T_{max}, \\
\text{for }\mathcal{V} &>&10^{12}\text{: \ \ \ }
T_{dom}<T_{D}<T_{f}^{LV}<T_{RH}^{0}<T_{max}.
\end{eqnarray}
As in case 1 above, the condition $T_D<T_{max}$ implies a
constraint on $\mathcal{V}$, that we will derive in Subsection
\ref{Sec:VolBound}. We underline again that this condition is
necessary but not sufficient, since for us $T_{RH}^0$ is an
undetermined parameter. In concrete models, in which one could
compute $T_{RH}^0$, the condition $T_{RH}^0 < T_{max}$ might lead
to further restrictions.
\end{enumerate}

Let us now prove our claim above that, when the modulus $\Phi$ is
not responsible for the initial reheating (case 2), it will decay
before its energy density begins to dominate the energy density of
the Universe. $\Phi$ will start oscillating around its VEV when
$H\sim m_{\Phi}$ at a temperature $T_{osc}$ given by:
\begin{equation}
T_{osc}\sim g_{*}^{-1/4}\sqrt{m_{\Phi}M_P}.  \label{Tosc}
\end{equation}
The energy density $\rho_{\Phi}$, stored by $\Phi $, and the ratio
between $\rho_{\Phi}$ and the radiation energy density at
$T_{osc}$ read as follows:
\begin{equation}
\left. \rho _{\Phi }\right\vert _{T_{osc}}\sim m_{\Phi
}^{2}\langle \tau_s \rangle ^{2}\text{ \ \ }\Rightarrow \text{ \ \
}\left. \left( \frac{\rho _{\Phi }}{\rho _{r}}\right) \right\vert
_{T_{osc}}\sim\frac{m_{\Phi }^{2}\langle \tau_s \rangle
^{2}}{g_{\ast }T_{osc}^{4}}\sim \frac{\langle \tau_s \rangle
^{2}}{M_{P}^{2}}.
\end{equation}
By definition, the temperature $T_{dom}$, at which $\rho_{\Phi}$
becomes comparable to $\rho_r$ and hence $\Phi$ begins to dominate
the energy density of the Universe, is such that:
\begin{equation}
\left. \left( \frac{\rho _{\Phi }}{\rho _{r}}\right) \right\vert
_{T_{dom}}\sim 1.
\end{equation}
Now, given that $\rho_{\Phi}$ redshifts as $T^3$ whereas $\rho_r$
scales as $T^4$, we can relate $T_{dom}$ with $T_{osc}$:
\begin{equation}
T_{dom}\left. \left( \frac{\rho _{\Phi }}{\rho _{r}}\right)
\right\vert _{T_{dom}}\sim T_{osc}\left. \left( \frac{\rho _{\Phi
}}{\rho _{r}}\right) \right\vert _{T_{osc}}\text{ \ \
}\Leftrightarrow \text{ \ \ }T_{dom}\sim
g_*^{-1/4}\frac{\langle\tau_s\rangle^2}{M_P^2}\sqrt{m_{\Phi}M_P}.
\label{Tdom}
\end{equation}
We shall show now that $T_{dom}<T_D$ with $T_D$ being the decay
temperature during radiation dominance, which is obtained by comparing
$H$ with $\Gamma_{\Phi\to XX}$:
\begin{equation}
T_D\sim g_{*}^{-1/4}\sqrt{\Gamma_{\Phi\to XX}M_{P}}. \label{TD}
\end{equation}
The ratio of (\ref{TD}) and (\ref{Tdom}) gives:
\begin{equation}
\frac{T_{D}}{T_{dom}}\sim \frac{\sqrt{\Gamma _{\Phi \rightarrow
XX}}}{\sqrt{ m_{\Phi }}}\frac{M_{P}^{2}}{\langle \tau _{s}\rangle
^{2}}.
\end{equation}
Using that $\Gamma _{\Phi \rightarrow XX}\sim \mathcal{V}m_{\Phi
}^{3}M_{P}^{-2}$ and $\langle \tau _{s}\rangle \sim 10 M_s \sim
10M_{P}\mathcal{V}^{-1/2}$, the last relation becomes:
\begin{equation}
\frac{T_{D}}{T_{dom}}\sim \frac{\left( \ln \mathcal{V}\right)
\sqrt{\mathcal{ V}}}{100}>1\text{ \ for \ }\mathcal{V}>10^{2.5}.
\end{equation}
Hence, we conclude that $T_D>T_{dom}$ and, therefore, $\Phi$ decays
before it can begin to dominate the energy density of the Universe.
The main consequence of this is that $\Phi$ cannot
dilute unwanted relics via its decay.

\subsection{Lower bound on $\mathcal{V}$}
\label{Sec:VolBound}

As we saw in the previous Subsection, there are two possible
scenarios for the cosmological evolution of the small modulus
$\Phi$. However, since the RHS of (\ref{222}) and (\ref{DUE})
coincide, in both cases the crucial quantity is the same, although
with a different physical meaning. Let us denote this quantity by
$T_*\sim (\Gamma_{\Phi}M_P)^{1/2}$. We shall impose that
$T_*<T_{max}$ and shall show below that from this requirement one
can derive a lower bound on the possible values of $\mathcal{V}$
in a general LVS. Before we begin, let us first recall that:
%\newpage
\begin{enumerate}
\item If $\Phi$ is responsible for the initial reheating via its
decay, then $T_*=T_{RH}^0$.

\item If $\Phi$ decays after the original reheating in a radiation
dominated era, then $T_*=T_D<T_{RH}^0$.
\end{enumerate}
Regardless of which of these two situations we consider, $T_*$ is
the temperature of the Universe after $\Phi$ decays. Then, in
order to prevent decompactification of the internal space, we need
to impose $T_*<T_{max}$. In general, this condition is necessary
but not sufficient because in case 2 one must ensure also that
$T_{RH}^0 < T_{max}$. This is a constraint that we cannot address
given that in this case $T_{RH}^0$ is an undetermined parameter
for us.

Let us now compute $T_*$ precisely. We start by using
the exact form of the decay rate $\Gamma_{\Phi\to XX}$:
\begin{equation}
\Gamma_{\Phi\to XX}=\frac{g_{\Phi XX}^2 m_{\Phi}^3}{64\pi M_P^2},
\label{gammaPhi}
\end{equation}
where
\begin{equation}
g_{\Phi
XX}=\frac{2^{5/4}\sqrt{3}}{\langle\tau_s\rangle^{3/4}}\sqrt{\mathcal{V}}.
\end{equation}
The mass of $\Phi$ is given by:
\begin{equation}
m_{\Phi}=\sqrt{P}\frac{2 a_s\langle\tau_s\rangle
W_0}{\mathcal{V}}M_P,
\end{equation}
where we are denoting with $P$ the prefactor of the scalar
potential: $P\equiv g_s e^{K_{cs}}/(8\pi)$. From the minimisation
of the scalar potential we have that
\begin{equation}
a_s\langle\tau_s\rangle=\ln\left(p\mathcal{V}\right)=\ln p+\ln
\mathcal{V},
\end{equation}
where
\begin{equation}
p\equiv \frac{12\sqrt{2}a_s A_s}{W_0\sqrt{\tau_s}}\sim
\mathcal{O}(1)\text{ \ \ }\Rightarrow\text{ \ \
}a_s\langle\tau_s\rangle\simeq\ln \mathcal{V},
\end{equation}
and so
\begin{equation}
m_{\Phi}=\sqrt{P}\frac{2W_0\ln \mathcal{V}}{\mathcal{V}}M_P.
\end{equation}
Therefore, the decay rate $\Gamma_{\Phi\to XX}$ turns out to be:
\begin{equation}
\Gamma_{\Phi\to XX}=P^{3/2}\frac{3W_0^3(\ln
\mathcal{V})^3}{\sqrt{2}\pi\langle\tau_s\rangle^{3/2}}\frac{M_P}{\mathcal{V}^2}.
\end{equation}
Finally, in order to obtain the total decay rate, we need to multiply
$\Gamma_{\Phi\to XX}$ by the total number of gauge bosons for the
MSSM $N_X=12$:
\begin{equation}
\Gamma_{\Phi\to XX}^{TOT}=P^{3/2}\frac{36W_0^3(\ln
\mathcal{V})^3}{\sqrt{2}\pi\langle\tau_s\rangle^{3/2}}\frac{M_P}{\mathcal{V}^2}.
\end{equation}
Now, we can find $T_*$ by setting $4\left(\Gamma_{\Phi\to
XX}^{TOT}\right)^2/3$ equal to $3 H^2$, with $H$ read off from the
Friedmann equation for radiation dominance:
\begin{equation}
T_*=\left(\frac{40}{\pi^2g_*}\right)^{1/4}\sqrt{\Gamma_{\Phi}^{TOT}M_P}=
P^{3/4}\frac{6}{\pi}\left(\frac{20}{g_*}\right)^{1/4}\frac{(W_0\ln
\mathcal{V})^{3/2}}{\langle\tau_s\rangle^{3/4}}\frac{M_P}{\mathcal{V}}.
\label{good}
\end{equation}
We are finally ready to explore the constraint $T_* < T_{max}$.
Recall that the maximal temperature is given by the
decompactification temperature (\ref{final}):
\begin{equation}
T_{max}=\left(\frac{P}{4\pi\kappa_1}\right)^{1/4}
\left[\frac{(\sqrt{2}-1)}{4\sqrt{2}}\right]^{1/4}
\frac{\sqrt{W_0}\langle\tau_s\rangle^{5/8}}
{\mathcal{V}^{3/4}}M_P.
\end{equation}
Let us now consider the ratio $T_{max}/T_*$ and impose that it is
larger than unity (using $g_*(MSSM)=228.75$):
\begin{equation}
R\equiv\frac{T_{max}}{T_*}=c\frac{\mathcal{V}^{1/4}}{(\ln
\mathcal{V})^{3/2}}\text{ \ \ with \ \ }c\equiv
J\left[\frac{(\sqrt{2}-1)g_*}{80\sqrt{2}}\right]^{1/4}
\frac{\pi\langle\tau_s\rangle^{11/8}}{6W_0}\simeq
\frac{\langle\tau_s\rangle^{11/8}}{2W_0}, \label{c}
\end{equation}
where we have defined:
\begin{equation}
J\equiv \left(4\pi\kappa_1
P^2\right)^{-1/4}=\frac{8.42}{\kappa_1^{1/4}}e^{-K_{\rm{cs}}/2}\text{
\ \ for \ \ }g_s=0.1, \label{J}
\end{equation}
and in (\ref{c}) we have set $J=1$. In fact, from (\ref{kappa1}),
we find that in the case of SQCD with $N_c=3$ and $N_f=6$,
$\kappa_1=2.625$. However for the MSSM we expect a larger value of
$\kappa_1$ which we assume to be of the order $\kappa_1=10$. Then
for natural values of $K_{\rm{cs}}$ like
$K_{\rm{cs}}=3$\footnote{The dependence of the K\"{a}hler
potential on the complex structure moduli can be worked out by
computing the different periods of the CY three-fold under
consideration. As derived in \cite{Periods}, for the simplest
example of a CY manifold with just one complex structure modulus
$U$ (the mirror of the quintic), for natural values of $U$,
$|U|\sim \mathcal{O}(1)\Rightarrow K_{\textrm{cs}}\sim
\mc{O}(1)$.}, from (\ref{J}), we find $J=1.05$. Let us consider
now the maximum and minimum values that the parameter $c$ can take
for natural values of $\langle\tau_s\rangle$ and $W_0$:
\begin{eqnarray}
\left\{
\begin{array}{c}
\langle \tau _{s}\rangle _{max }=100 \\
W_{0,min }=0.01
\end{array}
\right.  &\Longrightarrow &c_{max }\simeq 10^{4}, \label{cmax}\\
\left\{
\begin{array}{c}
\langle \tau _{s}\rangle _{min }=2 \\
W_{0,max }=100
\end{array}
\right.  &\Longrightarrow &c_{min }\simeq 10^{-2}. \label{cmin}
\end{eqnarray}

\begin{center}
\FIGURE[ht]{ \epsfig{file=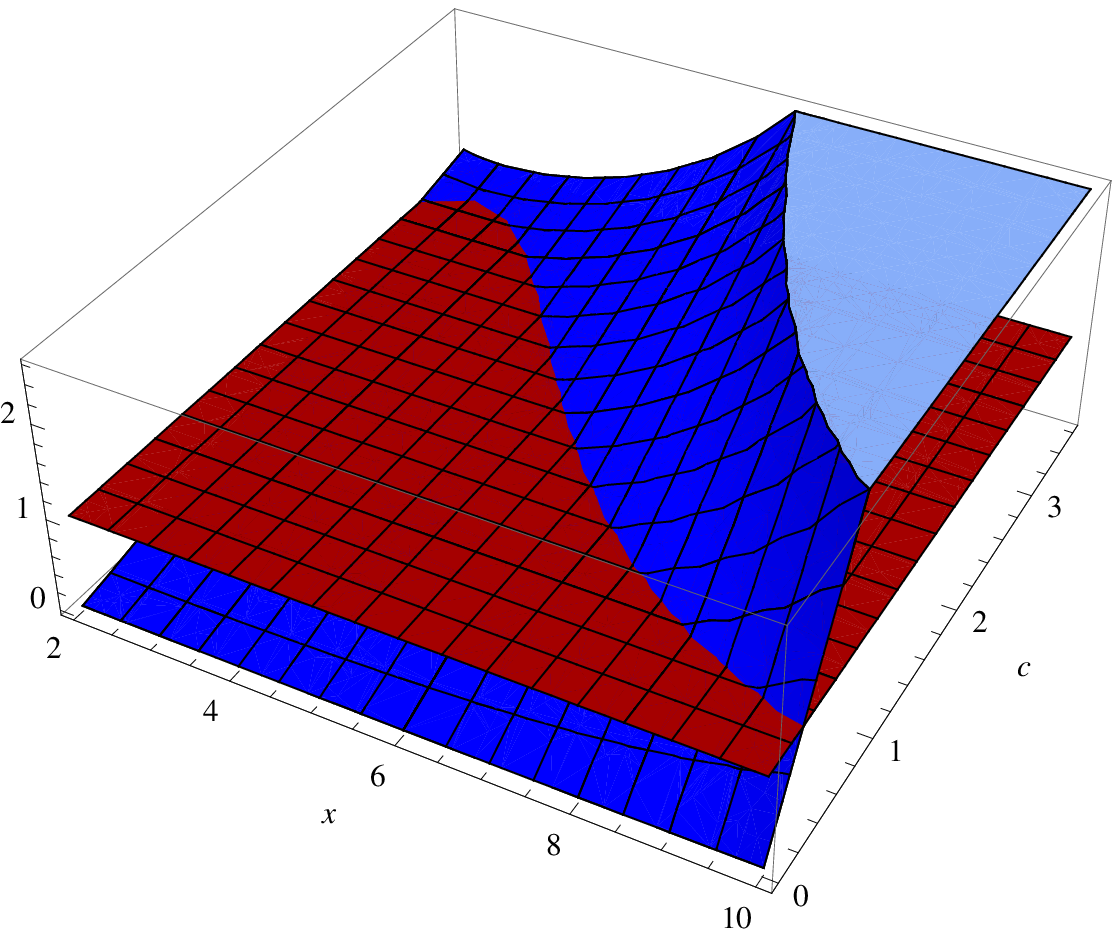, height=60mm,width=67mm}
\epsfig{file=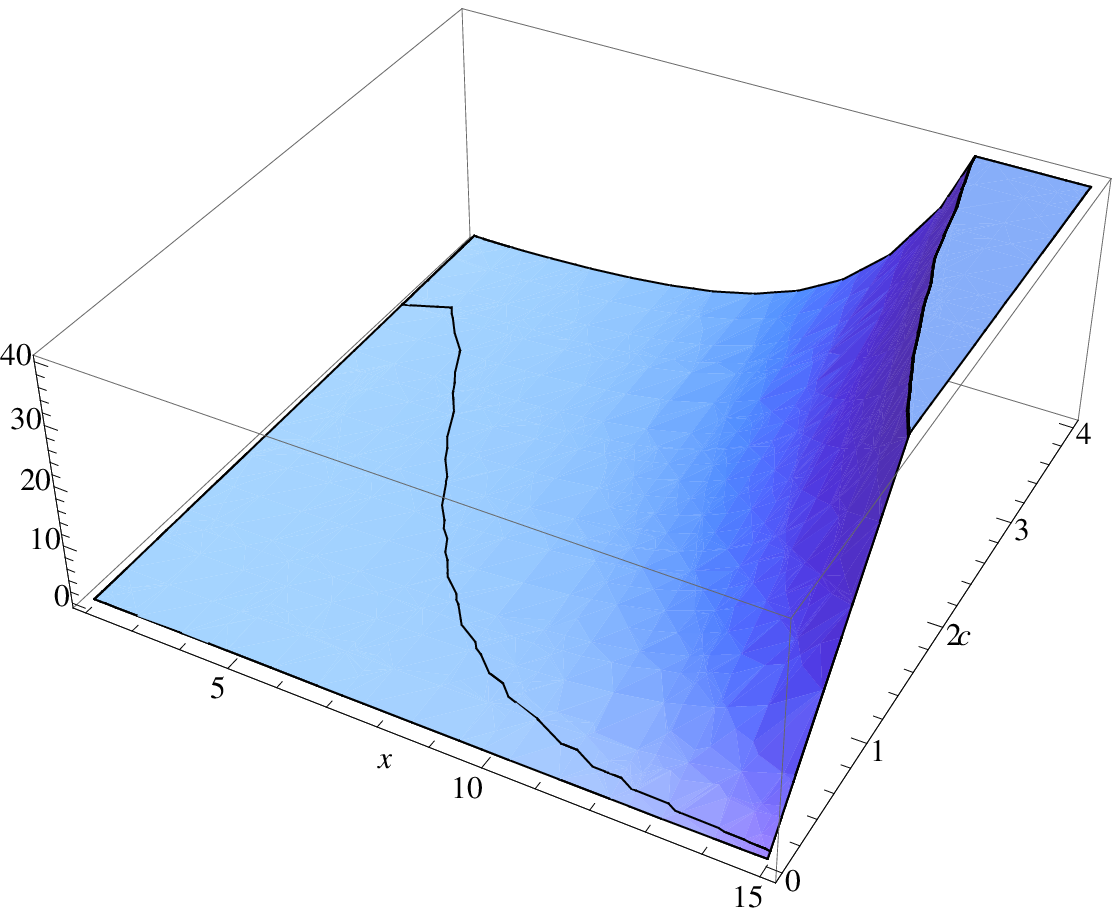, height=60mm,width=67mm}
\caption{Plots of the ratio $R\equiv T_{max}/T_*$ as a function of
$\mathcal{V}=10^x$ and the parameter $c_{min}<c<c_{max}$ as
defined in (\ref{c}), (\ref{cmax}) and (\ref{cmin}). In the left
plot, the red surface is the constant function $R=1$, whereas in
the right plot the black line denotes the curve in the
($x$,$c$)-plane for which $R=1$.} \label{VolConstraint1} }
\end{center}

\begin{figure}[t]
\begin{center}
\scalebox{0.9}{\includegraphics{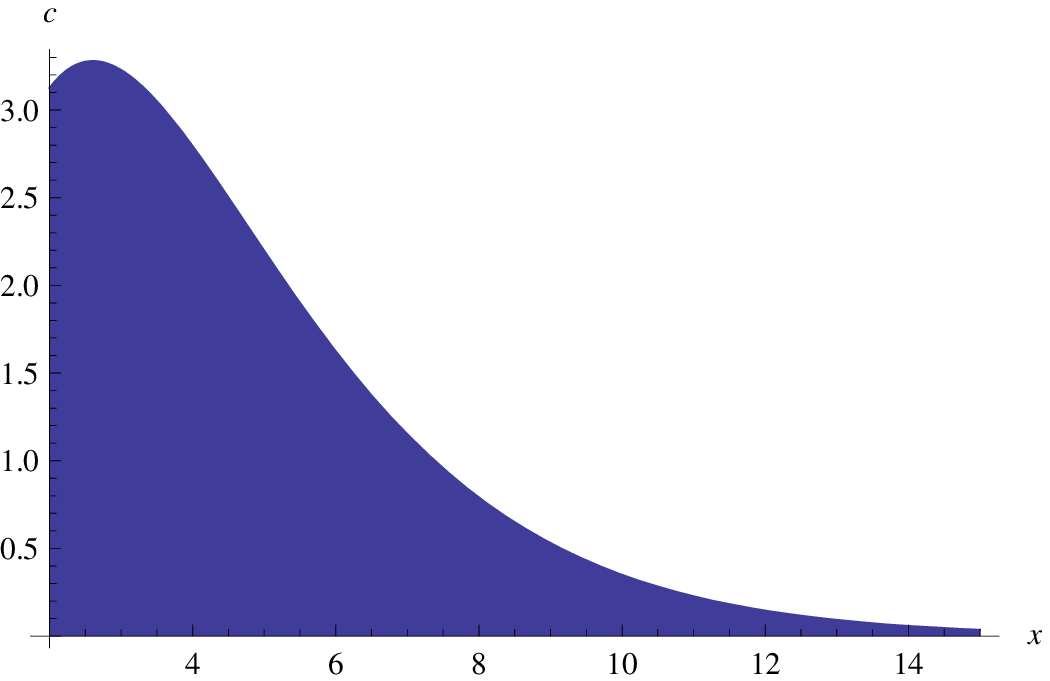}}
\end{center}
\vspace{-0.4cm} \caption{Plot of the $R=1$ curve in the
($x$,$c$)-plane. The shaded region represents the
phenomenologically forbidden area, in which the values of $x$ and
$c$ are such that $R<1$ $\Leftrightarrow$ $T_{max}<T_*$.}
\label{VolConstraint2}
\end{figure}

\TABULAR[ht]{l||l}
 { & $R>1$ $\Leftrightarrow$ $T_{max}>T_*$ \\
  \hline\hline
  $c=4$ & \hspace*{1.4cm}$\forall x$ \\
  $c=3$ & \hspace*{1cm}$x>2.1$ \\
  $c=2$ & \hspace*{1cm}$x>3.8$ \\
  $c=1$ & \hspace*{1cm}$x>5.9$ \\
  $c=0.5$ & \hspace*{1cm}$x>7.6$ \\
  $c=0.1$ & \hspace*{1cm}$x>11.3$ \\
  $c=0.05$ & \hspace*{1cm}$x>12.8$ \\
  $c=0.01$ & \hspace*{1cm}$x>16.1$
  \label{tabella}} {Lower bounds on the volume in the string frame
  $\mathcal{V}_s\sim 10^{x-3/2}$ for some benchmark scenarios.}

Now writing $\mathcal{V}\simeq 10^x$, $R$ becomes a function of
$x$ and $c$. Finally, we can make a 3D plot of $R$ with
$c_{min}<c<c_{max}$ and $2<x<15$, and see in which region $R>1$.
This is done in Figure \ref{VolConstraint1}. In order to
understand better what values of $\mathcal{V}$ are disfavoured, we
also plot in Figure \ref{VolConstraint2}, as the shaded region,
the region in the ($x$,$c$)-plane below the curve $R=1$, which
represents the phenomenologically forbidden area for which
$T_{max}<T_*$. We conclude that small values of the volume, which
would allow the standard picture of gauge coupling unification and
GUT theories, are disfavoured compared to larger values of
$\mathcal{V}$, that naturally lead to TeV-scale SUSY and are thus
desirable to solve the hierarchy problem. In Table \ref{tabella},
we show explicitly how the lower bound on the volume, for some
benchmark scenarios, favours LVS with larger values of
$\mathcal{V}$.

From the definition (\ref{c}) of the parameter $c$, it is
interesting to notice that for values of $\langle\tau_s\rangle$
far from the edge of consistency of the supergravity approximation
$\langle\tau_s\rangle\sim \mc{O}(10)$, $c$ should be fairly large,
and hence the bound very weak, for natural values of $W_0\sim
\mc{O}(1)$, while $c$ should get smaller for larger values of
$W_0$ that lead to a stronger bound. In addition, it is reassuring
to notice that for typical values of $\mathcal{V}\sim 10^{15}$,
$T_{max}>T_*$ except for a tiny portion of the ($x$,$c$)-space. It
also important to recall that the physical value of the volume as
seen by the string is the one expressed in the string frame
$\mathcal{V}_s$, while we are working in the Einstein frame where
$\mathcal{V}_s=g_s^{3/2}\mathcal{V}_E$. Hence if we write
$\mathcal{V}_E\sim 10^x$, then we have that
$\mathcal{V}_s=10^{x-3/2}$, upon setting $g_s=0.1$.

%\newpage
\medskip\noindent{\em General LARGE Volume Scenario}

\medskip\noindent
Let us now generalise our lower bound on $\mathcal{V}$ to the four
cases studied in Subsections \ref{Sec:MultHoleModTherm} and
\ref{Sec:K3FibrModTherm} for the multiple-hole Swiss-cheese and K3
fibration case (focusing on the small K3 fiber scenario)
respectively.

First of all, we note that, since in all the cases the 4-cycle
supporting the MSSM is stabilised by string loop corrections
\cite{ccq2}, we can estimate the actual height of the barrier seen
by this modulus as (see (\ref{ygfdo})):
\begin{equation}
V_b\sim\frac{W_0^2}{\mathcal{V}^3\sqrt{\tau}},
\end{equation}
where we are generically denoting any small cycle (either a
blow-up or a K3 fiber divisor) as $\tau$, given that the values of
the VEV of all these 4-cycles will have the same order of
magnitude. Then setting $V_b\sim T_{max}^4/\tau$, we obtain:
\begin{equation}
T_{max}^4\sim\frac{\sqrt{\tau}W_0^2}{\mathcal{V}^3}.
\label{newTmax}
\end{equation}
We notice that (\ref{newTmax}) is a bit lower than (\ref{final})
but the two expressions for $T_{max}$ share the same leading order
$\mathcal{V}$-dependence.

Let us now examine the 4 cases of Subsection
\ref{Sec:MultHoleModTherm} in more detail, keeping the same
notation as in that Subsection, and denoting as $\Phi$ the small
modulus of the single-hole Swiss-cheese scenario studied above:
\begin{enumerate}
\item The relevant decay is the one of $\tau_2$ to MSSM gauge bosons.
The order of magnitude of the mass of $\tau_2$ is:
\begin{equation}
m_{\tau_2}^2\sim\frac{\left(\ln\mathcal{V}\right)^2 W_0^2
}{\mathcal{V}^2\tau^2}, \label{masstau2}
\end{equation}
and so $\tau_2$ is lighter than $\Phi$, and, in turn, $T_*$ will
be smaller. In fact, plugging (\ref{masstau2}) in
(\ref{gammaPhi}), we end up with (ignoring numerical prefactors):
\begin{equation}
T_*\sim\frac{\left(\ln\mathcal{V}\right)^{3/2}
W_0^{3/2}}{\mathcal{V}\tau^{9/4}}. \label{tstarnew}
\end{equation}
Hence we obtain
\begin{equation}
R^{(1)}\equiv\frac{T_{max}}{T_*}=
c^{(1)}\frac{\mathcal{V}^{1/4}}{\left(\ln\mathcal{V}\right)^{3/2}}\text{
\ \ with \ \ }c^{(1)}\sim \frac{\tau^{19/8}}{W_0}.
\end{equation}
Comparing this result with (\ref{c}), we realise that $R^{(1)}\sim
R\,\tau$ and so the lower bound on $\mathcal{V}$ turns out to be
less stringent. The final results can still be read from Table
\ref{tabella} upon replacing $c$ with $c^{(1)}$.

\item The relevant decay is the one of $\tau_1$ to MSSM gauge
bosons since $m_{\tau_1}\sim m_{\Phi}$, and so $\tau_1$ is heavier
than $\tau_2$. Therefore $T_*$ will still be given by
(\ref{good}). Hence we obtain
\begin{equation}
R^{(2)}\equiv\frac{T_{max}}{T_*}=
c^{(2)}\frac{\mathcal{V}^{1/4}}{\left(\ln\mathcal{V}\right)^{3/2}}\text{
\ \ with \ \ }c^{(2)}\sim \frac{\tau^{7/8}}{W_0}.
\end{equation}
Comparing this result with (\ref{c}), we realise that $R^{(2)}\sim
R\,\tau^{-1/2}$ and so the lower bound on $\mathcal{V}$ turns out
to be more stringent. The final results can still be read from
Table \ref{tabella} upon replacing $c$ with $c^{(2)}$.

\item The relevant decay is the one of $\tau_1$ to hidden sector gauge
bosons. Hence we point out that the considerations of case 2 apply
also for this case.

\item The relevant decay is the one of $\tau_1$ to MSSM gauge
bosons, and so we can repeat the same considerations of case 2.
\end{enumerate}
The final picture is that for all cases the
$\mathcal{V}$-dependence of the ratio $T_{max}/T_*$ is the same as
in (\ref{c}). The only difference is a rescaling of the parameter
$c$. Thus we conclude that, as far as the lower bound on
$\mathcal{V}$ is concerned, the single-hole Swiss-cheese case
shows all the qualitative features of a general LVS.

Finally, we mention that in the case of a K3 fibration with small
K3 fiber, cases 2, 3, and 4 of Subsection \ref{Sec:K3FibrModTherm}
have the same behaviour as case 2 of the multiple-hole
Swiss-cheese, so giving a more stringent lower bound on
$\mathcal{V}$. We should note though that this lower bound does
not apply to case 1 of Subsection \ref{Sec:K3FibrModTherm}, since
both of the moduli have an $M_P$-suppressed, instead of
$M_s$-suppressed, coupling to MSSM gauge bosons. However, these
kinds of models tend to prefer larger values of $\mathcal{V}$ (due
to the fact that $a_s=2\pi$ for an ED3 instanton) which are not
affected by the lower bound that we derived.

\section{Discussion}
\label{Dis}

Let us now discuss some of the possible applications of these
results, as well as directions for future work. As we have
emphasized throughout the paper, there are two kinds of LVS,
depending on the magnitude of the value of the internal volume
$\mathcal{V}$. Their main cosmological characteristics are the
following:

\medskip\noindent{\em  \textbf{LV case}}

\medskip\noindent

In this case the volume is stabilised at large values of the order
$\mathcal{V}\sim 10^{15}$ which allows to solve the hierarchy
problem yielding TeV scale SUSY naturally. Here are the main
cosmological features of these scenarios:

\begin{itemize}
\item The moduli spectrum includes a light field $\chi$ related to the overall volume.
This field is a source for the cosmological moduli problem (CMP)
as long as $M_s<10^{13}$ GeV, corresponding to
$\mathcal{V}>10^{10}$. In fact, in this case the modulus $\chi$ is
lighter than $10$ TeV and, coupling with gravitational strength
interactions, it would overclose the Universe or decay so late to
ruin Big-Bang nucleosynthesis. There are two main possible
solutions to this CMP:
\begin{enumerate}
\item The light modulus $\chi$ gets diluted due to an increase in the entropy
that occurs when a short-lived modulus decays out of equilibrium
and while dominating the energy density of the Universe
\cite{Vafa,Acharya};

\item The volume modulus gets diluted due to a late period of low
energy inflation caused by thermal effects \cite{TI}.
\end{enumerate}
Assuming this problem is solved, the volume modulus becomes a dark
matter candidate (with a mass $m \sim 1$ MeV, if $\mathcal{V}\sim
10^{15}$) and its decay to $e^+ e^-$ could be one of the sources
that contribute to the observed $511$ KeV line, coming from the
centre of our galaxy.\footnote{However, recently it has been
discovered with the INTEGRAL spectrometer SPI \cite{INTEGRAL} that
the 511 KeV line emission appears to be asymmetric. This
distribution of positron annihilation resembles that of low mass
X-ray binaries, suggesting that these systems may be the dominant
origin of the positrons and so reducing the need for more exotic
explanations, such as the one presented in this paper.} The light
modulus $\chi$ can also decay into photons, producing a clean
monochromatic line that would represent a clear astrophysical
smoking-gun signal for these scenarios \cite{CQ}. We point out
that in the case of K3 fibrations, where the K3 fiber is
stabilised large \cite{ccq2}, the spectrum of moduli fields
includes an additional light field. This field is also a potential
dark matter candidate with a mass $m \sim 10$ keV, that could
produce another monochromatic line via its decay to photons.

\item At present, there are no known models of inflation in
LVS with intermediate scale $M_s$. However, the Fibre Inflation
model of \cite{fibInfl} can give rise to inflation for every value
of $\mathcal{V}$. The only condition, which fixes $\mathcal{V}\sim
10^3$, and so $M_s\sim M_{GUT}$, is the matching with the COBE
normalisation for the density fluctuations. Such a small value of
$\mathcal{V}$ is also necessary to have a very high inflationary
scale (close to the GUT scale) which, in turn, implies detectable
gravity waves. However, in principle it is possible that the
density perturbations could be produced by another scalar field
(not the inflaton), which is playing the role of a curvaton. In
such a case, one could be able to get inflation also for
$\mathcal{V}\sim 10^{15}$. In this way, both inflation and TeV
scale SUSY would be achieved within the same model, even though
gravity waves would not be observable. It would be interesting to
investigate whether such scenarios are indeed realisable.

\item As derived in Section \ref{Sec:DecTemp}, if the volume is stabilised such that
$\mathcal{V} \sim 10^{15}$, the decompactification temperature is
rather low: $T_{max}\sim 10^{7}$ GeV.
\end{itemize}

\medskip\noindent{\em  \textbf{SV case}}

\medskip\noindent

In this case the volume is stabilised at smaller values of the
order $\mathcal{V}\sim 10^{4}$, which allows to reproduce the
standard picture of gauge coupling unification with $M_s\sim
M_{GUT}$. Here are the main cosmological features of these
scenarios:

\begin{itemize}
\item Given that in this case $\mathcal{V}<10^{13}$, all the
moduli have a mass $m>10$ TeV, and so they decay before Big-Bang
nucleosynthesis. Hence these scenarios are not plagued by any CMP.

\item As we have already pointed out in the LV case above, smaller
values of $\mathcal{V}$ more naturally give rise to inflationary
models, as the one presented in \cite{fibInfl}. Here we observe
that the predictions for cosmological observables of Fiber
Inflation were sensitive to the allowed reheating temperature.
Since for $\mathcal{V}\sim 10^4$ GeV we have $T_{RH}^0<T_{max}\sim
10^{15}$ GeV and since in \cite{fibInfl} the authors considered
already a more stringent upper bound $T_{RH}^0<10^{10}$ GeV (in
order to avoid thermal gravitino overproduction), the presence of
a maximal temperature does not alter the predictions of that
inflationary scenario.

\item Fixing the volume at small values of the order
$\mathcal{V} \sim 10^3$, the decompactification temperature turns
out to be extremely high: $T_{max}\sim 10^{15}$ GeV.
\end{itemize}

According to the discussion above, it would seem that cosmology
tends to prefer smaller values of $\mathcal{V}$. The reason is
that in the SV case there is no CMP and robust models of inflation
are known, whereas for $\mathcal{V}\sim 10^{15}$ the light modulus
suffers from the CMP and no model of inflation has been found yet.
Interestingly enough, the lower bound on $\mathcal{V}$, derived in
this paper, suggests exactly the opposite. Namely, larger values
of $\mathcal{V}$ are favoured since, writing the volume as
$\mathcal{V}\sim 10^{x}$ and recalling the definition (\ref{c}) of
the parameter $c$, the constraint $T_*<T_{max}$ rules out a
relevant portion of the $(x,c)$-parameter space, that corresponds
to the SV case.

In view of this result, let us point out again that the LV case
has its advantages. For example, the decay of the light modulus
into $e^+ e^-$ could contribute to explain the origin of the $511$
KeV line. In addition, its decay to photons could produce a clean
smoking-gun signal of LVS. Furthermore, finding a realization of
inflation, that is compatible with the LV case, is not necessarily
an unsurmountable problem. In that regard, let us note that the
authors of \cite{CKLQ} proposed a model, which relates the LV to
the SV case. More precisely, the inflaton is the volume modulus
and inflation takes place at a high scale for small values of
$\mathcal{V}$. However, after inflation the modulus ends up at a
VEV located at $\mathcal{V}\sim 10^{15}$, thus obtaining TeV scale
SUSY. In fact, as we have already mentioned above, it could even
be possible to realize inflation directly in the LV case. A way to
achieve that would be to modify the the Fibre Inflation scenario
of \cite{fibInfl}, so that the density fluctuations are generated
by a field other than the inflaton. Such curvaton-like scenarios
would be very promising for the generation of non-gaussianities in
the CMB, as well as the realization of both low scale inflation
and low-energy SUSY. However, due to the low inflationary scale,
in these models gravity waves will be unobservable.

Now, even if inflation turns out not to be a problem for the LV
case, there is still the CMP due to the presence of the light
volume modulus. The results of this paper pose a challenge for the
solution of this problem. Indeed, as we have shown in Subsection
\ref{Sec:MaxTemp}, the CMP cannot be solved by diluting the volume
modulus via the entropy increase caused by the decay of the small
moduli. The reason is that the latter moduli decay before they can
begin to dominate the energy density of the Universe. So let us
now discuss in more detail the prospects of the other main
possible solution of the CMP in LVS, namely thermal inflation.

%\newpage
\medskip\noindent{\em  \textbf{Thermal Inflation}}

\medskip\noindent

Thermal inflation has been studied in the literature from the
field theoretic point of view \cite{TI}. The basic idea is that a
field $\phi$, whose VEV is much larger than its mass (and so is
called flaton) can be trapped by thermal corrections at a false
vacuum in the origin. At a certain temperature, its vacuum energy
density can start dominating over the radiation one, thus leading
to a short period of inflation. This period ends when the
temperature drops enough to destabilise the local minimum the
flaton was trapped in.

Since the flaton $\phi$ has to have a VEV $\langle\phi\rangle\gg
m_{\phi}$, it is assumed that the quartic piece in its potential
is absent. However in this way, the 1-loop thermal corrections
cannot trap the flaton in the origin because they go like
\begin{equation}
V_T\sim T^2 m_{\phi}^2=T^2 \frac{d^2 V}{d \phi^2},
\end{equation}
and there is no quartic term in $V$ that would give rise to a term
like $T^2\phi^2$. Hence, it is usually assumed that there is an
interaction of the flaton with a very massive field, say a scalar
$\psi$, of the form $g\psi^2\phi^2$, where $g\sim 1$ so that
$\psi$ thermalises at a relatively low temperature. At this point,
a 1-loop thermal correction due to $\psi$ would give the required
term
\begin{equation}
V_T\sim T^2 m_{\psi}^2=gT^2\phi^2.
\end{equation}
When $\phi$ gets a nonzero VEV, the interaction term
$g\psi^2\phi^2$ generates a mass term for $\psi$ of the order
$m_{\psi}\sim \langle\phi\rangle$. Hence, when $\phi$ is trapped
in the origin at high $T$, $\psi$ becomes very light. Close to the
origin, the potential looks like:
\begin{equation}
V=V_0+(g T^2-m_{\phi}^2)\phi^2+...\,,
\end{equation}
where $V_0$ is the height of the potential in the origin. A period
of thermal inflation takes place in the temperature window
$T_c<T<T_{in}$, where $T_{in}\sim V_0^{1/4}$ is the temperature at
which the flaton starts to dominate the energy density of the
Universe (beating the radiation energy density $\rho_r\sim T^4$)
and $T_c\sim m_{\phi}/g$ is the critical temperature at which the
flaton undergoes a phase transition rolling towards the $T=0$
minimum. The number of e-foldings of thermal inflation is given
by:
\begin{equation}
N_e\sim
\ln\left(\frac{T_{in}}{T_{c}}\right)\sim\ln\sqrt{\frac{\langle\phi\rangle}{m_{\phi}}}.
\end{equation}

Let us see how the above picture relates to the LVS. In the case
of $\mathcal{V}\sim 10^{15}$, the modulus $\tau_s$ has the right
mass scale and VEV to produce $N_e\sim 10$ e-foldings of
inflation, which would solve the CMP without affecting the density
perturbations generated during ordinary high-energy inflation.
However, in Subsection \ref{Sec:ExplicitTPot} we derived the
relevant 1-loop temperature corrections to the scalar potential
and showed that they are always subleading with respect to the
$T=0$ potential, for temperatures below the Kaluza-Klein scale.
Hence, since thermal inflation requires the presence of new minima
at finite-temperature, we would be tempted to conclude that it
does not take place in the LVS. In fact, this was to be expected
also for the following reason. According to the field theoretic
arguments above, in order for thermal inflation to occur, it is
crucial that the flaton be coupled to a very massive field $\psi$.
However, in our model there is no particle, which is heavier than
the flaton candidate $\tau_s$. It is not so surprising, then, that
we are not finding thermal inflation.

Let us now discuss possible extensions of our model that could,
perhaps, allow for thermal inflation to occur, as well as the
various questions that they raise.

\begin{enumerate}
\item Since in our case $\tau_s$ is the candidate flaton field,
the necessary $\psi$ field would have a mass of the order
$m_{\psi}\sim\langle\tau_s\rangle M_s$, and so it is likely to be
a stringy mode. In such a case, it is not a priori clear how to
compute thermal corrections to $V_T$ due to the presence of $\psi$
in the thermal bath.

\item Even if we can compute $V_T$, it is not clear why these
corrections should trap $\tau_s$ at the origin. Note, however,
that this is not implausible, as the origin is a special point in
moduli space, where new states may become massless or the local
symmetry may get enhanced. Any such effect might turn out to play
an important role.

\item Even assuming that $V_T$ does trap $\tau_s$ in the origin,
one runs into another problem. Namely, the corresponding small
cycle shrinks below $M_s$ and so we cannot trust our low-energy
EFT. For a full description, we should go to the EFT that applies
close to the origin. The best known examples of these are EFTs for
blow-up fields at the actual orbifold point. In addition, one
should verify that $\mathcal{V}$ stays constant when the $\tau_s$
cycle shrinks to zero size.

\item When $\tau_s$ goes to zero, the field $\psi$ should
become massless, according to the comparison with the field
theoretic argument (if this comparison is valid). So possible
candidates for the role of the $\psi$ field could be winding
strings or D1-branes wrapping a 1-cycle of the collapsing 4-cycle.

\item If $\psi$ corresponds to a winding string, the interaction of the flaton
$\tau_s$ with $\psi$ cannot be seen in the EFT and it would be
very difficult to have a detailed treatment of this issue.

\item The field $\psi$ could also be a right handed neutrino,
or sneutrino, heavier than $\tau_s$. The crucial question would
still be if it would be possible to see $\psi$ in our EFT
description. In addition, one would need to write down $m_{\psi}$
as a function of $\tau_s$ and $\mathcal{V}$. It goes without
saying that this issue is highly dependent on the particular
mechanism for the generation of neutrino masses.

\item Besides the small modulus $\tau_s$, another possible flaton
candidate could be a localised matter field such as an open string
mode. However we notice that the main contribution to the scalar
potential of this field should come from D-terms, and that a
D-term potential usually gives rise to a mass of the same order of
the VEV. Hence it may be difficult to find an open string mode
with the typical behaviour of a flaton field.
\end{enumerate}
In general, all of the above open questions are rather difficult
to address. This poses a significant challenge for the derivation
of thermal inflation in LVS and the corresponding solution of the
CMP. However, let us note that the CMP could also be solved by
finding different models of low-energy inflation, which do not
rely on thermal effects.

\section{Conclusions}
\label{Con}

In this paper, we studied how finite-temperature corrections
affect the $T=0$ effective potential of type IIB LVS and what are
the subsequent cosmological implications in this context.

We showed that the small moduli and modulini can reach thermal
equilibrium with the MSSM particles. Despite that, we were able to
prove that their thermal contribution to the effective potential
is always subleading compared to the $T=0$ potential, for
temperatures below the Kaluza-Klein scale. As a result, the
leading temperature-dependent part of the effective potential is
due only to the MSSM thermal bath and it turns out to have runaway
behaviour at high $T$. We derived the decompactification
temperature $T_{max}$, above which the $T=0$ minimum is completely
erased and the volume of the internal space starts running towards
infinity. Clearly, in this class of IIB compactifications the
temperature $T_{max}$ represents the maximal allowed temperature
in the early Universe. Hence, in particular, it gives an upper
bound on the initial reheating temperature after inflation:
$T_{RH}^0<T_{max}$. {}\footnote{Note, however, that it may be
possible to relax this constraint to a certain degree by studying
the dynamical evolution of the moduli in presence of finite
temperature corrections, in the vein of the considerations of
\cite{barreiro} for the KKLT set-up.} The temperature $T_{RH}^0$
is highly dependent on the details of the concrete inflationary
model and re-heating process, and so in principle its
determination is beyond the scope of our paper. Nevertheless, we
can compute the temperature of the Universe after the small moduli
decay. They are rather short-lived and their decay can either be
the main source of initial reheating (in which case the
temperature after their decay is exactly $T_{RH}^0$) or it can
occur during a radiation dominated epoch, after initial reheating
has already taken place. In both cases, the resulting temperature
of the Universe $T_*$ has to satisfy $T_* < T_{max}$\,, which
implies a lower bound on the allowed values of $\mathcal{V}$. We
were able to derive this bound and show that it rules out a large
range of smaller ${\cal V}$ values (which lead to standard GUT
theories), while favouring greater values of ${\cal V}$ (that lead
to TeV scale SUSY). Note though, that the condition $T_*<T_{max}$
is both necessary and sufficient in the case the decay of the
small moduli is the origin of initial reheating, whereas it is
just necessary but not sufficient in the case the small moduli
decay below $T_{RH}^0$.

Finally, we discussed possible cosmological applications of our work.
In particular, we argued that, to realize thermal inflation in this
type of compactifications, one needs to go beyond the current effective
field theory description of the closed string moduli sector.

\section*{Acknowledgements}

We would like to thank Joe Conlon, Anshuman Maharana, Nelson Nunes
and Michael Ratz for useful conversations. We are especially
grateful to Fernando Quevedo for fruitful discussions and valuable
comments on the paper. L.A. is supported by DOE grant
FG02-84-ER40153. V.C. is supported by a Queen Mary Westfield Trust
Scholarship. M.C. is partially funded by St John's College, EPSRC
and CET.

\appendix
\section{Moduli couplings}
\label{App:ModuliCouplings}

We shall now assume that the MSSM is built via magnetised $D7$
branes wrapping an internal 4-cycle within the framework of 4D
$\mathcal{N}=1$ supergravity. The full Lagrangian of the system
can be obtained by expanding the superpotential $W$, the
K\"{a}hler potential $K$ and the gauge kinetic functions $f_i$ as
a power series in the matter fields:
\begin{eqnarray}
W &=&W_{mod}(\varphi )+\mu (\varphi
)H_{u}H_{d}+\frac{Y_{ijk}(\varphi
)}{6}C^{i}C^{j}C^{k}+..., \\
K &=&K_{mod}(\varphi ,\bar{\varphi} )+\tilde{K}_{i\bar{j}}(\varphi
,\bar{\varphi} )C^{i}C^{\bar{j}}+
\left[ Z(\varphi ,\bar{\varphi})H_{u}H_{d}+h.c.\right] +..., \\
f_{i} &=&\frac{T_{MSSM}}{4\pi }+h_{i}(F)S. \label{gaugeKinfunc}
\end{eqnarray}
In the previous expressions, $\varphi$ denotes globally all the
moduli fields, and $W_{mod}$ and $K_{mod}$ are the superpotential
and the K\"{a}hler potential for the moduli, which we have
discussed in depth in Section \ref{Sec:Review}. $H_u$ and $H_d$
are the two Higgs doublets of the MSSM, and the $C$'s denote
collectively all the matter fields. In the expression for the
gauge kinetic function (\ref{gaugeKinfunc}), $T_{MSSM}$ is the
modulus related to the 4-cycle wrapped by the MSSM $D7$ branes,
and $h_i(F)$ are 1-loop topological functions of the world-volume
fluxes $F$ on different branes (the index $i$ runs over the three
MSSM gauge group factors). Finally the moduli scaling of the
K\"{a}hler potential for matter fields
$\tilde{K}_{i\bar{j}}(\varphi ,\bar{\varphi})$ and $Z(\varphi
,\bar{\varphi})$, for LVS with the small cycle $\tau_s$ supporting
the MSSM, has been derived in \cite{ConlonCremadesQuevedo} and
looks like:\footnote{Note that, in the case of more than one small
cycle supporting the MSSM, these expressions would be more
complicated.}
\begin{equation}
\tilde{K}_{i\bar{j}}(\varphi
,\bar{\varphi})\sim\frac{\tau_s^{1/3}}{\tau_b}k_{i\bar{j}}(U)\text{
\ \ and \ \ }Z(\varphi
,\bar{\varphi})\sim\frac{\tau_s^{1/3}}{\tau_b}z(U).
\label{chiralMetric}
\end{equation}

\subsection{Moduli couplings to ordinary particles}
\label{Sec:ModCouplOrdPart}

We now review the derivation of the moduli couplings to gauge
bosons, matter particles and Higgs fields for high temperatures
$T>M_{EW}$. In this case all the gauge bosons and matter fermions
are massless.

\bigskip

$\bullet \textbf{ Couplings to Gauge Bosons}$

\bigskip

The coupling of the gauge bosons $X$ to the moduli arise from the
moduli dependence of the gauge kinetic function
(\ref{gaugeKinfunc}). We shall assume that the MSSM $D7$ branes
are wrapping the small cycle\footnote{The large cycle would yield
an unrealistically small gauge coupling: $g^2\sim
\langle\tau_b\rangle^{-1}\sim 10^{-10}$.}, and so we identify
$T_{MSSM}\equiv T_{s}$. We also recall that the gauge couplings of
the different MSSM gauge groups are given by the real part of the
gauge kinetic function, and that one obtains different values by
turning on different fluxes. Thus the coupling of $\tau_s$ with
the gauge bosons is the same for $U(1)$, $SU(2)$ and $SU(3)$. We
now focus on the $U(1)$ factor without loss of generality. The
kinetic terms read (neglecting the $\tau_s$-independent 1-loop
contribution)
\begin{equation}
\mathcal{L}_{gauge}=-\frac{\tau_s}{M_P}F_{\mu\nu}F^{\mu\nu}.
\end{equation}
We then expand $\tau_s$ around its minimum and go to the
canonically normalised field strength $G_{\mu\nu}$ defined as
\begin{equation}
G_{\mu\nu}=\sqrt{\langle\tau_s\rangle}F_{\mu\nu}, \label{redef}
\end{equation}
and obtain
\begin{equation}
\mathcal{L}_{gauge}=-G_{\mu\nu}G^{\mu\nu}-\frac{\delta\tau_s}
{M_P\langle\tau_s\rangle}G_{\mu\nu}G^{\mu\nu}. \label{FmunuFmunu}
\end{equation}
Now by means of (\ref{small}) we end up with the following
\textit{dimensionful} couplings
\begin{eqnarray}
\mathcal{L}_{\chi XX}&\sim&\left(\frac{1}
{M_P\ln{\mathcal{V}}}\right)\chi G_{\mu\nu}G^{\mu\nu}, \\
\mathcal{L}_{\Phi XX}&\sim&\left(\frac{\sqrt{\mathcal{V}}}
{M_P}\right)\Phi G_{\mu\nu}G^{\mu\nu}. \label{ImpModuliCoupl}
\end{eqnarray}

%\newpage

$\bullet \textbf{ Couplings to matter fermions}$

\bigskip

The terms of the supergravity Lagrangian which are relevant to
compute the order of magnitude of the moduli couplings to an
ordinary matter fermion $\psi$
%for temperatures below the EW symmetry breaking scale,
are its kinetic and mass
terms\footnote{Instead of the usual 2-component spinorial
notation, we are using here the more convenient 4-component
spinorial notation.}:
\begin{equation}
\mathcal{L}=\tilde{K}_{\bar{\psi}\psi}\bar{\psi}i\gamma^{\mu}\partial_{\mu}\psi+
e^{K/2}\lambda H\bar{\psi}\psi,
\end{equation}
where $H$ is the corresponding Higgs field (either $H_u$ or
$H_d$). The moduli scaling of $\tilde{K}_{\bar{\psi}\psi}$ is
given in (\ref{chiralMetric}), whereas $e^{K/2}=\mathcal{V}^{-1}$.
Expanding the moduli and the Higgs around their VEVs, we obtain
\begin{equation}
\mathcal{L}=\frac{\langle\tau_s\rangle^{1/3}
}{\langle\tau_b\rangle }\left(1+\frac{1}{3}\frac{\delta\tau_s}
{\langle\tau_s\rangle}-\frac{\delta\tau_b}{\langle\tau_b\rangle}+...\right)
\bar{\psi}i\gamma^{\mu}\partial_{\mu}\psi+
\frac{1}{\langle\tau_b\rangle^{3/2}}\left(1-\frac{3}{2}\frac{\delta\tau_b}
{\langle\tau_b\rangle}+...\right)\lambda\left(\langle
H\rangle+\delta H\right) \bar{\psi}\psi.
\end{equation}
We now canonically normalise the $\psi$ kinetic terms
($\psi\to\psi_c$) and rearrange the previous expression as
\begin{eqnarray}
\mathcal{L} &=&\bar{\psi}_{c}\left(i\gamma^{\mu }\partial _{\mu
}+m_{\psi }\right) \psi _{c}+\left( \frac{1}{3}\frac{\delta \tau
_{s}}{ \langle \tau _{s}\rangle }-\frac{\delta \tau _{b}}{\langle
\tau _{b}\rangle } \right) \bar{\psi}_{c}\left( i\gamma^{\mu
}\partial _{\mu }+m_{\psi
}\right) \psi _{c}  \notag \\
&&-\left( \frac{1}{3}\frac{\delta \tau _{s}}{\langle \tau
_{s}\rangle }+ \frac{1}{2}\frac{\delta \tau _{b}}{\langle \tau
_{b}\rangle }\right) m_{\psi }\bar{\psi}_{c}\psi
_{c}+\mathcal{L}_{\delta H},  \label{sec}
\end{eqnarray}
where
\begin{equation}
m_{\psi}\equiv\frac{\lambda\langle
H\rangle}{\langle\tau_s\rangle^{1/3}\langle\tau_b\rangle^{1/2}},\text{
\ and \ }\mathcal{L}_{\delta
H}=\left(\frac{\lambda}{\langle\tau_b\rangle^{1/2}\langle\tau_s\rangle^{1/3}}\right)\delta
H\bar{\psi}_c\psi_c-\left(\frac{3\lambda}{2\langle\tau_b\rangle^{3/2}\langle\tau_s\rangle^{1/3}}
\right)\delta
\tau_b\delta H \bar{\psi}_c\psi_c. \label{secc}
\end{equation}
The second term of (\ref{sec}) does not contribute to the moduli
interactions since Feynman amplitudes vanish for on-shell final
states satisfying the equations of motion. Writing everything in
terms of $\Phi$ and $\chi$, we end up with the following
\textit{dimensionless} couplings
\begin{eqnarray}
\mathcal{L}_{\chi\bar{\psi}_c\psi_c}&\sim&\left(\frac{m_{\psi}}{M_P}\right)\chi\bar{\psi}_c\psi_c, \label{ein}\\
\mathcal{L}_{\Phi\bar{\psi}_c\psi_c}&\sim&\left(\frac{m_{\psi}\sqrt{\mathcal{V}}}{M_P}\right)
\Phi\bar{\psi}_c\psi_c. \label{zwei}
\end{eqnarray}
Moreover the first term in the Higgs Lagrangian (\ref{secc}) gives
rise to the usual Higgs-fermion-fermion coupling, whereas the
second term yields a modulus-Higgs-fermion-fermion vertex:
\begin{eqnarray}
\mathcal{L}_{\delta H\bar{\psi}_c\psi_c}&\sim&
\left(\frac{1}{\mathcal{V}^{1/3}}\right)\delta H\bar{\psi}_c\psi_c, \\
\mathcal{L}_{\chi\delta H\bar{\psi}_c\psi_c}&\sim&
\left(\frac{1}{M_P\mathcal{V}^{1/3}}\right)\chi\delta H\bar{\psi}_c\psi_c, \\
\mathcal{L}_{\Phi\delta
H\bar{\psi}_c\psi_c}&\sim&\left(\frac{1}{M_P\mathcal{V}^{5/6}}\right)
\Phi\delta H\bar{\psi}_c\psi_c.
\end{eqnarray}
We notice that for $T>M_{EW}$ the fermions are massless since
$\langle H\rangle=0$, and so the two direct moduli couplings to
ordinary matter particles (\ref{ein}) and (\ref{zwei}) are absent.
\bigskip

$\bullet \textbf{ Couplings to Higgs Fields}$

\bigskip

The form of the un-normalised kinetic and mass terms for the Higgs
from the supergravity Lagrangian, reads:
\begin{equation}
\mathcal{L}_{Higgs}=\tilde{K}_{\bar{H}H}\partial_{\mu}H
\partial^{\mu}\bar{H}
-\tilde{K}_{\bar{H}H} \left(\hat{\mu}^2+m_0^2\right)H\bar{H},
\label{LagraHiggs}
\end{equation}
where $H$ denotes a Higgs field (either $H_u$ or $H_d$), and
$\hat{\mu}$ and $m_0$ are the canonically normalised
supersymmetric $\mu$-term and SUSY breaking scalar mass
respectively. Their volume dependence, in the dilute flux limit,
is \cite{SoftSUSY}:
\begin{equation}
|\hat{\mu}|\sim m_0\sim\frac{M_P}{\mathcal{V}\ln\mathcal{V}}.
\label{mu}
\end{equation}
In addition to (\ref{LagraHiggs}), there is also a mixing term of
the form
\begin{equation}
\mathcal{L}_{Higgs\text{ \ }mix}=Z\left(\partial_{\mu}H_d
\partial^{\mu}H_u+\partial_{\mu}\bar{H}_d
\partial^{\mu}\bar{H}_u\right)-\tilde{K}_{\bar{H}H}B\hat{\mu}\left(H_d
H_u+\bar{H}_d \bar{H}_u\right), \label{LagraHiggsMix}
\end{equation}
with
\begin{equation}
B\hat{\mu}\sim m_0^2. \label{Bmu}
\end{equation}
However given that we are interested only in the leading order
volume scaling of the Higgs coupling to the moduli, we can neglect
the $\mathcal{O}(1)$ mixing of the \textit{up} and \textit{down}
components, and focus on the simple Lagrangian:
\begin{eqnarray}
\mathcal{L}_{Higgs}&=&\tilde{K}_{\bar{H}H}\left(\partial_{\mu}H
\partial^{\mu}\bar{H}
-\frac{M_P^2}{\left(\mathcal{V}\ln\mathcal{V}\right)^2}H\bar{H}\right)
\notag \\
&=&-\frac{1}{2}\tilde{K}_{\bar{H}H}\left[ \bar{H}\left( \square
+\frac{M_P^2}{\left( \mathcal{V}\ln \mathcal{V}\right)
^{2}}\right) H+H\left( \square +\frac{M_P^2}{\left( \mathcal{V}\ln
\mathcal{V} \right) ^{2}}\right) \bar{H}\right],
\label{LagrangianHiggs}
\end{eqnarray}
where we have integrated by parts. We now expand
$\tilde{K}_{\bar{H}H}$ and
$\left(\mathcal{V}\ln\mathcal{V}\right)^{-2}$ and get:
\begin{eqnarray}
\mathcal{L}_{Higgs} &\simeq &-\frac{1}{2}K_{0}\left(
1+\frac{1}{3}\frac{ \delta \tau _{s}}{\langle \tau _{s}\rangle
}-\frac{\delta \tau _{b}}{\langle \tau _{b}\rangle }\right) \left[
\bar{H}\left( \square +m_{H}^{2}\left( 1-3\frac{\delta \tau
_{b}}{\langle \tau _{b}\rangle }
\right) \right) H+\right.   \notag \\
&&\left. H\left( \square +m_{H}^{2}\left( 1-3\frac{\delta \tau
_{b}}{\langle \tau _{b}\rangle }\right) \right) \bar{H} \right],
\end{eqnarray}
where
$K_0=\langle\tau_s\rangle^{1/3}\langle\mathcal{V}\rangle^{-2/3}$
and the Higgs mass is given by
\begin{equation}
m_{H}\simeq\frac{M_P}{\langle\mathcal{V}\rangle\ln\langle
\mathcal{V}\rangle}.
\end{equation}
Now canonically normalising the scalar kinetic terms $H\rightarrow
H_c=\sqrt{K_0}H$, we obtain
\begin{eqnarray}
\mathcal{L}_{Higgs} &=&-\frac{1}{2}\left[ \bar{H}_c\left( \square
+m_{H}^{2}\right) H_c+H_c\left( \square
+m_{H}^{2}\right) \bar{H}_c\right] \\
&&-\frac{1}{2}\left( \frac{1}{3}\frac{\delta \tau
_{s}}{\langle\tau _{s}\rangle }-\frac{\delta \tau _{b}}{\langle
\tau _{b}\rangle }\right)\left[ \bar{H}_c\left( \square
+m_{H}^{2}\right) H_c+H_c\left( \square +m_{H}^{2}\right)
\bar{H}_c\right] +3\frac{\delta \tau _{b}}{\langle \tau
_{b}\rangle }m_{H}^{2}\bar{H}_c H_c. \notag
\end{eqnarray}
The second term in the previous expression does not contribute to
scattering amplitudes since Feynman amplitudes vanish for final
states satisfying the equations of motion. Thus the
\textit{dimensionful} moduli couplings to Higgs fields arise only
from the third term once we express $\delta\tau_b$ in terms of
$\Phi$ and $\chi$ using (\ref{big}). The final result is
\begin{eqnarray}
\mathcal{L}_{\Phi \bar{H}_{c}H_{c}} &\sim &\left(
\frac{m_{H}^{2}}{M_{P} \sqrt{\mathcal{V}}}\right) \Phi
\bar{H}_{c}H_{c}\sim \left( \frac{M_{P}}{ \mathcal{V}^{5/2}(\ln
{\mathcal{V}})^{2}}\right) \Phi \bar{H}_{c}H_{c},
\label{Higgs1} \\
\mathcal{L}_{\chi \bar{H}_{c}H_{c}} &\sim &\left(
\frac{m_{H}^{2}}{M_{P}} \right) \chi \bar{H}_{c}H_{c}\sim \left(
\frac{M_{P}}{\mathcal{V}^{2}(\ln { \mathcal{V}})^{2}}\right) \chi
\bar{H}_{c}H_{c}.  \label{Higgs2}
\end{eqnarray}

\subsection{Moduli couplings to supersymmetric particles}
\label{Sec:ModCouplSUSY}

We shall now work out the moduli couplings to gauginos, SUSY
scalars and Higgsinos. Given that we are interested in thermal
corrections at high temperatures, we shall focus on $T>M_{EW}$.
Thus we can neglect the mixing of Higgsinos with gauginos into
charginos and neutralinos, which takes place at lower energies due
to EW symmetry breaking.

\bigskip

$\bullet \textbf{ Couplings to Gauginos}$

\bigskip

The relevant part of the supergravity Lagrangian involving the
gaugino kinetic terms and their soft masses looks like
\begin{equation}
\mathcal{L}_{gaugino}\simeq
\frac{\tau_s}{M_P}\bar{\lambda'}i\bar{\sigma}^{\mu}\partial_{\mu}\lambda'+\frac{F^s}{2}\left(\lambda'\lambda'+h.c.\right),
\label{Ploj}
\end{equation}
where in the limit of dilute world-volume fluxes on the D7-brane,
the gaugino mass is given by $M_{1/2}=\frac{F^{s}}{2\tau_s}$
\cite{SoftSUSY}. Now if the small modulus supporting the MSSM is
stabilised via non-perturbative corrections, then the
corresponding F-term scales as
\begin{equation}
F^{s}\simeq\frac{\tau_s}{\mathcal{V}\ln\mathcal{V}}. \label{F}
\end{equation}
Notice that the suppression factor
$\ln\mathcal{V}\sim\ln(M_P/m_{3/2})$ in (\ref{F}) would be absent
in the case of perturbative stabilisation of the MSSM cycle
\cite{ccq2}. Let us expand $\tau_s$ around its VEV and get:
\begin{equation}
\mathcal{L}_{gaugino}\simeq
\langle\tau_s\rangle\left[\bar{\lambda'}i\bar{\sigma}^{\mu}\partial_{\mu}\lambda'+\frac{1}{2}\frac{M_P}{\mathcal{V}\ln\mathcal{V}}
\left(\lambda'\lambda'+h.c.\right)\right]+\frac{\delta\tau_s}{M_P}
\left[\bar{\lambda'}i\bar{\sigma}^{\mu}\partial_{\mu}\lambda'+\frac{M_P}{\langle\mathcal{V}\rangle\ln\langle\mathcal{V}\rangle}
\frac{\left(\lambda'\lambda'+h.c.\right)}{2}\right].
\label{labello}
\end{equation}
We need now to expand also $\tau_b$ around its VEV in the first
term of (\ref{labello}):
\begin{equation}
\frac{1}{\mathcal{V}\ln\mathcal{V}}\simeq\frac{1}{\tau_b^{3/2}\ln\mathcal{V}}
\simeq\frac{1}{\langle\mathcal{V}\rangle\ln\langle\mathcal{V}\rangle}
\left(1-\frac{3}{2}\frac{\delta\tau_b}{\langle\tau_b\rangle}+...\right),
\end{equation}
and canonically normalise the gaugino kinetic terms
$\lambda'\rightarrow\lambda=\sqrt{\langle\tau_s\rangle}\lambda'$.
At the end we obtain:
\begin{equation}
\mathcal{L}_{gaugino}\simeq
\bar{\lambda}i\bar{\sigma}^{\mu}\partial_{\mu}\lambda+\frac{M_P}
{\langle\mathcal{V}\rangle\ln\langle\mathcal{V}\rangle}
\frac{\left(\lambda\lambda+h.c.\right)}{2} +
\frac{\left(\lambda\lambda+h.c.\right)}
{2\langle\mathcal{V}\rangle\ln\langle\mathcal{V}\rangle}\left(\frac{\delta\tau_s}
{\langle\tau_s\rangle}
-\frac{3}{2}\frac{\delta\tau_b}{\langle\tau_b\rangle}\right)
+\frac{\delta\tau_s}{\langle\tau_s\rangle M_P
}\bar{\lambda}i\bar{\sigma}^{\mu}\partial_{\mu}\lambda.
\label{labelloo}
\end{equation}
From (\ref{labelloo}) we can immediately read off the gaugino
mass:
\begin{equation}
M_{1/2}\simeq\frac{M_P}{\langle\mathcal{V}\rangle\ln\langle\mathcal{V}\rangle}
\simeq\frac{F^s}{\tau_s}\sim\frac{m_{3/2}}{\ln\left(M_P/m_{3/2}\right)}.
\end{equation}
Let us now rewrite (\ref{labelloo}) as:
\begin{equation}
\mathcal{L}_{gaugino}\simeq
\left(1+\frac{\delta\tau_s}{\langle\tau_s\rangle M_P}\right)\left[
\bar{\lambda}i\bar{\sigma}^{\mu}\partial_{\mu}\lambda
+\frac{M_{1/2}}{2}\left(\lambda\lambda+h.c.\right)\right]
-\frac{3}{4}\frac{\delta\tau_b}{\langle\tau_b\rangle}\frac{M_{1/2}}{M_P}
\left(\lambda\lambda+h.c.\right). \label{labellooo}
\end{equation}
We shall now focus only on the last term in (\ref{labellooo})
since it is the only one that contributes to decay rates. In fact,
Feynman amplitudes with on-shell final states that satisfy the
equations of motion, are vanishing. Using (\ref{big}), we finally
obtain the following \textit{dimensionless} couplings:
\begin{eqnarray}
\mathcal{L}_{\Phi \lambda\lambda } &\sim &\left(
\frac{M_{1/2}}{M_{P} \sqrt{\mathcal{V}}}\right) \Phi
\lambda\lambda \sim \left( \frac{1}{ \mathcal{V}^{3/2}\ln
{\mathcal{V}}}\right) \Phi \lambda\lambda ,
\label{rr} \\
\mathcal{L}_{\chi \lambda\lambda } &\sim &\left(
\frac{M_{1/2}}{M_{P}} \right) \chi \lambda\lambda \sim \left(
\frac{1}{\mathcal{V}\ln \mathcal{V}}\right) \chi \lambda\lambda .
\label{rrr}
\end{eqnarray}

\bigskip

$\bullet \textbf{ Couplings to SUSY Scalars}$

\bigskip

The form of the un-normalised kinetic and soft mass terms for SUSY
scalars from the supergravity Lagrangian, reads:
\begin{equation}
\mathcal{L}_{scalars}=\tilde{K}_{\alpha\bar{\beta}}\partial_{\mu}C^{\alpha}
\partial^{\mu}\bar{C}^{\bar{\beta}}
-\frac{\tilde{K}_{\alpha\bar{\beta}}}{\left(\mathcal{V}\ln\mathcal{V}\right)^2}
C^{\alpha}\bar{C}^{\bar{\beta}}.
\label{Lagra}
\end{equation}
Assuming diagonal K\"{a}hler metric for matter fields
\begin{equation}
\tilde{K}_{\alpha\bar{\beta}}=\tilde{K}_{\alpha}\delta_{\alpha\bar{\beta}},
\end{equation}
the initial Lagrangian (\ref{Lagra}) simplifies to
\begin{eqnarray}
\mathcal{L}_{scalars} &=&\tilde{K}_{\alpha }\left( \partial _{\mu
}C^{\alpha }\partial ^{\mu }\bar{C}^{\bar{\alpha}}-\frac{1}{\left(
\mathcal{V}\ln \mathcal{V}\right) ^{2}}C^{\alpha
}\bar{C}^{\bar{\alpha}}\right)
\notag \\
&=&-\frac{1}{2}\tilde{K}_{\alpha }\left[
\bar{C}^{\bar{\alpha}}\left( \square +\frac{1}{\left(
\mathcal{V}\ln \mathcal{V}\right) ^{2}}\right) C^{\alpha
}+C^{\alpha }\left( \square +\frac{1}{\left( \mathcal{V}\ln
\mathcal{V} \right) ^{2}}\right) \bar{C}^{\bar{\alpha}}\right].
\label{Lag}
\end{eqnarray}
We note that (\ref{Lag}) is of exactly the same form as the Higgs Lagrangian
(\ref{LagrangianHiggs}). This is not surprising since for
temperatures $T>M_{EW}$, the Higgs behaves effectively as a SUSY
scalar with mass of the order the scalar soft mass: $m_{H}\sim
m_0$. Thus we can read off immediately the \textit{dimensionful}
moduli couplings to the canonically normalised SUSY scalars
$\varphi$ from (\ref{Higgs1}) and (\ref{Higgs2}):
\begin{eqnarray}
\mathcal{L}_{\Phi \bar{\varphi}\varphi} &\sim &\left(
\frac{m_{0}^{2}}{M_{P} \sqrt{\mathcal{V}}}\right) \Phi
\bar{\varphi}\varphi\sim \left( \frac{M_{P}}{
\mathcal{V}^{5/2}(\ln
{\mathcal{V}})^{2}}\right) \Phi \bar{\varphi}\varphi, \\
\mathcal{L}_{\chi \bar{\varphi}\varphi} &\sim &\left(
\frac{m_{0}^{2}}{M_{P}} \right) \chi \bar{\varphi}\varphi\sim
\left( \frac{M_{P}}{\mathcal{V}^{2}(\ln {
\mathcal{V}})^{2}}\right) \chi \bar{\varphi}\varphi.
\end{eqnarray}

\bigskip

$\bullet \textbf{ Couplings to Higgsinos}$

\bigskip

The relevant part of the supergravity Lagrangian involving the
Higgsino kinetic terms and their supersymmetric masses looks like:
\begin{equation}
\mathcal{L}_{Higgsino}\simeq
\tilde{K}_{\bar{\tilde{H}}\tilde{H}}\left[
\bar{\tilde{H}}_ui\bar{\sigma}^{\mu}\partial_{\mu}\tilde{H}_u
+\bar{\tilde{H}}_di\bar{\sigma}^{\mu}\partial_{\mu}\tilde{H}_d
+\hat{\mu}\left(\tilde{H}_u\tilde{H}_d+h.c.\right)\right].
\end{equation}
After diagonalising the supersymmetric Higgsino mass term, we end
up with a usual Lagrangian of the form:
\begin{equation}
\mathcal{L}_{Higgsino}\simeq
\tilde{K}_{\bar{\tilde{H}}\tilde{H}}\left[
\bar{\tilde{H}}i\bar{\sigma}^{\mu}\partial_{\mu}\tilde{H}+\hat{\mu}\left(\tilde{H}\tilde{H}+h.c.\right)\right],
\end{equation}
where $\tilde{H}$ denotes collectively both the Higgsino mass
eigenstates, which are the result of a mixing between the
\textit{up} and \textit{down} gauge eigenstates. We recall also
that since we are focusing on temperatures above the EWSB scale,
we do not have to deal with any mixing between Higgsinos and
gauginos to give neutralinos and charginos. Expanding the
K\"{a}hler metric (\ref{chiralMetric}) and the $\mu$-term
(\ref{mu}), we obtain:
\begin{equation}
\mathcal{L}_{Higgsino} \simeq K_{0}\left( 1+\frac{1}{3}\frac{
\delta \tau _{s}}{\langle \tau _{s}\rangle }-\frac{\delta \tau
_{b}}{\langle \tau _{b}\rangle }\right)
\left[\bar{\tilde{H}}i\bar{\sigma}^{\mu}\partial_{\mu}\tilde{H}+
\frac{m_{\tilde{H}}}{2}\left( 1-\frac{3}{2}\frac{\delta \tau
_{b}}{\langle \tau _{b}\rangle } \right)
\left(\tilde{H}\tilde{H}+h.c.\right)\right],
\end{equation}
where
$K_0=\langle\tau_s\rangle^{1/3}\langle\mathcal{V}\rangle^{-2/3}$
and the physical Higgsino mass is of the same order of magnitude
of the soft SUSY masses:
\begin{equation}
m_{\tilde{H}}\simeq\frac{M_P}{\langle\mathcal{V}\rangle\ln\langle
\mathcal{V}\rangle}\simeq M_{1/2}.
\end{equation}
Now canonically normalising the scalar kinetic terms
$\tilde{H}\rightarrow\tilde{H}_c=\sqrt{K_0}\tilde{H}$, we end up
with:
\begin{eqnarray}
\mathcal{L}_{Higgsino} &=&\left(1+\frac{1}{3}\frac{ \delta \tau
_{s}}{\langle \tau _{s}\rangle }-\frac{\delta \tau _{b}}{\langle
\tau _{b}\rangle }\right)\left[
\bar{\tilde{H}}_{c}i\bar{\sigma}^{\mu }\partial _{\mu
}\tilde{H}_{c}+\frac{m_{\tilde{H}}}{2}\left(\tilde{H}_{c}\tilde{H}_c+h.c.\right) \right] \notag \\
&&-\frac{3}{4}\frac{\delta \tau _{b}}{\langle \tau _{b}\rangle}
m_{ \tilde{H}}\left(\tilde{H}_c\tilde{H}_c+h.c.\right).
\label{secco}
\end{eqnarray}
Writing everything in terms of $\Phi$ and $\chi$, from the last
term of (\ref{secco}), we obtain the following
\textit{dimensionless} couplings:
\begin{eqnarray}
\mathcal{L}_{\chi\tilde{H}_c\tilde{H}_c}&\sim&\left(\frac{m_{\tilde{H}}}{M_P}\right)
\chi\tilde{H}_c\tilde{H}_c
\sim\left(\frac{1}{\mathcal{V}\ln\mathcal{V}}\right)
\chi\tilde{H}_c\tilde{H}_c, \\
\mathcal{L}_{\Phi\tilde{H}_c\tilde{H}_c}&\sim&\left(\frac{m_{\tilde{H}}}{M_P\sqrt{\mathcal{V}}}\right)
\Phi\tilde{H}_c\tilde{H}_c\sim\left(\frac{1}{\mathcal{V}^{3/2}\ln\mathcal{V}}\right)
\Phi\tilde{H}_c\tilde{H}_c.
\end{eqnarray}

\subsection{Moduli self couplings}
\label{SELF}

In this Section we shall investigate if moduli reach thermal
equilibrium among themselves. In order to understand this issue,
we need to compute the moduli self interactions, which can be
obtained by first expanding the moduli fields around their VEV
\begin{equation}
\tau_{i}=\langle\tau_i\rangle+\delta\tau_i,
\end{equation}
and then by expanding the potential around the LARGE Volume vacuum
as follows:
\begin{equation}
V=V(\langle \tau _{s}\rangle ,\langle \tau _{b}\rangle
)+\frac{1}{2}\left. \frac{\partial ^{2}V}{\partial \tau
_{i}\partial \tau _{j}}\right\vert _{\min }\delta \tau _{i}\delta
\tau _{j}+\frac{1}{3!}\left. \frac{\partial ^{3}V}{\partial \tau
_{i}\partial \tau _{j}\partial \tau _{k}}\right\vert _{\min
}\delta \tau _{i}\delta \tau _{j}\delta \tau _{k}+....
\label{expandi}
\end{equation}
We then concentrate on the trilinear terms which can be read off
from the third term of (\ref{expandi}). We neglect the
$\mathcal{O}(\delta\tau_i^4)$ terms since the strength of their
couplings will be subleading with respect to the
$\mathcal{O}(\delta\tau_i^3)$ terms since one has to take a
further derivative which produces a suppression factor. Taking the
third derivatives and then expressing these self-interactions in
terms of the canonically normalised fields
\begin{eqnarray*}
\delta \tau _{b} &\sim &\mathcal{O}\left( \mathcal{V}^{1/6}\right)
\Phi +
\mathcal{O}\left( \mathcal{V}^{2/3}\right) \chi , \\
\delta \tau _{s} &\sim &\mathcal{O}\left( \mathcal{V}^{1/2}\right)
\Phi + \mathcal{O}\left(1\right) \chi,
\end{eqnarray*}
we end up with the following Lagrangian terms at leading order in
a large volume expansion:
\begin{eqnarray}
\mathcal{L}_{\Phi ^{3}} &\simeq
&\frac{M_{P}}{\mathcal{V}^{3/2}}\Phi ^{3}, \text{ \ \ \ \ \ \ \ \
\ }\mathcal{L}_{\Phi ^{2}\chi}\simeq \frac{M_{P}}{
\mathcal{V}^{2}}\chi \Phi ^{2}, \\
\mathcal{L}_{\chi^{2}\Phi} &\simeq
&\frac{M_{P}}{\mathcal{V}^{5/2}}\Phi \chi ^{2},\text{ \ \ \ \ \ \
\ \ \ }\mathcal{L}_{\chi ^{3}}\simeq \frac{M_{P}
}{\mathcal{V}^{3}}\chi ^{3}. \label{coupl}
\end{eqnarray}

\end{document}